\documentclass[
 onecolumn,
 amsmath,amssymb,
floatfix,superscriptaddress
]{revtex4-2}
\usepackage{graphicx}
\usepackage{dcolumn}
\usepackage{tikz}
\usepackage{placeins}
\usepackage{bm}
\usepackage{amsmath}
\usepackage{amssymb}
\usepackage{hyperref}

\begin{document}
\preprint{APS/123-QED}
\title{Stochastic elastohydrodynamics of contact and coarsening during membrane adhesion}
\author{Vira Dhaliwal}
\affiliation{Mechanics Division, Department of Mathematics, University of Oslo, 0316 Oslo, Norway}
\author{Jingbang Liu}
\affiliation{Mechanics Division, Department of Mathematics, University of Oslo, 0316 Oslo, Norway}

\author{Andreas Carlson}
\email{acarlson@math.uio.no}
\affiliation{Mechanics Division, Department of Mathematics, University of Oslo, 0316 Oslo, Norway}

\begin{abstract}
Contact between fluctuating, fluid-lubricated soft surfaces is prevalent in engineering and biological systems, a process starting with adhesive contact, which can give rise to complex coarsening dynamics. One representation of such a system, which is relevant to biological membrane adhesion, is a fluctuating elastic interface covered by adhesive molecules that bind and unbind to a solid substrate across a narrow gap filled with a viscous fluid. This flow is described by the stochastic elastohydrodynamics thin-film equation, which combines the effects of viscous nanometric thin film flow, elastic membrane properties, adhesive springs, and thermal fluctuations. The average time it takes the fluctuating elastic membrane to adhere is predicted by the rare event theory, increasing exponentially with the square of the initial gap height. Numerical simulations reveal a phase separation of membrane domains driven by the binding and unbinding of adhesive molecules. The coarsening process displays close similarities to classical Ostwald ripening; however, the inclusion of hydrodynamics affects power-law growth. In particular, we identify a new bending-dominated coarsening regime, which is slower than the well-known tension-dominated case.
\end{abstract}
\maketitle

\section{\label{sec:level1}Introduction}
Adhesion between soft fluctuating surfaces is found in a range of engineering applications and in biological systems. Complex lifeforms rely on cells adhering to each other, facilitated by the binding of membrane-anchored adhesive molecules across the gap between the membranes of the cells. The dynamics of adhesion involve a rich interplay of membrane deformation, chemical kinetics of the aforementioned molecular binding, as well as fluid flow in the narrow space between the membranes, which can give rise to intricate dynamical processes such as the coarsening of adhesion patches. A number of essential physiological processes depend on adhesion dynamics, such as cadherin-mediated adhesion \citep{RoyBerx_cadherin_2008}, lumen formation between cells in a growing embryo \citep{DumortierLVSTurlierMaitre2019} and the immune synapse which facilitates an immunological response \citep{grakoui1999,dustin_cooper_2000}. Experiments and numerical results have shown that the features of these phenomena can be passive processes only relying on the physical forces involved \citep{qi2001,Dustin2010,LeVerge-SerandourTurlier2021}. 

One way to describe the adhesion of membranes, is to adopt the Helfrich model with an adhesion potential~\citep{seifertAdhesionVesicles1990,smithEffectsPullingForce2003,smithForceinducedGrowthAdhesion2008,agrawalMechanicsMembraneMembrane2011,hillContactlineBendingEnergy2024}. This approach successfully predicts the equilibrium shape of the membrane, but neglects the motion of the extracellular fluid in the cleft between the two membranes. In such nanometrically thin but micrometrically wide channels, however, the forces required to squeeze the viscous extracellular fluid are not negligible, motivating a viscous thin film description of the flow~\citep{leongAdhesiveDynamicsLubricated2010}. The lubrication theory allows for a relatively simple inclusion of forces due to membrane deformation, molecule/protein binding, as well as thermal fluctuations \citep{carlson_mahadevan_synapse_2015,carlson2015_protadhesion_physfluids}. Membranes resist lateral deformation due to membrane tension $\gamma$~\citep{evansElasticAreaCompressibility1976}, as it increases their free energy in a manner analogous to interfacial surface tension, whose influence on the dynamics of thin liquid films has been studied extensively ~\citep{OronBankoff1997,craster2009}, particularly in the case of the dewetting of nanoscale thin liquid films ~\citep{ZhangLister1999,MeckeRauscher_2005,grun2006,nguyenCoexistenceSpinodalInstability2014,zhangMolecularSimulationThin2019,zhaoFluctuationdrivenDynamicsNanoscale2022}. In addition, membranes of finite width $d$ resist bending~\citep{evansBendingResistanceChemically1974}, as one side is compressed and the other side is stretched, which is characterized by a bending modulus $B=Ed^3/12(1-\nu^2)$, where $E$ is the Young's modulus and $\nu$ is the Poisson ratio. Both tension and bending resist the deformation of a flat membrane, but there are subtle differences in how they act. A bendocapillary length $l_{BC}=\sqrt{B/\gamma}$ can be obtained by balancing the forces from tension and  bending,  which is around $100$ nm for the properties of most cell membranes, and bending dominates for length scales smaller than the critical length scale~\citep{DesernoDiffCurv,RomanBico2010}.

For adhesion to occur, the membranes must first come close enough to each other to allow the adhesive molecules to start to form bonds. In the absence of directed motion due to active cytoskeletal forces or protein-membrane interactions~\citep{yuanMembraneBendingProtein2021,lieseMembraneShapeRemodeling2021}, the forces required to push out the extracellular fluid in the channel can be attributed to thermal fluctuations~\citep{aartsDirectVisualObservation2004,delgado-buscalioniHydrodynamicsNanoscopicCapillary2008} as well as the other fluctuations inherent to living matter \citep{GuoWeitz2014,GuptaGuo2017}. The fluctuations in the width of the channel, although small in amplitude and random in direction, can still bring membranes close enough to initiate adhesive molecule/protein binding if given sufficient time. This process is similar to the spontaneous thermal dewetting of a linearly stable thin viscous film coated on a solid substrate, where thermal fluctuations are needed to bring the liquid free surface close enough to the solid substrate for the disjoining pressure to rupture the film, for which the average waiting time for rupture can be predicted by rare-event theory \citep{SprittlesJBLGrafke2023,liuMeanFirstPassage2024}.

After the initial binding of adhesive molecules, the adhesion patches grow in size. If more than one type of adhesive molecule is involved, the adhesion patches can separate into distinct regions (phases) where one specific type of molecule is bound, as is seen in the immune synapse, receptor tyrosine kinases~\citep{janesConceptsConsequencesEph2012,caseRegulationTransmembraneSignaling2019} and simulation of membrane adhered to heterogenous substrate~\citep{shelbyMembranePhaseSeparation2023}. Excess fluid from the adhesion patches is squeezed into the regions where molecules are unbound, which further increases the distance between membranes, creating pockets of fluids known as lumens, that can be observed in mammalian embryo development \citep{DumortierLVSTurlierMaitre2019}. These lumens can also be reproduced in reconstituted systems by applying an osmotic shock to a giant unilamellar vesicle (GUV) that is adhered to a supported lipid bilayer \citep{DinetArroyoStaykova2023}. In these systems, the separated phases often undergo a passive, physically driven coarsening process where small patches of diminish, while larger patches grow. These dynamics are reminiscent of other phase separation processes occurring in the cooling of metal alloys \citep{AllenCahn1979,KomuraTakeda1985,Bray1994,LivetSutton2001}, liquid-liquid phase separation in biological systems \citep{SuParikh2024}, as well as in droplet aggregation when a thin liquid or polymer film is adhered to a solid substrate by an attractive potential due to intermolecular forces \citep{DerridaYekutelia1991,OttoRumpSlepcev2006,GrattonWitelski2008,LalSutton2020}. In those systems an effective interfacial tension drives the coarsening process, providing a well-established $t^{1/3}$ power law \citep{Bray1994} for the growth the characteristic length scale $L_c$. For the coarsening observed in biological membranes, while membrane tension could give rise to the same $1/3$ power law, the effect of membrane bending has not been studied and will potentially introduce different dynamics. Here, we will show that the fluid flow affects the predictions for the elastohydrodynamic coarsening process.

\section{Mathematical model and numerical methods} 
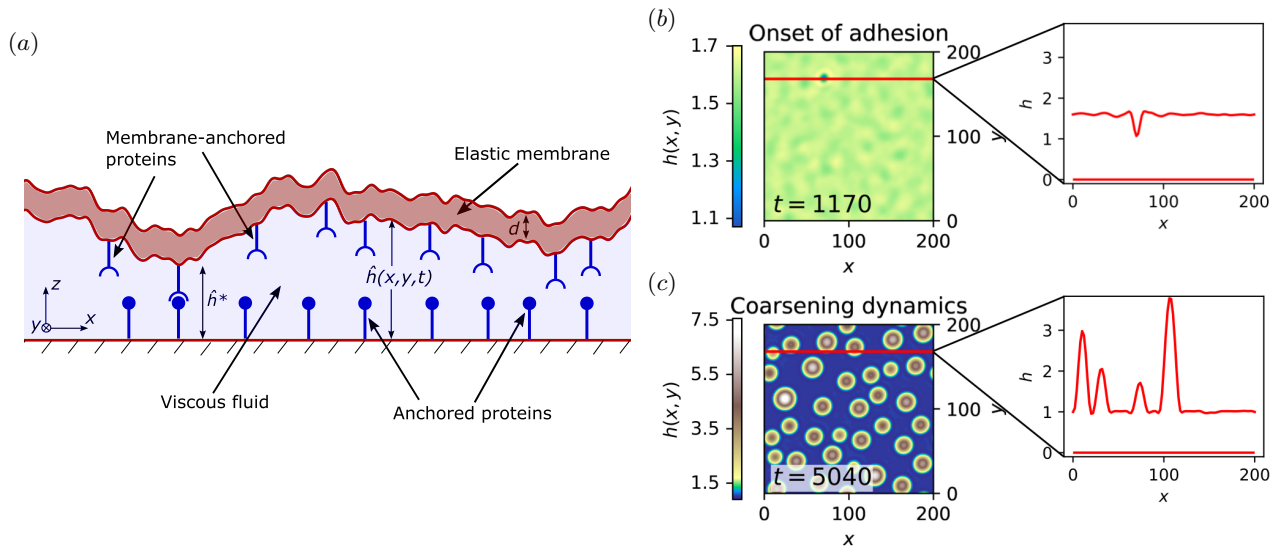
\begin{figure}
	 \centering
	\begin{tikzpicture}
   	\draw (0, 0) node[inner sep=0] (fig) {\includegraphics[width=0.45\textwidth]{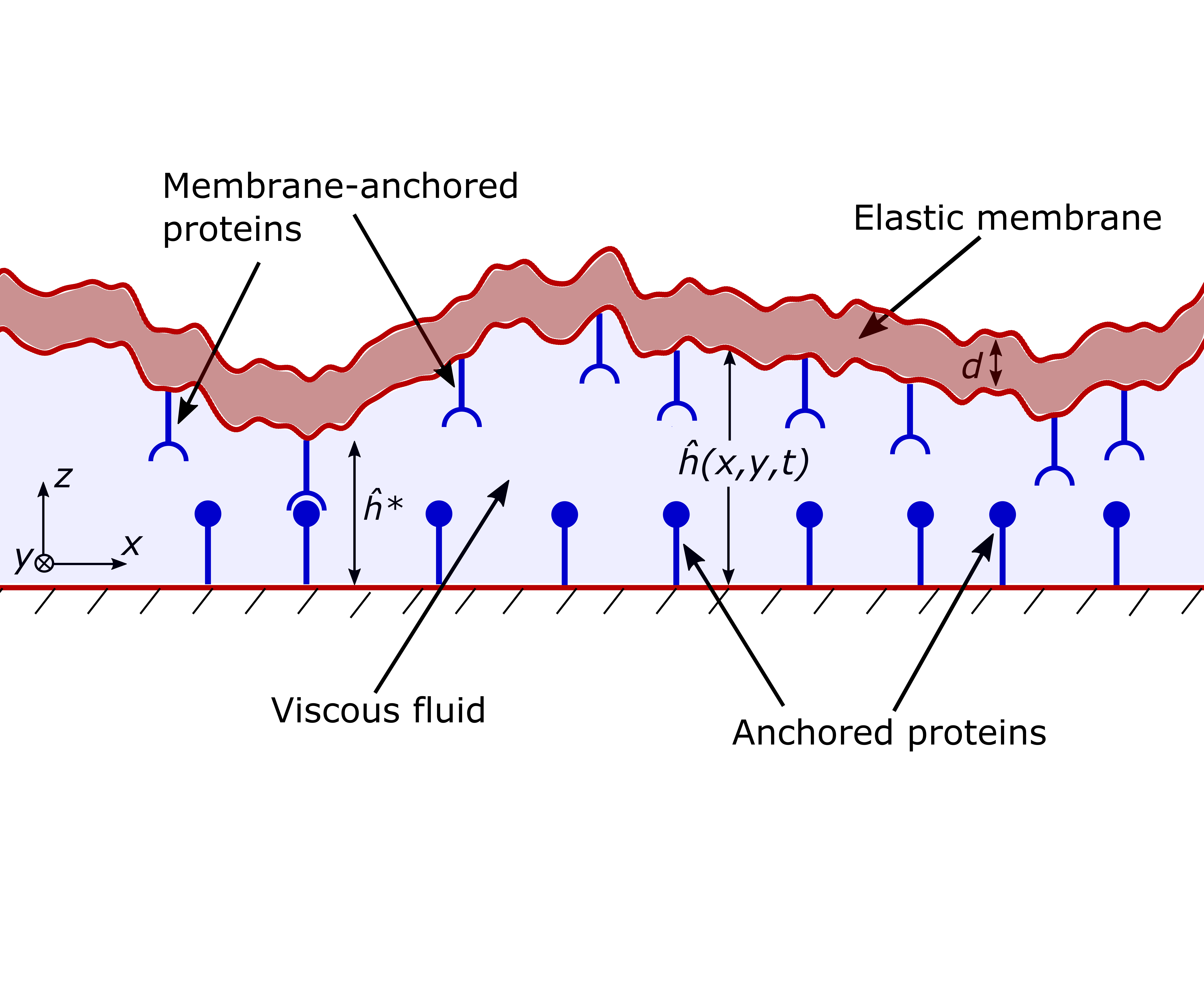}};
   	\node[] at (fig.north west){$(a)$};
 	\end{tikzpicture}
 	\begin{tikzpicture}
   	\draw (1, 1) node[inner sep=0] (fig) {\includegraphics[width=0.45\textwidth]{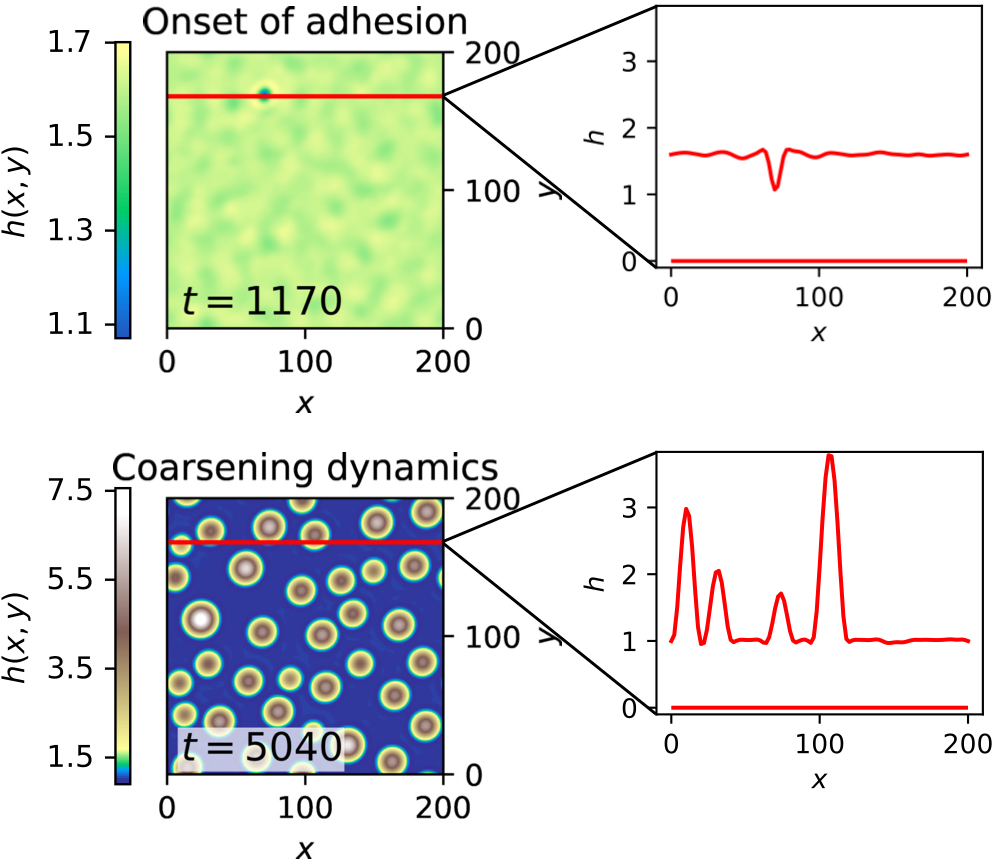}};
   	\node[] at (fig.north west){$(b)$};
    \node[] at (fig.west){$(c)$};
 	\end{tikzpicture}

	\caption{$(a)$ A sketch of an elastic membrane with thickness $d$ in close proximity to a rigid wall separated by a thin layer of viscous fluid of height $\hat{h}(\hat{x},\hat{y},\hat{t})$. Membrane molecules may bind across the channel only if their distance is below a critical value $h^*$. $(b)$ Left: a contour map of the non-dimensional height profile $h(x,y,t)$ at the time $t^*$, which is the onset of adhesion between the membrane and solid surface. Data are shown for a membrane with initial height $h_0=1.6$ with fluctuation intensity $Q_B=\frac{1}{l}\sqrt{\frac{2 k_B T }{B^{1/2}(\kappa c_0)^{3/2}}}=0.01$, with $k_BT$ the thermal energy, $B$ the bending stiffness, $c_0$ the equilibrium concentration and $\kappa$ the molecule spring stiffness coefficient. Right: a cross section of the profile along the red line in the contour map. $(c)$ Left: a contour map at a later time when most of the membrane is bound, but liquid is collected in unbound patches. Right: a cross section of the profile along the red line in the contour map, illustrating the formation of blisters during the coarsening process.}
	\label{fig:Schematics}
\end{figure}\label{sec:model}

To simultaneously study the fluid flow, elastic membrane bending, protein-like binding and stochastic fluctuations during membrane adhesion, we turn to the elastohydrodynamic thin-film equation~\citep{WilliamsDavis1982,carlson2016adhesiontouchdown}, describing the physical scenario of Fig. \ref{fig:Schematics}$(a)$, i.e., a thin layer of viscous liquid confined by an elastic membrane. For simplicity as well as reflecting in vitro experimental conditions in which a GUV interacts with a supported lipid bilayer \citep{FenzAna-Suncana2017,DinetArroyoStaykova2023}, the substrate is static while the upper membrane ``rests'' on the viscous fluid film at height $\hat{h}(\hat{x},\hat{y},\hat{t})$ (where the hats indicate dimensional variables) and can move vertically. As in the biological and synthetic systems described previously \citep{DumortierLVSTurlierMaitre2019,DinetArroyoStaykova2023}, the film height $\hat{h}$ (which is typically in the range of tens of nanometers), is much smaller than the length of the domain in the lateral directions, $L$ (typically several microns). Across the gap, membrane-bound proteins may bind or unbind to the solid surface. 




\subsection{\label{subsec:Equations}Stochastic thin film equation}
By assuming a small aspect ratio of the viscous channel, $\hat{h}/L\ll 1$, one can apply the lubrication approximation to describe the flow of a Newtonian fluid with viscosity $\mu$ in the channel, leading to a parabolic velocity profile \citep{batchelor_2000}. A random stress tensor is introduced in the momentum equations to account for the thermal fluctuations in the fluid, which are significant on the relevant nanometric scale~\citep{landau_lifshitz_fluids}. By imposing no-slip and kinematic boundary conditions at the membrane, one arrives at the following stochastic thin-film equation~\citep{duran2019,MeckeRauscher_2005,davidovitch2005,grun2006} that can be used to describe $\hat{h}(\hat{x},\hat{y},\hat{t})$:
\begin{equation}
    \frac{\partial \hat{h}(\hat{x},\hat{y},\hat{t})}{\partial \hat{t}} = \hat{\nabla}\cdot\left(\frac{\hat{h}^3(\hat{x},\hat{y},\hat{t})}{12\mu}\hat{\nabla} \hat{p}(\hat{x},\hat{y},\hat{t}) \right) + \sqrt{\frac{k_B T}{6\mu }}\hat{\nabla} \cdot\left(\hat{h}^{3/2}(\hat{x},\hat{y},\hat{t})\boldsymbol{\hat{\eta}}(\hat{x},\hat{y},\hat{t})\right)
    \label{eq:TF_dim}
\end{equation}
where $\hat{\nabla}$ represents the 2D gradient operator $(\partial/\partial \hat{x},\partial/\partial \hat{y})$ and the dependent variable $\hat h$ represents height in the third spatial direction.

The first term on the right hand side of Eq.~\eqref{eq:TF_dim} represents the change in film height due to a lateral flux driven by horizontal pressure gradient $\hat{\nabla} \hat{p}$. The second term on the right hand side incorporates the effect of thermal fluctuations at temperature $T$, by introducing a random flux in accordance with the fluctuation-dissipation theorem and averaged across the channel \citep{MeckeRauscher_2005,davidovitch2005,grun2006}. Here, $\boldsymbol{\hat{\eta}} (\hat{x},\hat{y},\hat{t})$, is a random vector with Gaussian white noise uncorrelated in both time and space, i.e. $\langle\hat{\eta}_i(\hat{x},\hat{y},\hat{t})\rangle=0$ and $\langle\hat{\eta}_i(\hat{x},\hat{y},\hat{t})\hat{\eta}_j(\hat{x}',\hat{y}',\hat{t}')\rangle=\delta_{ij}\delta(\hat{x}-\hat{x}')\delta(\hat{y}-\hat{y}')\delta(\hat{t}-\hat{t}')$, where $\langle\; \rangle$ is the ensemble average, $\delta_{ij}$ is the Kronecker symbol, $\delta$ is the Dirac distribution, and $k_B$ is the Boltzmann constant. 

Here, we note that the nonlinear prefactor $\hat{h}^3/(12\mu)$ arises from viscous resistance to the flow, describing the mobility of the fluid. In fact, Eq. \eqref{eq:TF_dim} can be re-written as a gradient flow \citep{duran2019,CatesLesHouches,SprittlesJBLGrafke2023} of the form
\begin{equation}
    \frac{\partial \hat{h}(\hat{x},\hat{y},\hat{t})}{\partial \hat{t}} = \hat{\nabla}\cdot\left(M(\hat{h})\hat{\nabla} \frac{\delta \hat{F}}{\delta \hat{h}}  + \sqrt{2 k_B TM(\hat{h})}\boldsymbol{\hat{\eta}}\right)
    \label{eq:GradFlow}
\end{equation}
where $M(\hat{h})$ is the mobility, and $\hat{F}[\hat{h}(\hat{x},\hat{y})]$ is an energy functional that gives rise to the pressure. Depending on the problem and assumptions, the mobility can take other forms~\citep{Glasner2008}. For example, in a crowded, narrow section of cytoplasm, one could use Darcy's Law to describe the flow through a porous medium, which reduces the mobility to $\sim \hat{h}$~\citep{GnannPetrache2018}. A slip boundary, on the other hand, would enhance mobility by introducing an additional $\sim \hat{h}^2$ term~\citep{ZhangSprittlesLockerby2020}. The $\sim M^{1/2}$ prefactor of the fluctuation term ensures that detailed balance is satisfied~\citep{duran2019,liuMeanFirstPassage2024}. The free energy $\hat{F}$ depends on what forces drive the flux in the channel, as described in the following sections.

\subsection{\label{subsec:membrane}Membrane dynamics}
Deformation of the membrane, as shown in Fig. \ref{fig:Schematics}$(a)$, results in tension and bending forces, which lead to a change in the pressure $\hat p(\hat{x},\hat{y},\hat{t})$ of the fluid in the channel. As the membrane changes shape, so does the fluid flux in accordance with equation \eqref{eq:TF_dim}. The membrane is idealized as an isotropic elastic solid with Young's modulus $E$, Poisson ratio $\nu$, and thickness $d$ \citep{landau_lifshitz_elasticity}. In the limit of small deflections, i.e., small spatial gradients in $\hat{h}(\hat{x},\hat{y},\hat{t})$, which is natural from the scale separation, the bending and the tension components in the pressure simplify to~\citep{Howell_Kozyreff_Ockendon_2008,Evans1985}:
\begin{equation}
  p_{\text{elastic}}(\hat{x},\hat{y},\hat{t}) =B \hat{\nabla} ^4\hat{h}(\hat{x},\hat{y},\hat{t}) - \gamma \hat{\nabla} ^2\hat{h}(\hat{x},\hat{y},\hat{t})
  \label{eq:membranepressures}
  \end{equation}  
where $B=Ed^3/12(1-\nu^2)$ is the bending rigidity of the membrane and $\gamma$ is the tension coefficient. Generally these two components are both present, but their relative strength will depend on the ratio of the characteristic horizontal length scale of the system, $L$, to the bendocapillary length $l_{BC}= \sqrt{B/\gamma}$~\citep{DesernoDiffCurv}. For $L/l_{BC}\ll 1$, bending should be the prominent driving force, whereas for $L/l_{BC}\gg 1$ we expect membrane tension to dominate. Note that the tension in a membrane can also be described by an integral constraint for the membrane length \citep{KodioVella2017}, and that if the limit of small deflections is exceeded, the Föppl-Von Kármán equations~\citep{audolyetPomeau} can be adapted to give a full description of membrane deformation.

The pressure terms in Eq. \eqref{eq:membranepressures} can be formulated as a gradient flow of Eq. \eqref{eq:GradFlow} as the two contributions can be rewritten as functional derivatives of free energies. The bending energy is $\hat{F}_{\text{bend}}[\hat{h}(\hat{x},\hat{y})] = \int  \frac{B}{2} \left|\hat{\nabla}^2 \hat{F}\right|^2 d\hat{A}$ and the surface energy is $\hat{F}_{\gamma}[\hat{h}(\hat{x},\hat{y})] = \int  \frac{\gamma}{2} \lvert\hat{\nabla} \hat{h}\rvert^2 d\hat{A}$.


\subsection{\label{subsec:Proteins}Dynamics of adhesive molecules}
When proteins or other adhesive molecules bind across the gap between the two membranes, additional forces are generated that contribute to the pressure $\hat p(\hat{x},\hat{y},\hat{t})$  \citep{carlson_mahadevan_synapse_2015,carlson2015_protadhesion_physfluids}. Here, we use a Hookean elastic spring model to describe how bound proteins resist stretching and compression. Given a concentration $\hat{c}(\hat{x},\hat{y},\hat{t})$ of the bound adhesive molecules, an additional contribution to the pressure $\hat p_{\text{adh}}$ takes the form:
\begin{equation}\label{eq:protspringpressure}
   \hat p_{\text{adh}}(\hat{x},\hat{y},\hat{t})=\kappa(\hat{h}(\hat{x},\hat{y},\hat{t})-l)\hat{c}(\hat{x},\hat{y},\hat{t}),
\end{equation}
where $l$ is the equilibrium bond length and $\kappa$ is the spring coefficient.

To determine $\hat{c}(\hat{x},\hat{y},\hat{t})$, we use a minimal description of the chemical kinetics of a single protein species, a full description of this model is given in Appendix~\ref{app:Protein kinetics} \citep{BellDemboBongrand,carlson_mahadevan_synapse_2015}. The adhesive molecules are assumed to be uniformly distributed on both surfaces at a constant concentration $c_0$. The concentration of pairs of unbound proteins is thus $c_0-\hat{c}(\hat{x},\hat{y},\hat{t})$. The molecules can bind or unbind at a rate of $(c_0-\hat c)K_{\text{on}}$ or $\hat cK_{\text{off}}$, respectively. The rate constant for binding, $K_{\text{on}}$, is a Gaussian distribution about the equilibrium length of the bond, $l$, with a standard deviation $\sigma$; whereas the rate constant for unbinding, $K_{\text{off}}$, is constant. Furthermore, we assume that the molecular dynamics of binding/unbinding are much faster than the time scale of viscosity-mediated membrane deformation, so that the adhesive molecule concentrations immediately adjust to the equilibrium values for a specific membrane height. Balancing the binding and unbinding rates then gives us the following expression for the concentration of a single species of bound molecules \citep{carlson2015_protadhesion_physfluids}:
\begin{equation}\label{eq:concentration}
    \hat{c}(\hat{x},\hat{y},\hat{t})=c_0\frac{\exp\left(-\left((\hat{h}(\hat{x},\hat{y},\hat{t})-l)/\sigma\right)^2\right)}{\exp\left(-\left((\hat{h}(\hat{x},\hat{y},\hat{t})-l)/\sigma\right)^2\right)+\tau_{\text{on}}/\tau_{\text{off}}},
\end{equation}
where a value of $\sigma/l=0.2$ is used in the results of this study, and the ratio of kinetic times for binding to unbinding is set to $\tau_{on}/\tau_{off}=1/3$. We note that with $\hat{c}(\hat{x},\hat{y},\hat{t})$ given by Eq. \eqref{eq:concentration}, the molecule kinetics can be reduced to a single function of $\hat{h}(\hat{x},\hat{y},\hat{t})$, meaning that we do not need to solve for $\hat{c}(\hat{x},\hat{y},\hat{t})$ as a variable in our system. The pressure given by Eqs. \eqref{eq:protspringpressure} and \eqref{eq:concentration} can conveniently be written as a free energy $\hat{F}_{\text{adh}}[h]$, as is described in Appendix~\ref{app:Protein kinetics}.

\subsection{Non-dimensional analysis\label{subsec:nondim}}

We work with non-dimensional versions of Eq. \eqref{eq:TF_dim} in the remainder of this article. Different scalings of the lengths and time are used in accordance with the dominant physical effects for the work presented in the subsequent sections. In each case Eq. \eqref{eq:TF_dim} is left with one non-dimensional parameter, $Q$, that is directly proportional to $\langle|\delta \hat{h}|\rangle$, the average amplitude of thermal fluctuations in the film \citep{dhaliwal2024instability}. Here, we outline the non-dimensionalisation used for each case; further details can be found in Appendix~\ref{app:Non-dimensionalization}.  

In section \ref{sec:AttachmentTime}, we study Eq. \eqref{eq:TF_dim} on a 1D domain, with $\gamma=0$ in Eq. \eqref{eq:membranepressures} and no adhesive molecules/proteins. In this case, the horizontal length $\hat{x}$ is scaled by the domain size $L$, as it is the only relevant horizontal length scale, the film height $\hat{h}(\hat{x},\hat{t})$ is scaled by the initial film height $\hat{h}_0$, and time $\hat{t}$ is scaled by $\tau_\mu=\frac{12\mu L^6}{B\hat{h}_0^3}$, which is the time scale of viscous relaxation of an elastohydrodynamic thin film, and can be obtained by scaling the left hand side of Eq. \eqref{eq:TF_dim} with the bending term on the right \citep{carlson_2018}. With these scalings, the dimensionless version of Eq. \eqref{eq:TF_dim} is:
\begin{equation}
    \frac{\partial h}{\partial t} =  \frac{\partial}{\partial x} \left(h^3\frac{\partial}{\partial x} \left(\frac{\partial^4}{\partial x^4} h\right)\right) + Q_{1D}\frac{\partial}{\partial x} \left(h^{3/2}\eta\right), 
\label{eq:nondimTF_Attachment}
\end{equation}
where $Q_{1D}=\frac{L}{\hat{h}_0}\sqrt{\frac{2k_B T L}{Bw}}$. Here, $w$ is the width of the quasi-1D film in the y-direction.

In section \ref{sec:Coarsening} we study Eq. \eqref{eq:TF_dim} on a 3D domain, with either $\gamma$ or $B$ set to zero in Eq. \eqref{eq:membranepressures} and the protein pressure contribution described in section \ref{subsec:Proteins}. In this case, a physical horizontal length scale can be obtained by balancing the relevant membrane pressure with the protein spring pressure \citep{carlson_mahadevan_synapse_2015}. When bending dominates, this gives $L_B=\left(B/c_0\kappa\right)^{1/4}$, and while tension dominates $L_\gamma=(\gamma/c_0\kappa)^{1/2}$. We scale $\hat{x}$ and $\hat{y}$ by this length scale, $\hat{h}$ by the protein equilibrium length $l$, the protein concentration $\hat{c}$ by $c_0$, and $\hat{t}$ by a timescale constructed as for the 1D case, but with the gradients scaled by the new horizontal length scale $L_\gamma$ or $L_B$. For the bending-driven case, inserting this scaling, i.e. $\hat{x}=L_Bx$, $\hat{y}=L_By$, and $\hat{t}=t\frac{12\mu B^{1/2}}{l^3(c_0\kappa)^{3/2}}$, gives the following dimensionless equation:
\begin{equation}
    \text{Bending ($\gamma=0$):} \quad \frac{\partial h}{\partial t} = \nabla\cdot\left(h^3\nabla \left(\nabla^4h+ (h-1)c\right) \right) + Q_{B}\nabla \cdot\left(h^{3/2}\boldsymbol{\eta}\right)
    \label{eq:nondimTF_bendcoarse},
\end{equation}
where $Q_{B}=\frac{1}{lL_B}\sqrt{\frac{2k_B T }{c_0\kappa}}$. For tension-driven films, the equivalent scalings, i.e. $\hat{x}=L_\gamma x$, $\hat{y}=L_\gamma y$, and $\hat{t}=t\frac{12\mu \gamma}{l^3(c_0\kappa)^{2}}$ give us the appropriate form of the non-dimensional thin film equation:
\begin{equation}
    \text{Tension ($B=0$):} \quad \frac{\partial h}{\partial t} = \nabla\cdot\left(h^3\nabla \left(-\nabla^2h+ (h-1)c\right) \right) + Q_{\gamma}\nabla \cdot\left(h^{3/2}\boldsymbol{\eta}\right)
    \label{eq:nondimTF_tenscoarse},
\end{equation}
where $Q_{\gamma}=\frac{1}{lL_{\gamma}}\sqrt{\frac{2k_B T }{c_0\kappa}}$.

\subsection{\label{subsec:Numerics}Finite element solver}
In this article, Eqns. \eqref{eq:nondimTF_Attachment}, \eqref{eq:nondimTF_bendcoarse}, and \eqref{eq:nondimTF_tenscoarse} are solved using the finite element method on a rectangular domain. In all simulations, periodic boundary conditions are imposed in the horizontal directions at the four boundaries. Since the film height $h$ is a dependent variable, simulations are performed in both 1D and 2D domains, which represent 2D and 3D films, respectively. The source code for the simulations in this article is available on GitHub \citep{GithubVira}.

To solve Eq.~\eqref{eq:TF_dim}, we set up a separate equation for the pressure $p$, with contributions coming from Eqs.~ \eqref{eq:membranepressures} and \eqref{eq:protspringpressure}.  Eq.~\eqref{eq:TF_dim} is divided into a system of two (plus an additional one for the film curvature $\nabla ^2 h(x,y,t)$ when there is a bending pressure) coupled second order partial differential equations. The system is reformulated into weak form where boundary terms can be neglected due to the periodic boundary conditions. The scalar fields $\nabla ^2 h(x,y,t)$, $p(x,y,t)$, and $h(x,y,t)$ are then discretised with linear elements, and the system of equations is solved using a Newton's method solver from the FEniCS finite element library~\citep{LoggMardalEtAl2012}. Time integration is performed using an implicit first order finite difference scheme. The domain and its discretisation in space and time are chosen according to the relevant non-dimensional form presented in section \ref{subsec:nondim}. For the 1D simulations presented in section \ref{subsec:AttachmentSim}, a domain of length $L=1$ is used with $\Delta x=0.01$ and $\Delta t =1\times10^{-8}$. For the 2D simulations presented in section \ref{subsec:Coarseningdesc}, a square $100\times100$-cell grid is used with both $\Delta x$ and $\Delta t$ varying in the range $1-3$ (which means that $L$ is in the range $100-300$) for the various results presented. 

The random vector $\boldsymbol{\eta}(x,y,t)$ is implemented by choosing random numbers using the ``normal'' function in the ``random'' class of  NUMPY~\citep{OliphantNumpy2006}. For the numerical solution at each time step, the two components of $\boldsymbol{\eta}(x,y,t)$ are each assigned a new value at every point in the mesh. Values are drawn from a Gaussian distribution with zero mean and a variance of $1/(\Delta x \Delta t)$ in 2D and $1/(\Delta x^2 \Delta t)$ in 3D. Due to the stochastic nature of the problem, each individual run is not to be considered as deterministic. For each set of parameters, we run multiple independent realisations and report the ensemble averages of the extracted/predicted parameters.

\section{Waiting time for the initiation of adhesion}\label{sec:AttachmentTime}
Membrane adhesion is facilitated by the binding of membrane-anchored adhesive molecules, which requires them to be in close range. It is then natural to ask: how long does it take for fluctuations to deform the membranes sufficiently so that they can bind to initiate adhesion? In this section, we use the model described in section \ref{sec:model} to study the average waiting time for initial contact to occur in a periodic 1D domain. The physical picture of our simplified model is shown in Fig. \ref{fig:Schematics}$(a)$; the two membranes are separated by a channel of viscous fluid with initial uniform thickness $h_0$. Although the channel height is initially only fluctuating about $h_0$, some part of the domain eventually reaches a critical thickness $h^*$ at which the two adhesive molecules come in contact.

\subsection{Rare-event theory}
\label{subsec:rareevent}
It is well known that thermal fluctuations can generate waves across the membrane, where the average wave amplitudes $\delta h_q$ of frequency $q$, or ``roughness'' of the membrane, depends on bending moduli and tension coefficient~\citep{helfrichUndulationsStericInteraction1984,rautuRoleOpticalProjection2017}. If the $h_0-h^*$ is of the same order as $\delta h_q$, then thermal fluctuations can easily initiate binding. If $h_0-h^*$ is sufficiently larger than $\delta h_q$, attachment of the molecules then requires the membrane profile $h(x,t)$ to attain a rather unlikely shape, which may take a long time. The process can thus be thought of as $h(x,y,t)$ fluctuating in an energy landscape that does not favour large deviations from $h_0$, until it eventually gets ``lucky'' and reaches $h^*$ at some point in the domain. Although the waiting time for protein binding is a random variable, the rare event theory can provide a prediction for the ensemble-averaged waiting time, known as the Eyring-Kramers law, which states that $\langle t_B \rangle \sim C \exp\left({2(F[h_B(x)]-F[h_0])/Q_{1D}^2}\right)$, where $\langle t_B \rangle$ is the ensemble-averaged binding time, $C$ is a prefactor, $F[h(x,t)]$ is the energy of a profile $h(x,t)$, and $h_B(x)$ is the final binding profile \citep{Eyring1935,KRAMERS1940,liuMeanFirstPassage2024}.

If the membrane primarily exhibits tension rather than bending, i.e. $L>l_{BC}$, the energy is determined by the second term in Eq. \eqref{eq:membranepressures}, and the problem is analogous to the rupture of a thin viscous liquid film with a free surface due to van-der Waals forces. This problem was recently studied by Sprittles et al., where the rare-event description based on the Eyring-Kramers formula was validated by both numerical simulations of the stochastic thin film equations and in molecular dynamics simulations~\citep{SprittlesJBLGrafke2023}. In this paper, these methods are adapted to determining the binding time between two membranes when rather the bending dominates the pressure term in Eq.~\eqref{eq:membranepressures}.

\begin{figure}
	 \centering
	\begin{tikzpicture}
   	\draw (0, 0) node[inner sep=0] (fig) {\includegraphics[width=0.46\textwidth]{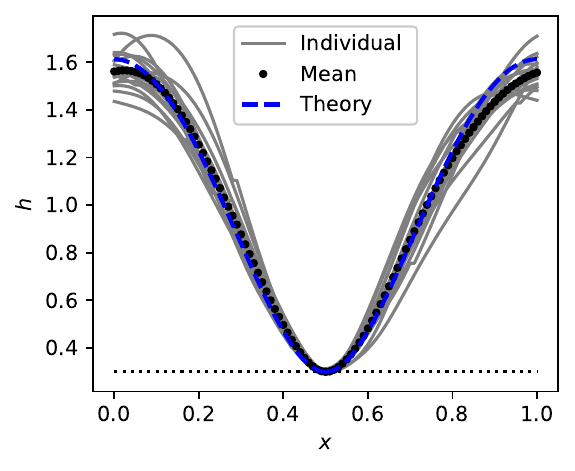}};
   	\node[] at (fig.north west){$(a)$};
 	\end{tikzpicture}
 	\begin{tikzpicture}
   	\draw (0, 0) node[inner sep=0] (fig) {\includegraphics[width=0.46\textwidth]{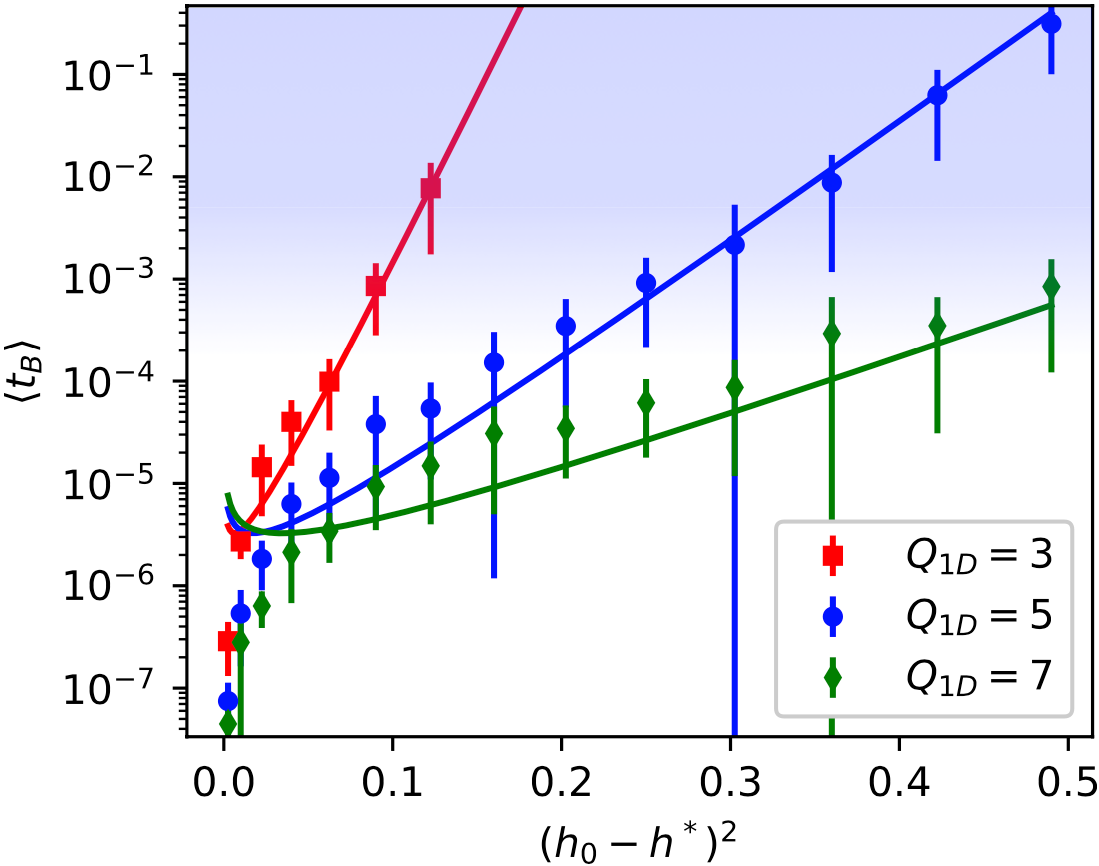}};
   	\node[] at (fig.north west){$(b)$};
 	\end{tikzpicture}

\caption{$(a)$ Film profiles at time of attachment obtained from 15 independent solutions (centered around the point of ``contact'') with parameters $Q_{1D}=5$, and $h^*=0.3$. The dotted black line represents the average of the individual simulations and the dashed blue line represents the theoretical prediction from the Euler-Lagrange equation. $(b)$ Average waiting time for adhesion  $\langle t^* \rangle$ as a function of $(h_0-h^*)^2$ for different values of the noise amplitude $Q_{1D}$. The lines represent the predicted value of $\langle t_B \rangle$ from Eq. \eqref{eq:Rareeventprediction} with no free parameter. Error bars represent the standard deviation for a set of $N=15$ simulations for each data point. The shaded blue color is intended as a guide to the eye to emphasize the region where rare-event theory is valid, i.e. attachment events are sufficiently unlikely.}
	\label{fig:Rare-event_multi}
\end{figure}

When bending dominates over tension, i.e. $L<l_{BC}$, the non-dimensional energy functional for the 1D membrane becomes
\begin{equation}
F[h(x)] = \int ^1_0 \frac{1}{2} \left(\frac{\partial^2 h}{\partial x ^2}\right)^2 dx.
\end{equation}
The pressure term in Eq.~\eqref{eq:membranepressures} can be recovered by taking a functional derivative of $F$ with respect to $h(x)$. Finding the average waiting time for adhesion requires finding the profile $h_B(x)$ that minimizes $F[h]$ while also maintaining conservation of mass, which enters as the constraint $\int_0^1(h-h_0)dx = 0$, and satisfying the periodic boundary condition. This constrained optimization problem can be solved using the Euler-Lagrange equation, which gives the fourth-order polynomial profile shown by the dashed blue line in Fig. \ref{fig:Rare-event_multi}$(a)$. The energy corresponding to this profile is $F_{B}= 360 (h_0-h^*)^2$. With a sharp asymptotics analysis (see Appendix~\ref{app:Rare-event} for details), the predicted average attachment time is given by
\begin{equation}\label{eq:Rareeventprediction}
\langle t_B \rangle = \frac{1}{\beta (h_B)}\sqrt{\frac{Q_{1D}^2}{(2\pi)^7}} \exp\left(720\left(\frac{h_0-h^*}{Q_{1D}}\right)^2\right),
\end{equation}
where the parameter $\beta$ is given by
 \begin{equation}\label{eq:beta}
\beta(h_B) = (h_0-h^*)(2\pi)^6\left(\frac{1}{2}h_0^3 +\frac{3}{8}h_0(h_0-h^*)^2\right).
\end{equation} 
We note that the expressions in Eqs. \eqref{eq:Rareeventprediction} and \eqref{eq:beta} predict the attachment time across the channel without any fitting parameters. They are valid in the limit of small noise, i.e. either large $h_0-h^*$ or small $Q_{1D}$, such that the trajectory of initial attachment must cross $h_B(x)$ in the energy landscape.

        
\subsection{Numerical predictions}\label{subsec:AttachmentSim}
Numerical simulations of Eq. \eqref{eq:nondimTF_Attachment} are used to test the prediction in Eq.~\eqref{eq:Rareeventprediction} considering only the effect of membrane bending. The simulations are initiated with a flat membrane of height $h_0=1$. Since there are no adhesive molecules, i.e. no force pulling/pushing on the membrane, the minimum film height fluctuates randomly with time, occasionally reaching a lower value than it has before. The simulations are run until the lowest point in the film reaches the cutoff height $h^*$, indicating that the molecules are now within binding range for initial attachment, and we denote this time as $t^*$. The point in the periodic domain at which attachment occurs is random. If the deformation required for film rupture, $h_0-h^*$, is small, attachment occurs rapidly and the profile shape at $t^*$ varies significantly from simulation to simulation. When $h_0-h^*$ is larger, however, the minimum film height fluctuates for a long time before reaching attachment (the number of time steps being many orders of magnitude), often coming close many times before finally reaching $h^*$. When this is the case, the individual profiles at $t^*$ are consistent, as shown in Fig. \ref{fig:Rare-event_multi}$(a)$, which suggests that attachment does indeed tend to occur at a specific minimum energy profile.


When the value of $h_0-h^*$ is large, as is the case for the profiles shown in Fig. \ref{fig:Rare-event_multi}$(a)$, one would expect that the final profile is similar to the fourth-order polynomial predicted theoretically in section \ref{subsec:rareevent}. The dashed blue line in Fig. \ref{fig:Rare-event_multi}$(a)$ shows that this is indeed the case if the profiles are centered about their minimum and then averaged. This indicates that it is truly rare for fluctuations to make the film reach $h^*$, and that this almost always occurs very close to the minimum energy profile that allows attachment. 

The average attachment time, $\langle t^* \rangle$, is plotted for various values of $Q_{1D}$ and $h^*$ in Fig. \ref{fig:Rare-event_multi}$(b)$. It is clear that the time increases rapidly when $h^*$ is reduced, and eventually grows exponentially with $(h_0-h^*)^2$ for constant $Q_{1D}$. The theoretical prediction for the attachment time across a membrane channel from Eq. \eqref{eq:Rareeventprediction} provides an excellent quantitative prediction for the average rupture time when $h_0-h^*$ is large. For smaller values of $h_0-h^*$, the rare-event theory is not valid, as the membrane profile does not neccesarily cross the saddle point on its way to attachment. We note that the $h_0-h^*$ required for the rare-event prediction to work is dependent on $Q_{1D}$, as smaller fluctuations are less likely to cause large deformations to the film profile. Generally, the rare-event prediction is valid when the attachment event is sufficiently rare, meaning that the average attachment time is sufficiently large. This can be achieved either by having a small value of $Q_{1D}$ or a large height difference $h_0-h^*$, as in the blue shaded region of Fig. \ref{fig:Rare-event_multi}$(b)$.

\section{Coarsening of adhesion patches}\label{sec:Coarsening}
We now turn our attention to the next stage in the dynamics, when the adhesive molecules have formed bonds ($t >\langle t_B\rangle$) that lead to a rich coarsening dynamics. To do this we study study Eq. \eqref{eq:GradFlow} with the protein binding model of Eqs. \eqref{eq:protspringpressure} and \eqref{eq:concentration} on a 2D domain, where we vary the membrane pressure term and the mobility factor. 

Once adhesive molecules start to bind across the thin film, the membrane is pulled towards the substrate at the binding site, leading to further binding of adjacent proteins \citep{BellDemboBongrand}. This process brings the surfaces closer to each other, squeezing liquid out laterally through the channel. Conservation of the liquid, however, may lead to distinct regions where the membrane separation is small and proteins are bound or where membrane separation is large and proteins unbound, as is depicted in Fig. \ref{fig:Schematics}$(c)$ \citep{FenzAna-Suncana2017,DinetArroyoStaykova2023}. The network of unbound domains containing excess liquid may then coarsen in order to reduce the deformation of the membrane, i.e., reducing the overall energy of the system. Smaller pockets/lumen-like structures shrink as liquid is transported toward larger pockets/lumens, which subsequently grow in size. Such coarsening is reminiscent of when liquid droplets form in a dewetting liquid thin film \citep{OttoRumpSlepcev2006}. It also falls into the broader category of coarsening dynamics in physical systems, which has been extensively studied both theoretically and experimentally \citep{HohenbergHalperin1977,Furukawa1985,Bray1994,CamleyBrown2011,CatesLesHouches}. We will now describe how our model fits into this context, demonstrating that elastohydrodynamic thin films display distinct coarsening behavior due to the combined effects of elasticity and viscosity.  

\subsection{Domain coarsening of adhered patches}\label{subsec:CoarseningTheory}

Equation~\eqref{eq:TF_dim} can be seen as the governing equation of a dynamical phase field, in which the film height $h(x,y,t)$ is the sole order parameter \citep{Bray2003,CatesLesHouches}. When coupled with the pressure term given by Eq.~\eqref{eq:membranepressures}, it is in fact quite similar to the widely-studied dynamical model B, which is used to describe the coarsening of a conserved order parameter
\citep{HohenbergHalperin1977}. Nevertheless, a number of features distinguish our system. First, a nonlinear $\sim h^3$ mobility (as described in section \ref{subsec:Equations}) arises from the lubrication flow in the narrow channel. Second, rather than the typical double-well potential,  our system imposes an energy landscape arising from the adhesive molecules/proteins described in section \ref{subsec:Proteins} and Appendix~\ref{app:Protein kinetics}. Finally, and perhaps most significantly, the classical Cahn-Hilliard Laplacian free energy term (which in this case represents isotropic membrane tension), is supplemented by the fourth order bending term as described in Eq. \eqref{eq:membranepressures}.

A quantitative understanding of phase separating systems is achieved through the dynamical scaling hypothesis, which states that in the later stage of a coarsening process, the domain structure of a system (as quantified by the structure function $S(\textbf{k},t)=\langle h_\textbf{k}(t)h_\textbf{-k}(t)\rangle$) is constant in time if one rescales the lengths by a single characteristic length scale $L_c(t)$ \citep{Bray1994,CamleyBrown2011}. This concept has been shown to be valid both numerically and experimentally in many systems, with power-law growth displayed when $L_c(t)$ is computed as a moment of $S(\textbf{k},t)$ or the radius of individual particles \citep{Furukawa1985,ShinozakiOono1993,SungHan1995,TatenoTanaka2021,SuParikh2024,SaiseauDelville2024}. For coarsening of the aforementioned model B system, $L_c(t)$ is known to grow according to a $\sim t^{1/3}$ power law, as has been validated by a number of numerical studies \citep{VladimirovaMauri1998,TiribocchiCates2015}, and can also be predicted theoretically using scaling arguments, renormalization group theory, or Lifshitz-Slyozov-Wagner theory (in the limit where the minority phase occupies a small volume fraction) \citep{Wagner1961,Bray1994,CamleyBrown2011,LIFSHITZSlyozov1961}.

For the case where the bending energy drives coarsening rather than surface tension it is unclear if the coarsening rate will follow the $\sim t^{1/3}$ power law. This essentially corresponds to replacing the surface tension contribution to the classical free energy functional, $F_{\gamma}(h(x,y)) = \int  \dfrac{1}{2} \lvert\nabla h\rvert^2 dA$, by the corresponding bending term $F_{\text{bend}}(h(x,y)) = \int  \dfrac{1}{2} \left(\nabla^2 h\right)^2 dA$. We now follow the derivation of \citet{Bray1994} to make a scaling prediction for how bending changes the coarsening dynamics (details are provided in Appendix~\ref{appsub:coarseningtheory}). We ignore the effects of the nonlinear mobility as well as the specifics of the potential function. We thus think of a system where the pressure acts as a chemical potential $\Phi \equiv\delta F/\delta h$, where $F$ consists of the aforementioned bending term and a symmetric double-well potential function $V(h)$. The chemical potential is thus 
\begin{equation}\label{eq:mu0}
    \Phi = \frac{\partial V}{\partial h} +\nabla^4h
\end{equation}
and the film height then evolves according to a continuity equation with flux $\boldsymbol{j}=-\nabla \Phi$. 

In late-stage coarsening when motion of the interface between two domains is slow, diffusion of the order parameter $h$ and chemical potential $\Phi$ in the bulk is fast, so both should satisfy Laplace's equation $\nabla^2h=\nabla^2\Phi=0$ in the bulk regions. The flux through the interface, and thus by continuity the velocity of the interface, can then be determined by finding $\Phi$ at the interface.

At an interface between the two phases with radius of curvature $R$, $\Phi$ can be re-expressed in terms of the coordinate $g$ representing distance along the unit vector  $\boldsymbol{\hat{g}}$ in the direction perpendicular to the interface ($g=\pm \infty$ in the bulk and $g=0$ at the centre of the interface). Noting that $\nabla h =\left(\partial h/\partial g\right)\boldsymbol{\hat{g}}$ and  $\nabla\cdot\boldsymbol{\hat{g}}=1/R$ near the interface, we find that 
\begin{equation}\label{eq:muexp}
    \Phi = \frac{\partial V}{\partial h} +\frac{\partial^4 h}{\partial g^4}+\frac{2}{R}\frac{\partial^3 h}{\partial g^3}-\frac{1}{R^2}\frac{\partial^2 h}{\partial g^2}+\frac{1}{R^3}\frac{\partial h}{\partial g}.
\end{equation}
The value of $\Phi$ at the interface is then determined by multiplying Eq. \eqref{eq:muexp} by $\partial h/\partial g$ and integrating across the interface from $g=-\infty$ to $g=\infty$. By imposing the boundary condition that $h$ is constant in the bulk and assuming that the chemical potential is a double well with equal potential on both sides, integration gives
\begin{equation}\label{eq:GTBC}
    \Phi \Delta h = -\frac{2}{R}\int^\infty_{-\infty}\left(\frac{\partial^2 h}{\partial g^2}\right)^2dg+\frac{1}{R^3}\int^\infty_{-\infty}\left(\frac{\partial h}{\partial g}\right)^2dg
\end{equation}
where $\Delta h$ is the height difference between the domains on the inside and outside of the interface and the integrals $-2\int^\infty_\infty\left(\frac{\partial^2 h}{\partial g^2}\right)^2dg$ and $\int^\infty_\infty\left(\frac{\partial h}{\partial g}\right)^2dg$ can be interpreted as effective interfacial tension-like coefficients $\Gamma_1$ and $\Gamma_2$ for the line tension between the two phases in 2D space. Eq. \eqref{eq:GTBC} thus represents a Gibbs-Thomson-like boundary condition that determines the value of $\Phi$ on interfaces with radius of curvature $R$. Since the flux of fluid is proportional to the gradient of the chemical potential (assuming a constant mobility), the velocity of the interface scales as $-\partial \Phi/\partial g$. Assuming that the domains are circular with radius $R$ corresponding to the macroscopic length scale $L_c$, this then gives the following growth law for bending-driven coarsening:
\begin{equation}\label{eq:growthrate}
    \frac{dL_c}{dt}\sim \frac{\Gamma_1}{L_c^2}+\frac{\Gamma_2}{L_c^4}.
\end{equation}
Eq. \eqref{eq:growthrate} stands in contrast to the model B case which has only a $\sim L_c^{-2}$ term. This suggests that bending-driven coarsening should have a coarsening rate with an exponent somewhere between $1/3$ and $1/5$ with time, with the value being determined by the relative sizes of $\Gamma_1$ and $\Gamma_2$, which in turn depend on the shape of the potential function. We note that this simple scaling prediction only describes the difference between surface tension-driven and bending-driven coarsening. It does not take into account the other differences between our model and the classical model B system described by \citet{Bray1994} such as the $\sim h^3$ mobility and the protein binding potential energy. Nevertheless, it provides a meaningful identification of the distinction between surface tension-driven and bending-driven coarsening, which will be investigated numerically in the subsequent sections.

\subsection{Numerical investigation of bending-driven coarsening dynamics}\label{subsec:Coarseningdesc}
\begin{figure}
     \begin{tikzpicture}
   	\draw (0, 0) node[inner sep=0] (fig) {\includegraphics[width=0.95\textwidth]{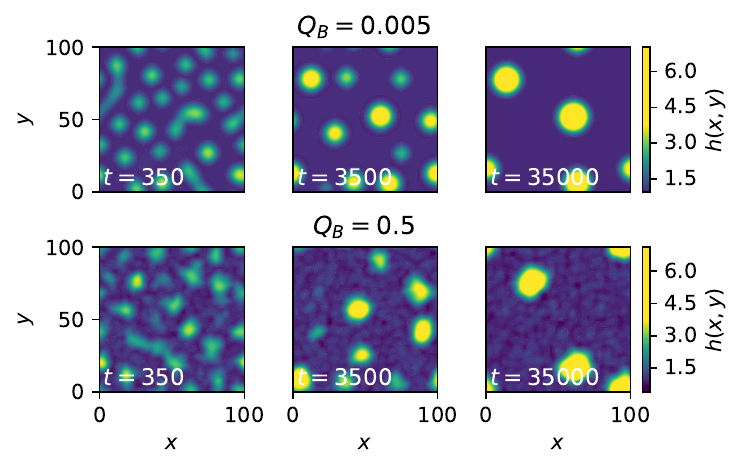}};
   	\node[] at (fig.north west){$(a)$};
    	\node[] at (fig.west){$(b)$};
 	\end{tikzpicture}

	\caption{Contour plots illustrating the height $h(x,y,t)$ of 3D thin films under the influence of protein binding and unbinding for different times $t$, with $h_0=1.4$. In $(a)$ the fluctuation parameter is $Q_{B}=0.005$, whereas in $(b)$ it is set to $Q_{B}=0.5$. Although similar coarsening dynamics occur at late times regardless of $Q_{B}$, the size of the domains at $t=350$ is somewhat larger for $Q_{B}=0.5$ due to early-time coalescence. }
	\label{fig:CoarseningContours_bend_varQ}
\end{figure}
We now present the results of numerical simulations of the phase separation of adhesive molecules/proteins during membrane adhesion when dominated by bending of the membrane, as described in Eq. \eqref{eq:nondimTF_bendcoarse}. The membrane is always initialised as a flat profile with non-dimensional initial height $h(x,y,0)= \hat{h}_0/l>1$ and the non-dimensional width of the protein binding rate distribution, $\sigma_{\text{on}}/l=0.2$. Fig.~\ref{fig:CoarseningContours_bend_varQ}($a$) shows the contour plots (a full video can be seen in Movie 1) of the film profile at a three different times when the initial film height is just at the edge of the range in which the proteins can bind. Fluctuations are clearly present for the first time steps, where the film is pulled to the substrate by the adhesive molecules. Most of the film profile moves toward the equilibrium length of the adhesive molecules ($h=1$ in non-dimensional terms) in order to reduce the energy of the system. Due to conservation of liquid mass, excess liquid must flow somewhere else, but periodic boundary conditions prevent fluid from leaving the domain. Thus, in some regions the film is pushed up, forming circular ``pockets'' or ``lumens'' in which proteins are unbound and the film height is significantly larger than the initial height, as can be seen in the first panel to the left in Fig.~\ref{fig:CoarseningContours_bend_varQ}($a$).



With time, the detached domains coarsen in an Ostwald ripening-like process, resulting in fewer, larger domains as can be seen in the second and third panels of Fig. \ref{fig:CoarseningContours_bend_varQ}($a$). Once the circular pockets form, their centres remain essentially fixed as fluid is slowly transported from smaller to larger domains in a diffusion-like manner. When the size of a single pocket drops below a critical threshold, it suddenly ``implodes'', which causes a minor horizontal rearrangement of nearby domains (see supplementary videos). Fig. \ref{fig:coarseninggrowth_example_bend} ($a$) shows the time evolution of the characteristic size $L_c$ for these domains, as computed using the method of Shinozaki and Oono \citep{ShinozakiOono1993}, averaging the structure factor $S(\textbf{k},t)$ across 15 independent realizations (details of how we calculte $L_c$ are provided in Appendix~\ref{appsub:Lcalc}). Although the distinct ``implosion'' events lead to discrete jumps in the growth for an individual simulation, the ensemble-averaged $L_c$ grows smoothly with time following a power law. The growth is significantly slower than the $\sim t^{1/3}$ power law observed for a model B system \citep{Bray1994,TiribocchiCates2015}, instead, the exponent is closer to $1/5$, which can be expected based on our analysis in section \ref{subsec:CoarseningTheory}.
 \begin{figure}
 	 \centering
 \includegraphics[width=0.95\textwidth]{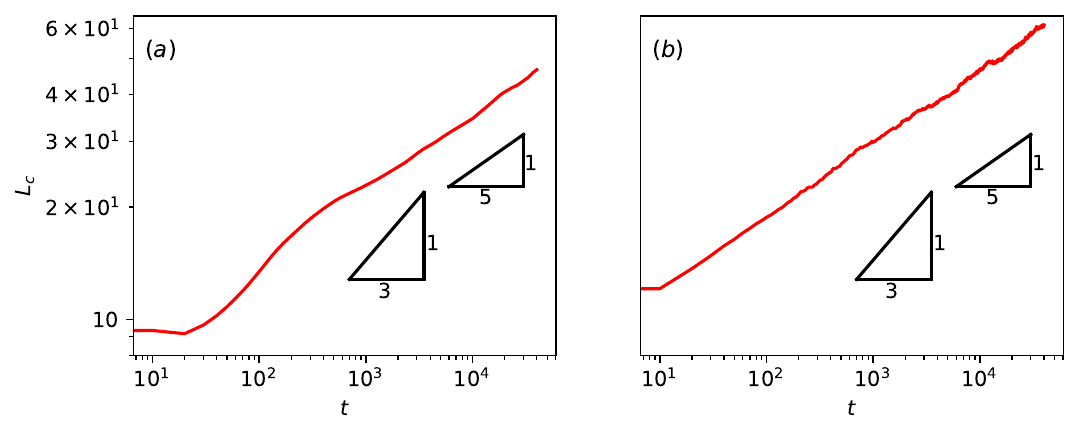}

 	\caption{ The length scale $L_c$ computed using Eq. \eqref{eq:app_SOexp} from the average of $N=15$ individual simulations under the same conditions as the data in Fig. \ref{fig:CoarseningContours_bend_varQ}. In $(a)$ the fluctuation parameter is $Q_{B}=0.005$, whereas in $(b)$ it is set to $Q_{B}=0.5$. Power law coarsening with an exponent well below $1/3$ is observed in both cases, but starts off with larger domain size when $Q_{B}=0.5$.
 	\label{fig:coarseninggrowth_example_bend}}
 \end{figure}

The contour plots in figure \ref{fig:CoarseningContours_bend_varQ}($b$) show the evolution of height profiles of a coarsening film when the amplitude of the thermal fluctuations, represented by the parameter $Q_{B}$, is increased by two orders of magnitude. With higher noise amplitude, the lumens are irregular in shape and constantly deforming (see supplementary video 2). The fluctuations also cause significant lateral motion of the domains, which leads to some instances of coalescence at very early times. Despite this, at the later time, large droplets seem to repel each other, and domain growth is primarily caused by fluid flow through adhered patches, as was observed in the low noise amplitude case. Interestingly, the power law for domain growth is relatively unaffected by the strength of the fluctuations, as shown in Fig. \ref{fig:coarseninggrowth_example_bend}($b$). The size of the domains before the late-stage power law growth is somewhat higher when the fluctuations are larger, perhaps due to coalescence events at early times, as can be seen in the first panel of Fig. \ref{fig:CoarseningContours_bend_varQ}.

To better understand how and why the coarsening rate of our system differs from the more familiar model B system, we perform additional simulations in which we vary the initial membrane height $h_0$, the fluctuation strength $Q$, the mobility prefactor in the thin film equation, and  which term in Eq. \eqref{eq:membranepressures} we use. In general, we find that coarsening behaviour is a preserved feature of the system even when these parameters are changed. In order to ensure an initial configuration with many small pockets in the domain, $h_0$ should be close to the edge of the binding range of the adhesive molecules. If $h_0$ is too small, the entire domain is already in a bound state and few pockets form. If, on the other hand, $h_0$ is too large, it takes a very long time for contact to occur, which is typically at only one point in the domain, leading to a small number of initial pockets when coarsening starts. 
\begin{figure}
     \begin{tikzpicture}
   	\draw (0, 0) node[inner sep=0] (fig) {\includegraphics[width=0.95\textwidth]{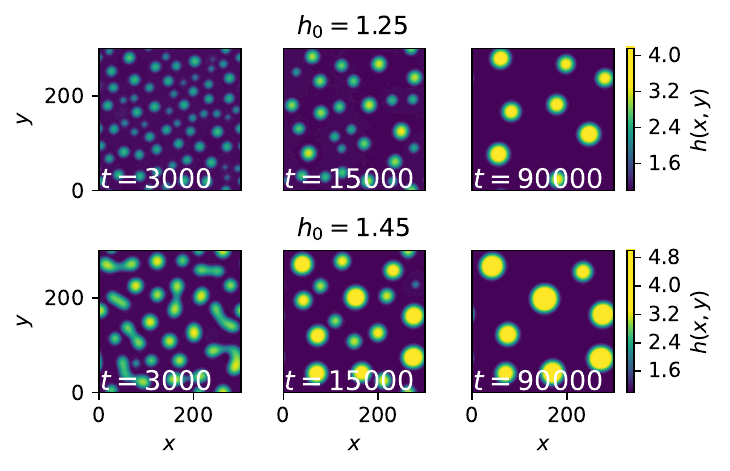}};
   	\node[] at (fig.north west){$(a)$};
    	\node[] at (fig.west){$(b)$};
 	\end{tikzpicture}

	\caption{Contour plots illustrating the height $h(x,y,t)$ of 3D thin films under the influence of protein binding and unbinding for different times $t$, with $Q_{\gamma}=0.01$. In $(a)$ the initial height is $h_0=1.25$, whereas in $(b)$ it is set to $h_0=1.45$. Increasing $h_0$ leads to a larger initial domain size, but slower coarsening at late times. }
	\label{fig:CoarseningContours_int_varh0}
\end{figure}

To begin with, we simulate coarsening using the tension term of Eq. \eqref{eq:membranepressures} (corresponding to the well-studied Cahn-Hilliard free energy) instead of the bending term, as in Eq. \eqref{eq:nondimTF_tenscoarse}. As expected, coarsening behaviour is observed, as shown in Fig. \ref{fig:CoarseningContours_int_varh0}. Comparing the first snapshots of figures \ref{fig:CoarseningContours_int_varh0}$(a)$ and \ref{fig:CoarseningContours_int_varh0}$(b)$ demonstrates that larger initial heights lead to fewer domains at early times. At later times, we observe that the coarsening actually appears to be slower when $h_0$ is increased.
\begin{figure}
	 \centering
	\begin{tikzpicture}
   	\draw (0, 0) node[inner sep=0] (fig) {\includegraphics[width=0.46\textwidth]{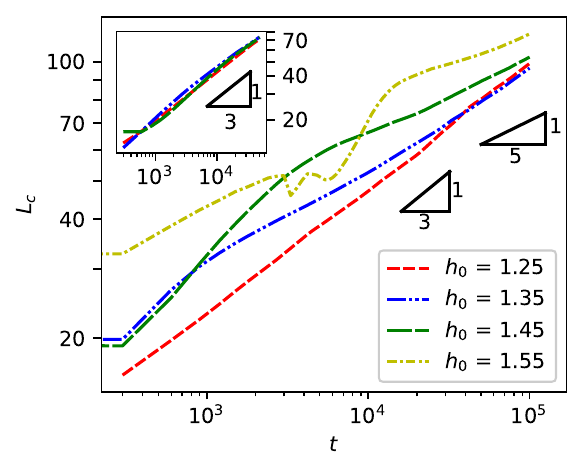}};
   	\node[] at (fig.north west){$(a)$};
 	\end{tikzpicture}
 	\begin{tikzpicture}%
   	\draw (0, 0) node[inner sep=0] (fig) {\includegraphics[width=0.46\textwidth]{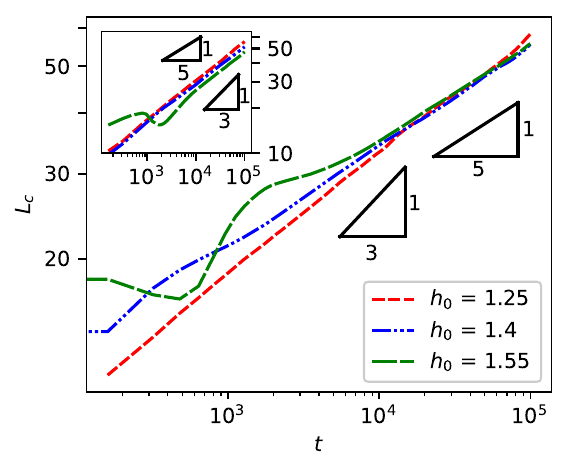}};
   	\node[] at (fig.north west){$(b)$};
 	\end{tikzpicture}

\caption{$(a)$ Scaling length $L_c$ as a function of time for tension-driven coarsening in a viscous film with varying $h_0$ when $Q_{\gamma}=0.01$. $L_c$ is calculated from the average $S(\textbf{k},t)$ for $N=22$ independent simulations. The inset shows the results when a constant mobility is used instead. $(b)$ Scaling length $L_c$ as a function of time for bending-driven coarsening in a viscous film with varying $h_0$ when $Q_B=0.01$. $L_c$ is calculated from the average $S(\textbf{k},t)$ for $N=20$ independent simulations. The inset shows the results when a constant mobility is used instead.}

	\label{fig:varhcoarsening}
\end{figure}

Fig. \ref{fig:varhcoarsening}$(a)$ shows how the characteristic length scale $L_c$ of a viscous film profile grows with time when the membrane exhibits only interfacial tension. When $h_0$ is small, the growth exponent is close to $\sim t^{1/3}$, in accordance with what one expects for the well-known model B system. For higher values of $h_0$, the growth rate for late-stage coarsening decreases. This seems to be caused by the nonlinear $\sim h^3$ term arising from the viscous resistance to flow in the governing equation. To confirm this, we perform additional simulations in which we replace the nonlinear mobility by a constant mobility with value $1$. The inset in fig. \ref{fig:varhcoarsening}$(a)$ shows that $L_c\sim t^{1/3}$ growth is observed regardless of $h_0$ when the mobility is constant, as expected.

In Fig. \ref{fig:varhcoarsening}$(b)$, similar results are presented, but this time the tension term in Eq. \eqref{eq:membranepressures} is replaced by the bending term, i.e., we set $\gamma>0$ and $B=0$ and solve Eq. \ref{eq:nondimTF_tenscoarse}. As suggested by the theory in section \ref{subsec:CoarseningTheory}, the power law for bending-driven coarsening is reduced from $L_c\sim t^{1/3}$. The inset in Fig \ref{fig:varhcoarsening}$(b)$ shows that this is indeed the case when the mobility is kept constant, with a growth exponent around $1/4$ consistently observed. For the viscous case, there is a decrease in the growth exponent when the value of $h_0$ is increased, but this effect is not as pronounced as it is for the tension-driven films. 

The decreased coarsening rate as $h_0$ is increased may seem counterintuitive at first glance, as a thicker film should lead to less viscous resistance to fluid flow and thus a higher mobility. In the coarsening process, however, the fluid flow between pockets must pass through the adhered region, where the film height is fixed at the equilibrium height of the adhesive molecules ($h=1$), thus restricting the mobility of the flow through this region. Increasing $h_0$ thus only increases the depth of the pockets, which  decreases the flux through the interface between the pocket and the attached region.
 
\begin{figure}
	 \centering
	\begin{tikzpicture}
   	\draw (0, 0) node[inner sep=0] (fig) {\includegraphics[width=0.46\textwidth]{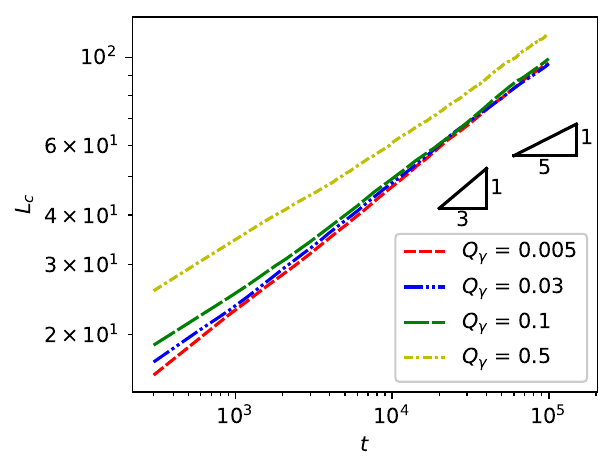}};
   	\node[] at (fig.north west){$(a)$};
 	\end{tikzpicture}
 	\begin{tikzpicture}
   	\draw (0, 0) node[inner sep=0] (fig) {\includegraphics[width=0.46\textwidth]{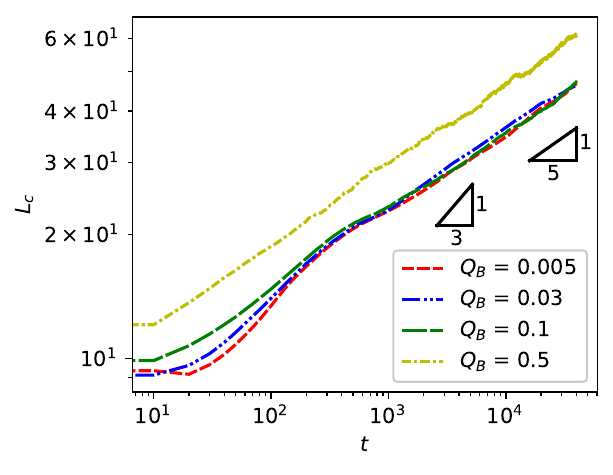}};
   	\node[] at (fig.north west){$(b)$};
 	\end{tikzpicture}

\caption{$(a)$ Scaling length $L_c$ as a function of time for tension-driven coarsening in a viscous film with varying $Q_{\gamma}$. These results are for $h_0=1.25$ and the $L_c$ is calculated from the average $S(\textbf{k},t)$ for $N=22$ independent simulations. $(b)$ Scaling length $L_c$ as a function of time for bending-driven coarsening in a viscous film with varying $Q_B$. These results are for $h_0=1.4$ and the $L_c$ is calculated from the average $S(\textbf{k},t)$ for $N=15$ independent simulations.}
	\label{fig:varQcoarsening}
\end{figure}

Fig. \ref{fig:varQcoarsening} shows the coarsening behavior for both tension-driven and bending-driven viscous films as $Q$ is varied by two orders of magnitude. In both cases, the value of $L_c$ at the beginning of coarsening is larger when the fluctuations are increased to $Q=0.5$, due to increased coalescence at very early times. In the bending case, large fluctuations also lead to an earlier onset of the power-law coarsening regime, as can be seen in Fig. \ref{fig:varQcoarsening}$(b)$. During the late coarsening stage, however, fluctuations do not play a significant role. In Fig. \ref{fig:varQcoarsening}$(a)$ we observe that increasing the magnitude of fluctuations leads to a slight decrease in the power law for tension-dominated films, while Fig. \ref{fig:varQcoarsening}$(b)$ shows that changing $Q_B$ does not seem to significantly affect the power law of bending-dominated films. This seems to suggest that the main role of the fluctuations is simply to provide perturbations at early times which initiate the adhesion process.

\section{Discussion}
\label{sec:Discussion}
We have demonstrated that the waiting time for membranes to come into contact across a viscous channel can be effectively predicted by the rare-event theory. This demonstrates the utility of rare-event theory in predicting the average time for important but unlikely events to occur in stochastic systems. Such statistical techniques are relevant in many biological systems where rare events caused by stochastic fluctuations may be necessary for important biological processes \citep{ChouDorsogna2014}. Although the adhesion of live cells often involves active cellular machinery such as actomyosin and filopodia to help cells bind, our results may be useful for estimating the likelihood that cells may come into contact when such active mechanisms are absent and may help to distinguish between actively and passively driven dynamics.  

The membrane coarsening results of section \ref{sec:Coarsening} present a minimal mathematical description of coarsening behaviour similar to the recent experiments of \citet{DinetArroyoStaykova2023}. In their experiments, an osmotic shock is applied to GUVs initially attached through biotin-neutravidin bonds to a supported lipid bilayer, pushing liquid from inside the GUV into in the membrane-membrane channel, forming small unattached lumens. As the system evolves, the smaller lumens shrink and eventually disappear, while larger lumens grow. Our work suggests a physical model to explain the intrinsic coarsening process in those experiments. Based on the results of section \ref{sec:Coarsening}, one would expect to observe power-law coarsening of the domains in such experiments. Unfortunately, the data in the experiments of Dinet et al. is not quite amenable to study the coarsening law because the separate processes of fluid flux through the membrane into the channel, flux of fluid from pocket to pocket, and flux of fluid from the pockets to the exterior region are all occurring simultaneously. Nevertheless, it points to experimental realisations allowing for the observation of the predicted power-law coarsening dynamics by isolating the interluminal coarsening process, e.g., by using a larger GUV so the center of the adhesion patch is not as affected by flux to the outside.

Our work provides a framework for understanding coarsening exponents in thin films. The results of section \ref{subsec:Coarseningdesc} provide a scaling-based prediction that the power law exponent is smaller for bending-driven than for tension-driven coarsening. A more rigorous prediction of the growth rate for bending-driven coarsening might be feasible applying renormalization group theory, as is described by \citet{Bray1994}. Whether a physical system falls under bending- or tension-dominated regime depends on the ratio of the characteristic length scale of the domains to the bendocapillary length scale, $L_{bc} = \sqrt{B/\gamma} $. An estimate  of $L_{bc}$ based on typical values for biological membranes is smaller than the size of lumens in the experiments of \citet{DinetArroyoStaykova2023}, which suggests that tension should dominate for that system. Even so, we would expect a growth exponent with time that is smaller than $\sim t^{1/3}$ due to the nonlinear mobility coming from the fluid viscosity. It would also be interesting to see if the coarsening rate is altered if one were to change the membrane properties to have a larger value of $L_{bc}$.

Another observation to highlight is that the noise amplitude seems to have no effect on the power law for the coarsening process in both tension and bending dominated regimes. This is not true, however, for the spreading of droplets on a flat substrate, which can be seen as the coarsening of a single lumen. In tension dominated regime, as the droplet spreads, the growth of its radius follows the classical Tanner's Law~\citep{lhtannerSpreadingSiliconeOil1979} if the system is deterministic, and follows a fluctuation-enhanced Tanner's Law~\citep{davidovitch2005} in the presence of noise. In the bending-dominated regime, a different power law for growth was also found for deterministic~\citep{listerViscousControlPeeling2013} and stochastic~\citep{carlson_2018,pedersen_niven_salez_dalnoki-veress_carlson_2019} systems. One key difference between our setup and droplet spreading is the presence of bound adhesive molecules, which keeps the membrane fixed at a constant height in the adhered region. Another difference is that instead of a single droplet, we have a network of pockets connected by adhesion patches. 

Our numerical results predict that coarsening is slower for larger initial heights in viscous, tension-driven coarsening, which is directly relevant to coarsening in dewetting liquid films. Experiments on dewetting nanometric polymer films by \citet{LimaryGreen2003} indeed showed that the coarsening exponent was smaller when the initial film height was increased. Although the driving potential in such a system comes from the surface energies of the materials rather than protein-like binding, the governing equations should otherwise be the same, and we would expect similar dynamics. Further work on this problem could lead to a more complete understanding of how and why an increased initial film height delays coalescence.

Finally, a future interesting avenue to explore would be looking into experimental systems of elastohydrodynamic coarsening beyond a biological context. If the molecular binding-based adhesive potential in our model is replaced by some other physical mechanism, the results of our work should still hold. For example, if liquid is trapped between two thin elastic sheets grafted with polymer brushes that are attracted to each other \cite{ManiMaha2012}, one might observe coarsening of non-adhered pockets quite similar to the observations in our simulations.\\

{\em Declaration of Interests. The authors report no conflict of interest}
\begin{acknowledgments}

\end{acknowledgments}

\FloatBarrier

\appendix
\section{Adhesive molecule kinetics}\label{app:Protein kinetics}
In this section, we present the simple model used for the kinetics of the adhesive molecule bond occuring across the intermembrane channel. In biological systems this typically corresponds to the bond between two membrane-anchored proteins across the extracellular gap. The proteins are assumed to be present at all times on both sides of the channel at a constant concentration $c_0$. The concentration of bound proteins at a particular location is denoted as $\hat{c}(\hat{x},\hat{y},\hat{t})$. If proteins bind with a rate constant $K_{\text{on}}$ and unbind with a rate constant $K_{\text{off}}$, the dynamics of $\hat{c}(\hat{x},\hat{y},\hat{t})$ are governed by (assuming that diffusion and the flow are slow compared to the binding kinetics)
\begin{equation}\label{eq:app_proteinkinetics}
    \frac{\partial \hat{c}}{\partial \hat{t}} = (c_0-\hat{c})K_{\text{on}}-\hat{c}K_{\text{off}}.
\end{equation}
The binding rate constant $K_{\text{on}}$ is represented as  a probability of crossing an energy barrier, and is dependent on the film height $\hat{h}$ with a maximum likelihood of binding at $\hat{h}=l$ \citep{BellDemboBongrand}. The off rate, $K_{\text{off}}$, is taken as a constant.
\begin{align}\label{eq:app_rateconstants}
    K_{\text{on}}=\frac{1}{\tau_{\text{on}}}\exp\left(-\left((\hat{h}(\hat{x},\hat{y},\hat{t})-l)/\sigma\right)^2\right), && K_{\text{off}}=\frac{1}{\tau_{\text{off}}}.
\end{align}
where $l$ is the equilibrium length of the  intermembrane protein bond, $\sigma$ is the width of the kinetic binding zone, and $\tau_{\text{on}}$ and $\tau_{\text{off}}$ are kinetic times.

For this system, if the protein kinetics are allowed to relax much faster than the timescale of film deformation, we can assume the bound protein concentration $\hat{c}$ to be quasi-constant \citep{carlson2015_protadhesion_physfluids}. At chemical equilibrium, Eq. \eqref{eq:app_proteinkinetics} leads us to express the equilibrium concentration $\hat{c}^{eq}$ as
  \begin{equation}\label{eq:app_ceq}
    \hat{c}^{eq}(\hat{h}(\hat{x},\hat{y},\hat{t})) = c_0\frac{ K_{\text{on}}}{ K_{\text{on}}+\frac{\tau_{\text{on}}}{\tau_{\text{off}}}}, 
  \end{equation}
which gives us the expression for $\hat{c}(\hat{h})$ in Eq. \eqref{eq:concentration}. In our simulations, we take the dimensionless binding distribution width $\sigma/l$ to be 0.2, and the binding/unbinding timescale ratio $\tau_{\text{on}}/\tau_{\text{off}}$ to be $1/3$. Although this is rather arbitrary, we observe little change in our results when these parameters are altered.

\begin{figure}
	 \centering
\includegraphics[width=0.65\textwidth]{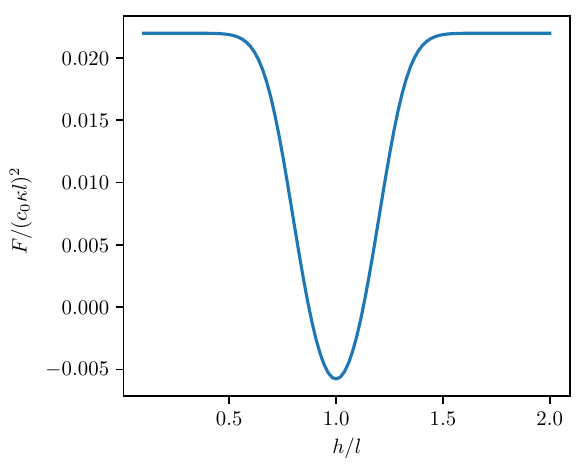}

	\caption{Plot of the energy density from Eqn. \eqref{eq:app_proteinenergy}. The energy density $F$ is non-dimensionalised by $c_0\kappa l^2$, and the height by $l$. The binding distribution width is set to $\sigma/l=1/5$ and the timescale ratio is $\tau_{\text{on}}/\tau_{\text{off}}=1/3.$
	\label{app:fig_protein energy}}
\end{figure}

Finally, we note that the pressure contribution from the bound proteins as described in Eq. \eqref{eq:protspringpressure} with $\hat{c}(\hat{h})$ given by Eq. \eqref{eq:app_ceq} can be written as the functional derivative of a free energy with respect to the film profile $\hat{h}(\hat{x},\hat{y})$.

\begin{equation}
    F_{\text{adh}}[h] = \int- \frac{ c_0\kappa \sigma^2}{2}\left(\ln\left[\exp\left(-\frac{(\hat{h}(\hat{x},\hat{y},\hat{t})-l)^2}{\sigma^2}\right)+\frac{\tau_{\text{on}}}{\tau_{\text{off}}}\right]\right)dA,
\label{eq:app_proteinenergy}
\end{equation}
This corresponds to a single energy well, as is shown in Fig. \ref{app:fig_protein energy}.

\section{Non-dimensionalization of equations}\label{app:Non-dimensionalization}

As mentioned in section \ref{subsec:Numerics}, we non-dimensionalize the governing equations before computing our numerical solutions. In this section, we provide the details of how we have done so for each of the cases discussed in sections \ref{sec:AttachmentTime} and \ref{sec:Coarsening}.

\subsection{Attachment simulations}\label{app:Non-dimensionalization_att}
For the simulations to estimate the binding time of a fluctuating film, a quasi-1D domain of length $L$ in the x-direction and width $w$ in the y-direction is considered. Since we are focused on the effects of bending, tension is not included. We also ignore the protein binding term since we are interested in how long it takes to reach the values of $h$ where these become significant (similar results are observed when they are included, but the theoretical calculation is more challenging). Under these circumstances and considering the bending-dominated case, the 1D version of Eq. \eqref{eq:TF_dim} simplifies to
\begin{equation}
    \frac{\partial \hat{h}(\hat{x},\hat{t})}{\partial \hat{t}} =  \frac{\partial}{\partial \hat{x}} \left(\frac{\hat{h}(\hat{\hat{x}},\hat{t})^3}{12\mu}\frac{\partial}{\partial \hat{x}} \left(B\frac{\partial^4}{\partial \hat{x}^4} \hat{h}(\hat{x},\hat{t})\right)\right) + \sqrt{\frac{k_B T}{6\mu w }}\frac{\partial}{\partial \hat{x}} \left(\hat{h}(\hat{x},\hat{t})^{3/2}\hat{\eta}(\hat{x},\hat{t})\right).
\label{eq:app_dimTF_Attachment}
\end{equation}
 The finite width $w$ must be included here since thermal flucutations are inherently a 3D phenomenon, but will in the end be absorbed into the prefactor of the stochastic term~\citep{grun2006}. We nondimensionalize Eq. \eqref{eq:app_dimTF_Attachment} by introducing the scaling relations:
\[\hat{h}=hh_0, \hspace{10pt} \hat{x}=xL
\hspace{10pt}
\hat{t}=t\frac{12 \mu L^6}{Bh_0^3},
\hspace{10pt}\hat{\eta}=\eta\sqrt{\frac{B h_0^3}{12 \mu L^7}}\]
where the dimensionless variables are without hats. The time is scaled by a characteristic time scale on which bent elastohydrodynamic thin films relax  \citep{carlson_2018} (which can be obtained by scaling the left hand side of Eq. \eqref{eq:app_dimTF_Attachment} with the first term on the right hand side), and $\boldsymbol{\eta}$ is scaled by a dimensionally correct combination of the time and horizontal length scales. When these relations are inserted into Eq. \eqref{eq:app_dimTF_Attachment}, the dimensionless thin film equation becomes 
\begin{equation}
    \frac{\partial h}{\partial t} =  \frac{\partial}{\partial x} \left(h^3\frac{\partial}{\partial x} \left(\frac{\partial^4}{\partial x^4} h\right)\right) + Q_{1D}\frac{\partial}{\partial x} \left(h^{3/2}\eta\right), 
\label{eq:app_nondimTF_Attachment}
\end{equation}
The non-dimensional number $Q_{1D}=\frac{L}{h_0}\sqrt{\frac{2k_B T L}{Bw}}$ represents the strength of the thermal fluctuations in the domain. In fact, $Q_{1D}$ is directly proportional to the thermal roughness of the film by the relation $Q_{1D}=(\langle|\delta \hat{h}|\rangle/h_0)(360)^{-1/2}$, where $\langle|\delta \hat{h}|\rangle$ is the thermal roughness of a film.

\subsection{Coarsening simulations}\label{app:Non-dimensionalization_coarse}
For the simulations of coarsening in thin films, a 3D domain was used, and the non-dimensionalization of the equations is different as opposed to the attachment simulations. In this section, we decribe how Eq. \eqref{eq:TF_dim} is non-dimensionalized when the protein dynamics of Eq. \eqref{eq:protspringpressure} are included. The approach is different in the bending-driven and tension-driven cases because in each case the horizontal length scale is non-dimensionalised by the characteristic domain size obtained by balancing the protein forces with the driving membrane force \citep{carlson_mahadevan_synapse_2015}.
\subsubsection{Bending-driven}

When coarsening is driven by bending, Eq. \eqref{eq:TF_dim} becomes
\begin{equation}
    \frac{\partial \hat{h}}{\partial \hat{t}} = \hat{\nabla}\cdot\left(\frac{\hat{h}^3}{12\mu}\hat{\nabla} \left(B\hat{\nabla}^4\hat{h}+\kappa (\hat{h}-l)c\right) \right) + \sqrt{\frac{k_B T}{6\mu }}\hat{\nabla} \cdot\left(\hat{h}^{3/2}\boldsymbol{\hat{\eta}}\right)
    \label{eq:app_dimTF_bendcoarse}.
\end{equation}
We nondimensionalize Eq. \eqref{eq:app_dimTF_bendcoarse} by introducing the scaling relations

\[\hat{h}=hl, \hspace{10pt} \hat{x}=x\left(\frac{B}{c_0\kappa}\right)^{1/4},
\hspace{10pt} \hat{y}=y\left(\frac{B}{c_0\kappa}\right)^{1/4},
\hspace{10pt}
\hat{t}=t\frac{12\mu B^{1/2}}{l^3(c_0\kappa)^{3/2}}, \hspace{10pt}
\hat{c}=cc_0, \hspace{10pt}
\boldsymbol{\hat{\eta}}=\boldsymbol{\eta}\frac{c_0\kappa l^{3/2}}{(12B\mu)^{1/2}}
\]
where the dimensionless variables are without hats. The characteristic horizontal length scale by which $x$ and $y$ are non-dimensionalized, $(B/c_0\kappa)^{1/4}$, comes from a balance between the protein spring pressure and the bending pressure, and describes the characteristic length scale of protein domains \citep{carlson_mahadevan_synapse_2015}. The timescale comes from a balance of the left-hand side term in Eq. \eqref{eq:app_dimTF_bendcoarse} with the bending pressure term, using the aforementioned horizontal length scale to scale the gradient $\hat{\nabla}$. When these scalings are introduced into Eq. \eqref{eq:app_nondimTF_bendcoarse}, the dimensionless thin film equation becomes
\begin{equation}
    \frac{\partial h}{\partial t} = \nabla\cdot\left(h^3\nabla \left(\nabla^4h+ (h-1)c\right) \right) + Q_{B}\nabla \cdot\left(h^{3/2}\boldsymbol{\eta}\right)
    \label{eq:app_nondimTF_bendcoarse},
\end{equation}
The non-dimensional number $Q_{B}=\frac{1}{l}\sqrt{\frac{2k_B T }{B^{1/2}(\kappa c_0)^{1/2}}}$ represents the strength of the thermal fluctuations in the domain. In fact, $Q_{B}$ is directly proportional to the average amplitude of thermal fluctuations of the film by the relation $Q_{B}\sim(\langle|\delta \hat{h}|\rangle/l)((B/(c_0\kappa))^{1/4}/L)$, where $\langle|\delta \hat{h}|\rangle$ is the thermal roughness of a film. A physical interpretation for $Q_{B}$ is thus a non-dimensionless thermal roughness of the film, although the amplitude is dependent on the domain size, which is an inherent and poorly studied feature of thermal fluctuations in such films.

\subsubsection{Tension-driven}

When coarsening is driven by tension, Eq. \eqref{eq:TF_dim} becomes
\begin{equation}
    \frac{\partial \hat{h}}{\partial \hat{t}} = \hat{\nabla}\cdot\left(\frac{\hat{h}^3}{12\mu}\hat{\nabla} \left(-\gamma\hat{\nabla}^2\hat{h}+\kappa (\hat{h}-l)\hat{c}\right) \right) + \sqrt{\frac{k_B T}{6\mu }}\hat{\nabla} \cdot\left(\hat{h}^{3/2}\boldsymbol{\hat{\eta}}\right)
    \label{eq:app_dimTF_intcoarse}.
\end{equation}
We nondimensionalize Eq. \eqref{eq:app_dimTF_intcoarse} by introducing the scaling relations

\[\hat{h}=hl, \hspace{10pt} \hat{x}=x\left(\frac{\gamma}{c_0\kappa}\right)^{1/2},
\hspace{10pt} \hat{y}=y\left(\frac{\gamma}{c_0\kappa}\right)^{1/2},
\hspace{10pt}
\hat{t}=t\frac{12\mu \gamma}{l^3(c_0\kappa)^{2}}, \hspace{10pt}
\hat{c}=cc_0, \hspace{10pt}
\boldsymbol{\hat{\eta}}=\boldsymbol{\eta}\frac{(c_0\kappa)^{3/2} l^{3/2}}{(12\mu)^{1/2}\gamma}
\]

where the dimensionless variables are without hats. The characteristic horizontal length scale by which $x$ and $y$ are non-dimensionalized, $(\gamma/c_0\kappa)^{1/2}$, comes from a balance between the protein spring pressure and the interfacial tension, and describes the characteristic length scale of protein domains. The timescale comes from a balance of the left-hand side term in Eq. \eqref{eq:app_dimTF_intcoarse} with the tension term, using the aforementioned horizontal length scale to scale the gradient $\hat{\nabla}$. When these scalings are introduced into Eq. \eqref{eq:app_dimTF_intcoarse}, the dimensionless thin film equation becomes
\begin{equation}
    \frac{\partial h}{\partial t} = \nabla\cdot\left(h^3\nabla \left(-\nabla^2h+ (h-1)c\right) \right) + Q_{\gamma}\nabla \cdot\left(h^{3/2}\boldsymbol{\eta}\right)
    \label{eq:app_nondimTF_tenscoarse},
\end{equation}
 The non-dimensional number $Q_{\gamma}=\frac{1}{l}\sqrt{\frac{2k_B T }{\gamma}}$ represents the strength of the thermal fluctuations in the domain. In fact, $Q_{\gamma}$ is directly proportional to the average amplitude of thermal fluctuations of a freely fluctuating film subjected to interfacial tension by the relation $Q_{\gamma}\sim(\langle|\delta \hat{h}|\rangle/l)$, where $\langle|\delta \hat{h}|\rangle$ is the thermal roughness of a film without the protein binding term. A physical interpretation for $Q_{\gamma}$ is thus a non-dimensionless thermal roughness of the film, which for a tension-dominated film is not dependent on $L$. This has in fact been verified numerically in a previous work where the length scale $\langle|\delta \hat{h}|\rangle$ was calculated directly \citep{dhaliwal2024instability}.

\section{Rare-event theory}\label{app:Rare-event}

To predict the average waiting time (mean first passage time) for proteins to bind, we apply the rare-event theory for a gradient flow following the procedure outlined in~\citep{liuMeanFirstPassage2024}, with a simple modification in the asymptotic since the system is not bistable, that is, the transition is not from one local minimum to another. For the mean first passage time of a non-gradient system, the reader can refer to~\citep{grafkeSharpAsymptoticEstimates2024}. To make the derivation more general and easier to follow, we will first derive the formula of the mean first passage time for a general gradient flow, and then demonstrate the application to the elastohydrodynamic thin film equation.

\subsection{Mean first passage time for gradient flow}\label{appsub:fpttheory}

Let us first formalise the problem. Consider a general stochastic differential equation describing a gradient flow
\begin{equation}
\label{eq:app_rare_events_SDE}
    dX_t = -M(X_t)\nabla F(X_t) dt + \sqrt{2\varepsilon} M_{1/2}(X_t)dW_t\,,
\end{equation}
where $X_t\in\mathbb{R}^{n}$ is a random variable, $M(X_t):\mathbb{R}^n\to\mathbb{R}^{n\times n}$ is the semi-positive definite mobility matrix (or mobility operator), $F(X_t)$ is some free energy of the system, $\nabla$ is the gradient with respect to $X_t$, $\varepsilon$ is the noise amplitude, $M_{1/2}(X_t):\mathbb{R}^{n}\to\mathbb{R}^{n\times n}$ is the ``square root'' of the mobility, namely $M_{1/2}M^{T}_{1/2}=M$ ($^{T}$ stands for transpose or Hermitian adjoint), and $W_t$ is the $n$-dimensional Brownian motion. It describes a system driven by the negative gradient of energy, i.e. a force that minimize the energy locally, while being disturbed by a Gaussian white noise. For simplicity, let $-M(X_t)\nabla F(X_t)= b(X_t)$ and $\sqrt{2}M_{1/2}(X_t)=\xi(X_t)$, Eq.~\eqref{eq:app_rare_events_SDE} is then
\begin{equation}
\label{eq:app_rare_events_SDE2}
    dX_t = b(X_t)dt + \sqrt{\varepsilon}\xi(X_t)dW_t\,.
\end{equation}
Assume there is a stable fixed point $x_A$ such that $b(x_A)=0$, and the time it takes for the system to first exit a basin of attraction $D$ of $x_A$ starting at $x\in D$ is denoted as
\begin{equation*}
    T_B(x) = \inf\{t>0|X_t\notin D\}.
\end{equation*}
We are interested in the the mean first passage time, $w_B(x)=\mathbb{E}[T_B(x)]$, that is, the expectation of $T_B$,  which fulfills the inhomogeneous stationary Kolmogorov equation~\citep{gardiner2009stochastic}
\begin{equation}
\label{eq:app_komogorov}
    \begin{cases}
        \mathcal{L}w_B(x)=-1,& \text{for } x\in D\\
        w_B(x)=0,         ,& \text{for } x\in\partial D,
    \end{cases}
\end{equation}
where $\partial D$ is the boundary of the basin of attraction $D$. Here, we choose $D$ such that there is a global minimum $x_B$ of $F(x)$ on $\partial D$, and the normal vector $\hat n$ of $\partial D$ pointing outward at $x_B$ aligns with $\nabla F$, that is, $\partial D$ is tangent to the contour line of the energy landscape only at $x_B$. $\mathcal{L}$ is the generator of Eq.~\eqref{eq:app_rare_events_SDE2},
\begin{equation}
\label{eq:app_generator}
    \mathcal{L} = b(x)\cdot\nabla+\frac{1}{2}\varepsilon a(x):\nabla\nabla,
\end{equation}
where $:$ is the scalar product. From the generator we can deduce the invariant distribution $\rho_\infty(x)$ through the stationary Fokker-Planck equation
\begin{equation*}
    \mathcal{L}^\dagger\rho_\infty=0,
\end{equation*}
where $\mathcal{L}^\dagger$ is the $L^2$-adjoint of the generator
\begin{equation}
    \mathcal{L}^\dagger\circ =-\nabla\cdot(b(x)\,\circ)+\frac{1}{2}\varepsilon\nabla\nabla :(a(x)\,\circ).
\end{equation}
In the case of a gradient flow Eq.~\eqref{eq:app_rare_events_SDE}, one can show that the invariant distribution is given as the Gibbs distribution
\begin{equation}
    \rho_\infty(x) = C e^{-F(x)/\varepsilon},
\end{equation}
where $C$ is some constant. From the large deviation theory~\citep{freidlinRandomPerturbationsDynamical2012} we know that for $\varepsilon\to 0$,
\begin{equation}
\label{eq:app_wB_tau}
    w_B(x_A)\asymp e^{\Delta F/\varepsilon},
\end{equation}
where $\Delta F=(F(x_B)-F(x_A))$. In another word, in the limit of small noise amplitude $\varepsilon$, the process almost certainly exit the basin of attraction $D$ at the saddle point $x_B$, and the mean first passage time scales exponentially with the energy barrier between $x_A$ and $x_B$ with some unknown prefactor. To calculate the prefactor, we define a new random variable
\begin{equation*}
    \tau (x) = e^{-\Delta F/\varepsilon} w_B(x),
\end{equation*}
and the Kolmogorov equation~\eqref{eq:app_komogorov} becomes
\begin{equation}
\label{eq:app_komogorov2}
    \begin{cases}
        \mathcal{L}\tau(x)=-e^{-\Delta F/\varepsilon},& \text{for } x\in D\\
        \tau(x)=0,         ,& \text{for } x\in\partial D.
    \end{cases}
\end{equation}
Consider a point $x\in D$ near the boundary $\partial D$, we can expand asymptotically in the direction of $\hat n$
\begin{equation}
    x = u - \varepsilon\eta\hat n,
\end{equation}
where $u\in\partial D$ and $\eta>0$. Note that this is different from the asymptotic expansion used in~\citep{liuMeanFirstPassage2024}. This lead to
\begin{align}
\label{eq:app_symp_grad}
    \nabla f &= \frac{\partial f}{\partial x_i} \mathbf{e}_i = \frac{d f}{d \eta}\frac{\partial \eta}{\partial x_i}\mathbf{e}_i = -\frac{1}{\varepsilon}\frac{df}{d\eta}\hat n,\\
    a:\nabla\nabla f &= a_{ij}\frac{\partial}{\partial x_i}\frac{\partial }{\partial x_j}f = \frac{1}{\varepsilon^2}a_{ij}\frac{d^2 f}{d\eta^2}\hat n_i\hat n_j = \frac{1}{\varepsilon^2}\hat n\cdot (a\hat n)\frac{d^2 f}{d\eta^2}.
\end{align}
We can then expand the generator to the leading order and rewrite Eq.~\eqref{eq:app_generator} as
\begin{align}
\notag
    \mathcal{L} &= b(x)\cdot \nabla + \frac{1}{2}\varepsilon a(x):\nabla\nabla \approx b(u)\cdot\nabla + \frac{1}{2}\varepsilon a(u) :\nabla\nabla\\
    \label{eq:app_generator_expand}
    &=\frac{1}{\varepsilon}\Big[ \underbrace{-b(u)\cdot \hat n(u)}_{\beta(u)}\frac{d}{d\eta}+\underbrace{\frac{1}{2}\hat n \cdot(a(u)\hat n)}_{\alpha(u)}\frac{d^2}{d\eta^2}  \Big].
\end{align}
As $\varepsilon\to 0$, the leading term of the right hand side of the Kolmogorov equation~\eqref{eq:app_komogorov2} is zero, and we arrive at
\begin{equation*}
    0 = \mathcal{L}\tau(\eta) = \frac{1}{\varepsilon}\Big[\beta(u)\frac{\partial \tau}{\partial \eta}+\alpha(u)\frac{\partial^2\tau}{\partial \eta^2}\Big],
\end{equation*}
which can be solved with the boundary condition~\eqref{eq:app_komogorov2} to give
\begin{equation}
\label{eq:app_tau_solved}
    \tau(\eta) = C_0\left[1-\exp\left(-\frac{\beta}{\alpha}\eta\right)\right],
\end{equation}
where $C_0$ is some constant that we will determine next. Integrating the Kolmogorov equation $\mathcal{L}\tau(x) = -\exp(-\Delta F/\varepsilon)$ against the invariant density $\rho_\infty(x)$, and apply the divergence theorem repeatedly we get
\begin{align*}
    -\exp(-\Delta F/\varepsilon)\int_B\rho_\infty(x)dx &= \int_B\mathcal{L}\tau(x)\rho_\infty(x) dx\\
    &=\int_B b(x)\cdot\nabla\tau(x)\rho_\infty(x) + \frac{1}{2}\varepsilon a(x):\nabla\nabla\tau (x)\rho_\infty dx\\
    &\overset{\makebox[0pt]{\mbox{\normalfont \eqref{eq:app_symp_grad}}}}{=}\;\;-\frac{1}{2}\int_{\partial D}\rho_\infty(u)\alpha(u)\frac{d\tau}{d\eta} du\quad
    \overset{\makebox[0pt]{\mbox{\normalfont \eqref{eq:app_tau_solved}}}}{=}\;\;-\frac{1}{2}C_0\int_{\partial D}\beta(u)\rho_\infty(u) du.
\end{align*}
This then lead to
\begin{equation*}
    C_0 = 2\exp(-\Delta F/\varepsilon)\dfrac{\displaystyle\int_B\rho_\infty(x) dx}{\displaystyle\int_{\partial D}\beta(u)\rho_\infty(u) du},
\end{equation*}
and the mean first passage time by Eq.~\eqref{eq:app_wB_tau} is given by
\begin{equation*}
    w_B = 2\dfrac{\displaystyle\int_B\rho_\infty(x) dx}{\displaystyle\int_{\partial D}\beta(u)\rho_\infty(u) du}.
\end{equation*}
If the mobility operator has conserved quantity~\citep{liuMeanFirstPassage2024}, the integrations must be performed over the hyperplane,
\begin{equation}
\label{eq:app_wb}
    w_B = 2\dfrac{\displaystyle\int_{D/K}\rho_\infty(x) dx}{\displaystyle\int_{\partial D/K}\beta(u)\rho_\infty(u) du},
\end{equation}
where $K$ represents the dimensions perpendicular to the conserved quantities.
\subsection{Average waiting time of adhesion for elastic membrane}
We now apply the formula for the first passage time~(\ref{eq:app_wb}) to the 1D elastohydrodynamic
thin film equation~\eqref{eq:nondimTF_Attachment} with only bending and thermal noise on a periodic domain $x\in[0,1]$. Its gradient flow form is given by
\begin{equation*}
    \frac{\partial h}{\partial t} = -m[h]\frac{\delta F}{\delta h} + \sqrt{2\varepsilon}m_{1/2}[h]\bm{\eta},
\end{equation*}
where $F[h]$ is the energy functional
\begin{equation}
\label{eq:app_membrane_U}
    F[h] = \int_0^1 \frac{1}{2}\left(\frac{\partial^2 h}{\partial x^2}\right)^2 dx,
\end{equation}
$\delta F/\delta h$ is the functional derivative of the energy functional\footnote{For calculation of the functional derivative, see~\citep{SprittlesJBLGrafke2023,liuMeanFirstPassage2024}.},
\begin{equation}
    \label{eq:app_membrane_dU}
    \frac{\delta F}{\delta h}[h] = \frac{\partial^4 h}{\partial x^4},
\end{equation}
$m[h]$ is the mobility operator (acting on a test-function $\xi$)
\begin{equation}
\label{eq:app_membrane_M}
    m[h]\xi = -\frac{\partial}{\partial x}\left(h^3\frac{\partial}{\partial x}\xi\right),
\end{equation}
$m_{1/2}[h]$ is the ``square root'' of the mobility operator (acting on a test-function $\xi$)
\begin{equation*}
    m_{1/2}[h]\xi = \sqrt{h^3}\frac{\partial}{\partial x}\xi,
\end{equation*}
$\varepsilon$ is the noise amplitude
\begin{equation*}
    \varepsilon = Q_{1D}^2/2,
\end{equation*}
and $\bm \eta$ is a Gaussian white noise. It is obvious that the mobility operator conserves mass given that a constant function is a zero eigenfunction~\citep{liuMeanFirstPassage2024}, and the formula for the average waiting time must respect the conserved quantity, which is Eq.~\eqref{eq:app_wb}. Without loss of generality we let the mass to be $h_0$, and we would like to calculate the average time it takes for thermal fluctuations to drive a flat membrane $h_A(x) = h_0$ to bend into a shape $h(x)$ with minimum height $h^*$. This defines the boundary of the attractive basin $\partial D$, namely, all the membrane shapes that have minimum height $h^*$. 
Since the invariant density $\rho_\infty$ is exponential, the integrals in Eq.~\eqref{eq:app_wb} can be approximated using the Laplace method
\begin{align*}
    \int_{D/K}\rho_\infty dh &= C\int_{D/K}\exp\left(-\frac{F[h]}{\varepsilon}\right) dh\\
    &\approx C\exp\left(-\frac{F[h_0]}{\varepsilon}\right)\int_{D/K}\exp\left(-\frac{1}{2\varepsilon}(h-h_0)^TH[h_0](h-h_0)\right)dh,\\
    \int_{\partial D/K}\beta(h)\rho_\infty dh &= C\int_{\partial D/K}\beta(h)\exp\left(-\frac{F[h]}{\varepsilon}\right)dh \\
    &\approx C\beta(h_B)\exp\left(-\frac{F[h_B]}{\varepsilon}\right)\int_{\partial D/K}\exp\left(-\frac{1}{2\varepsilon}(h-h_B)^T H[h_B](h-h_B)\right)dh,
\end{align*}
where $H[h] = \delta^2 F/\delta h^2$ is the Hessian operator of the energy $F(h)$, and 
\begin{equation}
\label{eq:app_Hessian_product}
    h^T H[h] h = \int_0^1 h\frac{\partial^4 h}{\partial x^4} dx,
\end{equation}
can be interpreted as the inner product with respect to the Hessian~\citep{liuMeanFirstPassage2024}. Apply these to Eq.(~\ref{eq:app_wb}), and the average waiting time is given by
\begin{equation}
    w_B = 2\frac{\int_{D/K}\exp\left(-\frac{1}{2\varepsilon}(h-h_0)^TH[h_0](h-h_0)\right)dh}{\int_{\partial D/K}\exp\left(-\frac{1}{2\varepsilon}(h-h_B)^T H[h_B](h-h_B)\right)dh}\exp\left(\frac{F[h_B]-F[h_0]}{\varepsilon}\right),\label{app:eq_wb_biig}
\end{equation}
whose form agrees with the result of the Large deviation theory~(\ref{eq:app_wB_tau}), and next we need to determine the energy barrier $F[h_B]-F[h_0]$ and evaluate the integrals.

\subsubsection{Minimum energy profile for binding}

If the noise amplitude is very small, $\varepsilon\ll 1$, it is almost certain that the system will reach $\partial D$ while increasing the least energy possible. So finding the energy barrier $F[h_B]-F[h_0]$ is reduced to finding the $h_B$ with minimum energy $F$, that has a minimum height $h^*$, while also conserving mass and satisfying the boundary conditions of the problem. Mathematically, this corresponds to a constrained optimization problem in which we seek to minimize the energy functional $F[h]$. In this section, we use the Lagrange multiplier method~\citep{Boas2006} to find $h_B(x)$ and its corresponding energy $F[h_B]$. The Lagrangian taking into account the conservation of mass is
\begin{equation}
    \tilde{F}=\frac{1}{2}\left(\frac{\partial\tilde{F}}{\partial x}\right)^2+\lambda(h-h_0),
\end{equation}
where $\lambda$ is the Lagrange multiplier. The Euler-Lagrange equation is then given by
\begin{equation}
    \lambda+\frac{\partial^4 h}{\partial x^4}=0,
\end{equation}
for which the general solution is a fourth order polynomial. In addition, the solution must satisfy our boundary and symmetry conditions, written as:
\begin{gather}\label{app:eq_ELBC_h*}
    h(0)=h(1)=h^*\\ \label{app:eq_ELBC_sym}
\frac{\partial h}{\partial x}\Bigr|_{x=1/2}=\frac{\partial^3 h}{\partial x^3}\Bigr|_{x=1/2}=0  \\  \label{app:eq_ELBC_smooth}
\frac{\partial h}{\partial x}\Bigr|_{x=0}=\frac{\partial h}{\partial x}\Bigr|_{x=1}=0 .
\end{gather}
Eq. \eqref{app:eq_ELBC_smooth} specifies that the profile must be smooth at the rupture point. This is not the case for tension-dominated films \citep{SprittlesJBLGrafke2023}, but this condition was included here since numerical results show only smooth profiles. 

When the boundary conditions of Eqs. \eqref{app:eq_ELBC_h*}-\eqref{app:eq_ELBC_smooth} and conservation of mass are applied to a fourth-order polynomial, we obtain the following profile:
\begin{equation} \label{app:eq_profile}
    h_B(x)=h^*+30(h_0-h^*)(x^4-2x^3+x^2)
\end{equation}
This is the profile we predict as the average profile at the moment of binding when the parameters $Q_{1D}$ and $h_0-h^*$ are selected such that the binding event is sufficiently rare. This prediction is confirmed by the data presented in Fig. \ref{fig:Rare-event_multi}, where the dashed blue line represents Eq. \eqref{app:eq_profile} and the black dots represent the average profile for 15 individual simulations. 

The energy barrier is then given by
\begin{equation}
F[h_B]-F[h_0]= 360 (h_0-h^*)^2.
\end{equation}

\subsubsection{Evaluation of the integrals}
We now turn to the integrals in Eq. \eqref{app:eq_wb_biig}. Due to periodicity, membrane shapes can be decomposed into Fourier modes. At the cost of a small error, this allows us to evaluate the integrals analytically. Decomposing $h-h_0$ into Fourier modes gives
\begin{equation*}
    h(x)-h_0 = \sum_{n=1}^\infty a_n\cos(2\pi nx) + b_n\sin(2\pi nx),
\end{equation*}
and so
\begin{equation*}
    (h-h_0)^T H[h_0](h-h_0) = \sum_{n=1}^\infty (2\pi n)^8\frac{1}{2}(a_n^2+b_n^2),
\end{equation*}
which lead to
\begin{align}
\notag
    \int_{D/K}\exp\left(-\frac{1}{2\varepsilon}(h-h_0)^TH[h_0](h-h_0)\right)dh &=\int_{-\infty}^\infty\exp\left(-\frac{1}{2\varepsilon}\sum_{n=1}^{\infty}(2\pi n)^8\frac{1}{2}(a_n^2+b_n^2)\right) \prod_{p=1}^\infty da_n db_n\\
    \label{eq:app_wb_frac_top}
    &=\prod_{n=1}^\infty\frac{4\varepsilon\pi}{(2\pi n)^8}.
\end{align}
We can also decompose $h-h_B$ into Fourier modes, however, care must be taken when integrating over $\partial D/K$ by retaining only the modes parallel to $\partial D$ at $h_B$, or equivalently, by removing the modes perpendicular to $\partial D$ at $h_B$. Due to the way we constructed $\partial D$, the only mode to remove is $\hat n(h_B)$, that is
\begin{equation}
    \hat n(h_B) = \frac{\delta F[h_B]/\delta h}{|\delta F[h_B]/\delta h|} = 2\cos(2\pi x).
\end{equation}
And so the decomposition is given by
\begin{equation*}
    h(x)-h_B(x) = d_1\sin(2\pi x) + \sum_{n=2}^\infty c_n\cos(2\pi nx)+d_n\sin(2\pi nx),
\end{equation*}
which lead to 
\begin{equation*}
    (h-h_B)^T H[h_B](h-h_B) = \frac{1}{2}(2\pi)^8 d_1^2 + \sum_{n=2}^\infty(2\pi n)^8\frac{1}{2}(a_n^2+b_n^2)
\end{equation*}
and so
\begin{align}
\notag
    &\int_{\partial D/K}\exp\left(-\frac{1}{2\varepsilon}(h-h_B)^TH[h_B](h-h_B)\right)dh \\
    \notag
    =& \int_{-\infty}^\infty \exp\left(-\frac{1}{2\varepsilon}(2\pi)^8\frac{1}{2}d_1^2\right) dd_1\int_{-\infty}^\infty\exp\left(-\frac{1}{2\varepsilon}\sum_{n=2}^\infty(2\pi n)^8\frac{1}{2}(c_n^2+d_n^2)\right)\prod_{n=2}^\infty dc_n dd_n\\
    \label{eq:app_wb_frac_bot}
    =& \sqrt{\frac{2\varepsilon\pi}{(2\pi)^8}}\prod_{n=2}^\infty\frac{4\varepsilon\pi}{(2\pi n)^8}.
\end{align}
\subsubsection{Final result}
Combining Eq.~\eqref{eq:app_wb}\eqref{eq:app_wb_frac_top}\eqref{eq:app_wb_frac_bot}, we get
\begin{equation}
\label{eq:app_wb_membrane_almost_final}
    w_B = \frac{2}{\beta(h_B)}\exp\left(\frac{F[h_B]-F[h_0]}{\varepsilon}\right)\frac{4\varepsilon\pi}{(2\pi)^8},
\end{equation}
and the only missing part is $\beta(h_B)$. By Eq.~\eqref{eq:app_generator_expand}\eqref{eq:app_membrane_M}\eqref{eq:app_membrane_dU}, we have
\begin{align*}
    \beta(h_B) &= -\hat n(h_B)\cdot \left(M(h_B)\frac{\delta F}{\delta h}(h_B)\right)=-(h-h^*)(2\pi)^6\left(\frac{1}{2}h_0^3+\frac{3}{8}h_0(h_0-h^*)^2\right),
\end{align*}
where the inner product is interpreted as integral, same as Eq.~\eqref{eq:app_Hessian_product}. Finally, with Eq.~\eqref{eq:app_membrane_U}, we arrive at the expression for the average waiting time used in the main text Eq.~\eqref{eq:Rareeventprediction}
\begin{equation}
    \label{eq:app_wb_membrane_final}
    \langle t_B \rangle = w_B = \frac{1}{\beta (h_B)}\sqrt{\frac{Q_{1D}^2}{(2\pi)^7}} \exp\left(720\left(\frac{h_0-h^*}{Q_{1D}}\right)^2\right).
\end{equation}

\section{Domain coarsening theory}\label{app:Coarsening}
In section \ref{sec:Coarsening}, we present the results of simulations in which lumens are formed in the space between two membranes, and then coarsen with time. Here, we provide more details about how we quantify and rationalize the coarsening behavior. First, we will discuss how we compute the characteristic length scale $L_c$ in light of previous work on phase separating systems. Second, we will provide more details on the theoretical description from section \ref{subsec:CoarseningTheory} which we use to rationalize the decreased growth rate for bending-driven coarsening.

\subsection{Computation of the scaling length \texorpdfstring{$L_c$}{Lc}}\label{appsub:Lcalc}

The separation of two phases in a 2D domain is always a complex process, where the fluxes giving rise to coarsening are inherently local phenomena that depend on the specific morphology of the profile. For non-uniformly distributed profiles such as those shown in Fig. \ref{fig:Schematics}$(a-b)$, it is not straightforward to describe the morphology of the profile. Nevertheless, the system does exhibit a clear and obvious change with respect to time, as shown in Figs. \ref{fig:CoarseningContours_bend_varQ} and \ref{fig:CoarseningContours_int_varh0}. To gain a quantitative understanding of these phenomena, a statistical approach can be used to gain insight into the ensemble-averaged behavior of such profiles over time. Specifically, the well-established scaling hypothesis for phase separating dynamical systems states that during the late-stage of coarsening, the domain structure is self-similar with respect to time when the length is rescaled by a single length scale $L_c(t)$ \citep{Furukawa1985,Bray1994}. When systems demonstrate this type of behaviour, the morphologies of individual realizations of the system are still distinct, but multiple realizations will look similar to the eye at a single timestep. If one then zooms out so that the length scale increases in accordance with $L_c(t)$, the morphologies of multiple trajectories will be indistinguishable even as time increases \citep{CamleyBrown2011}. Many systems have been shown to demonstrate this type of behaviour both numerically and experimentally \citep{Furukawa1999,KomuraTakeda1985,TatenoTanaka2021,LalSutton2020,LivetSutton2001,SuParikh2024,SungHan1995}.

The scaling hypothesis is thus a powerful concept that allows us to meaningfully understand coarsening domains. The question that remains, however, is how the length scale $L_c(t)$ can be calculated from a height profile $h(x,y,t)$. The scaling hypothesis suggests that the domain structure should be independent of time except for a dependence on $L_c(t)$. The structure can by represented by its equal-time correlation function, which is defined as
\begin{equation}
C(\boldsymbol{r},t)=\langle h(\boldsymbol{x}+\boldsymbol{r},t)h(\boldsymbol{x},t)\rangle, 
\end{equation}
where $\boldsymbol{x}$ is the position vector $(x,y)$, $\boldsymbol{r}$ is a displacement vector, and the angular brackets represent an ensemble average. If the scaling hypothesis is valid, $C(\boldsymbol{r},t)$ should behave according to
\begin{equation}
C(\boldsymbol{r},t)=f\left(\frac{r}{L_c(t)}\right).
\end{equation}
The equal-time structure factor, $S(\textbf{k},t)$ is defined as
\begin{equation}
    S(\textbf{k},t)=\langle h_\textbf{k}(t)h_\textbf{-k}(t)\rangle,
\end{equation}
where $\boldsymbol{k}$ is now a wavevector and $h_k$ is the 2D Fourier transform of $h(\boldsymbol{x)}$. $S(\textbf{k},t)$ is simply the Fourier transform of $C(\boldsymbol{r},t)$, and must then have the scaling form 
\begin{equation}
    S(\textbf{k},t)=L_c^2g(k,L_c),
\end{equation}
where $g$ is the Fourier transform of $f$ and \citep{Bray1994}. To calculate $L_c$ one needs to extract a length scale from $S(\textbf{k},t)$. This is commonly done by taking a moment of spherically averaged structure function $S(k,t)$ \citep{CamleyBrown2011,ShinozakiOono1993,Furukawa1999}. In this paper, we found more consistent results by calculating the characteristic length using the following expression as suggested by Shinozaki and Oono \citep{ShinozakiOono1993}: 
\begin{equation}
    L_c =2\pi\frac{\sum\limits_{k\ne0} |\boldsymbol{k}|^{-2}S(\boldsymbol{k})}{\sum\limits_{k\ne0} |\boldsymbol{k}|^{-1}S(\boldsymbol{k})}. \label{eq:app_SOexp}
\end{equation}

Our procedure for calculating $L_c$ thus consisted of the following. First $h_\textbf{k}$ was computed for an individual profile by taking a 2D fast Fourier transform of the height profile $h(x,y)$. Next, $h_\textbf{k}(t)h_\textbf{-k}$ was computed for each profile, since it is equal to the Fourier transform of $C(\textbf{r})$. Then, $S(\textbf{k})$ is computed as the ensemble average of $h_\textbf{k}h_\textbf{-k}$. Finally, Eq. \eqref{eq:app_SOexp} is used to compute $L_c$.

\subsection{Coarsening rate for bending-driven coarsening}\label{appsub:coarseningtheory}
\begin{figure}
	 \centering
\includegraphics[width=0.65\textwidth]{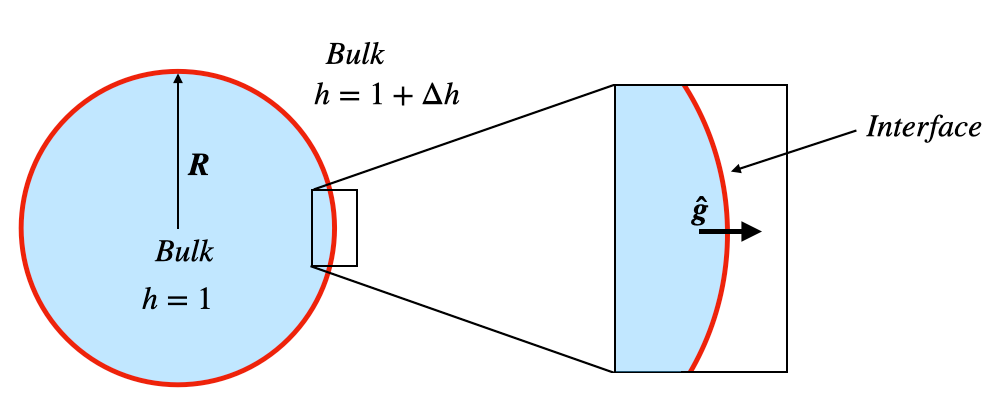}

	\caption{Schematic showing the physical picture of ``diffusive'' domain coarsening. Two bulk phases are separated by a curved interface. The unit vector $\boldsymbol{\hat{g}}$
	\label{app:fig_interfaceschematic}}
\end{figure}

In order to predict how the coarsening rate will change when the tension term is replaced by the bending term in Eq. \eqref{eq:membranepressures}, we follow the scaling logic of \citet{Bray1994}. We note that the following theory only takes into account the change of the pressure term. It does not account for the effects of nonlinear mobility or a complicated single-well potential which are included in our mathematical model. The simplified system we study in this section is thus the following:
\begin{equation}\label{simpTF}
    \frac{\partial h}{\partial t}= \nabla^2 \frac{\delta F }{\delta h},
\end{equation}
where the free energy $F$ is given by
\begin{equation}\label{bendenerggy}
    F=\int \left(\frac{1}{2}\left(\nabla^2h\right)^2+V(h)\right).
\end{equation}
Our goal will be to find out the motion of domain walls between two bulk regions, as depicted in Fig. \ref{app:fig_interfaceschematic}. We consider $V(h)$ to be a symmetric double-well potential with wells of even depth located at $h=1$ and $h=1+\Delta h$. Although this obviously does not match the energy from protein binding described in Eq. \eqref{eq:app_proteinenergy}, we expect the scaling rule to be insensitive to the exact form of $V(h)$. This is indeed justified by the numerical results in section \ref{subsec:Coarseningdesc}, which validate the $1/3$ power law predicted by Bray for constant mobility films even when the single-well potential of \eqref{eq:app_proteinenergy} is used.

When the energy from Eq. \eqref{bendenerggy} is inserted into Eq. \eqref{simpTF}, we get
\begin{equation}
    \frac{\partial h}{\partial t} = \nabla^2\left(\nabla^4h+ V'(h)\right) 
    \label{eq:app_nondimTF_coarsesimp}.
\end{equation}
We start by investigating the bulk phases where $h$ is in the bound state, and thus only slightly deviates from its equilibrium value. We linearize Eq. \eqref{eq:app_nondimTF_coarsesimp} by introducing $h=1+\tilde{h}$. This gives us the following linearized equation:
\begin{equation}
    \frac{\partial \tilde{h}}{\partial t} = \nabla^6\tilde{h}+ V''(1)\nabla^2\tilde{h} 
    \label{eq:app_linTFh},
\end{equation}
Since the characteristic domain size $L_c$ is large during late-stage coarsening, the $\sim \nabla^6$ term can be neglected, which reduces Eq. \eqref{eq:app_linTFh} to a diffusion equation for $\tilde{h}$ with diffusion coefficient $V''(1)$. Now, we expect that during late-stage coarsening, the diffusion field relaxes much faster than the motion of the domain walls. We can thus assume that this diffusion field is always in a quasi-equilibrium with the location of the interface, i.e., 
\begin{equation}
    \nabla^2h=0
\end{equation}
in the bulk regions away from the interface.

Before we start investigating the interfaces between the bulk phases, we will reformulate Eq. \eqref{eq:app_nondimTF_coarsesimp} in terms of a flux $\boldsymbol{j}$ given by the gradient of a chemical potential $\Phi$ (which is equivalent to the pressure in our system):
\begin{align}  \label{eq:app_contmu}
    \frac{\partial h}{\partial t}&= -\nabla\cdot \boldsymbol{j}\\
    \boldsymbol{j}&=-\nabla \Phi \label{eq:app_fluxmu}\\
    \Phi &= \frac{\partial V}{\partial h} +\nabla^4h \label{eq:app_mudef}.
\end{align} 
If we now introduce the linearised $h=1+\tilde{h}$ into Eq. \eqref{eq:app_mudef}, we get:
\begin{equation}
    \Phi = \nabla^4\tilde{h}+ V''(1)\tilde{h} 
    \label{eq:app_linmu}.
\end{equation}
Again, during the latter stages of coarsening the length scale is large, meaning that the $\sim \nabla^4$ term in Eq. \eqref{eq:app_linmu} is negligible. This gives us $\Phi \sim \tilde{h}$ in the bulk phases, meaning that $\Phi$ also satisfies the Laplace equation
\begin{equation}
    \nabla^2\Phi=0.
\end{equation}

To find out what $\Phi$ is at the boundary between the bulk phases, we first introduce the alternate coordinate system shown in Fig. \ref{app:fig_interfaceschematic} in order to simplify the vector calculus calculations. The unit vector $\boldsymbol{\hat{g}}$ points in the direction perpendicular to the interface ($g=\pm \infty$ in the bulk and $g=0$ at the centre of the interface). In the following we will study an interface with the circular geometry shown in Fig. \ref{app:fig_interfaceschematic}, meaning that $\nabla \phi =\left(\partial\phi/\partial g\right)\boldsymbol{\hat{g}}$ and  $\nabla\cdot\boldsymbol{\hat{g}}=1/R$ near the interface. This allows us to compute the Laplacian as
$\nabla^2\phi=\partial^2\phi/\partial g^2 + (\partial\phi/\partial g) \nabla\cdot\boldsymbol{\hat{g}}$. When these identities are inserted into Eq. \eqref{eq:app_mudef} we find that 
\begin{equation}\label{app:eq_muexp}
    \Phi = \frac{\partial V}{\partial h} +\frac{\partial^4 h}{\partial g^4}+\frac{2}{R}\frac{\partial^3 h}{\partial g^3}-\frac{1}{R^2}\frac{\partial^2 h}{\partial g^2}+\frac{1}{R^3}\frac{\partial h}{\partial g}.
\end{equation}
To find the value of $\Phi$ at an interface with radius of curvature $R$, we then multiply Eq. \eqref{app:eq_muexp}by $\partial h/\partial g$ and integrate across the interface from $g=-\infty$ to $g=\infty$. This gives us
\begin{equation}\label{app:eq_GTBC_mellomsteg}
    \Phi \Delta h = \Delta V
    +\int^\infty_{-\infty}\frac{\partial^4 h}{\partial g^4}\frac{\partial h}{\partial g}dg
    +\frac{2}{R}\int^\infty_{-\infty}\frac{\partial^3 h}{\partial g^3}\frac{\partial h}{\partial g}dg
    -\frac{1}{2R^2}\int^\infty_{-\infty}\frac{\partial}{\partial g}\left(\frac{\partial h}{\partial g}\right)^2dg
    +\frac{1}{R^3}\int^\infty_{-\infty}\left(\frac{\partial h}{\partial g}\right)^2dg.
\end{equation}
The $\Delta V$ term on the right hand side of Eq. \eqref{app:eq_GTBC_mellomsteg} is zero because the potential wells we are considering have equal depth. The third term on the right hand side disappears to to the bulk boundary condition $\frac{\partial h}{\partial g}\Big\rvert_{\pm \infty}=0$. The second term on the right hand side disappears upon using integration by parts and implementing the boundary conditions $\frac{\partial h}{\partial g}\Big\rvert_{\pm \infty}=\frac{\partial^2 h}{\partial g^2}\Big\rvert_{\pm \infty}=0$. Eq. \eqref{app:eq_GTBC_mellomsteg} then reduces to:
\begin{equation}\label{app:eq_GTBC}
    \Phi \Delta h = -\frac{2}{R}\int^\infty_{-\infty}\left(\frac{\partial^2 h}{\partial g^2}\right)^2dg+\frac{1}{R^3}\int^\infty_{-\infty}\left(\frac{\partial h}{\partial g}\right)^2dg,
\end{equation}
which is the equivalent of the Gibbs-Thompson boundary condition for an interface with bending energy. The factors $-2\int^\infty_\infty\left(\frac{\partial^2 h}{\partial g^2}\right)^2dg$ and $\int^\infty_\infty\left(\frac{\partial h}{\partial g}\right)^2dg$ can be interpreted as effective ``line tension'' coefficients $\Gamma_1$ and $\Gamma_2$ as they represent an energy per unit length associated with the circular interface between the two phases. Eq. \eqref{app:eq_GTBC} can thus be rewritten as:
\begin{equation}\label{app:eq_GTBC_sigm}
    \Phi \Delta h = \frac{\Gamma_1}{R}+\frac{\Gamma_2}{R^3}.
\end{equation}

Having found the chemical potential of a curved interface, we can now use Eq. \eqref{eq:app_fluxmu} to find the flux $\boldsymbol{j}$. The velocity $v_{int}$ at which the interface moves van be found from the difference in flux leaving the interface and flux entering the interface:
\begin{equation}\label{app:eq_fluxbal}
    v_{int}\Delta h =\textbf{j}_{out}-\textbf{j}_{in}=-\left(\left(\frac{\partial\Phi}{\partial g}\right)_{R+\epsilon}-\left(\frac{\partial\Phi}{\partial g}\right)_{R-\epsilon}\right).
\end{equation}
Setting $v_{int}$ equal to the rate of the change of the characteristic length scale $L_c$ (in this case the radius $R$), and utilizing the chemical potential from Eq. \eqref{app:eq_GTBC_sigm}, we can get a prediction for how $L_c$ will grow with time:
\begin{equation}\label{app:eq_growthrate}
    \frac{dL_c}{dt}\sim \frac{\Gamma_1}{L_c^2}+\frac{\Gamma_2}{L_c^4}.
\end{equation}
This growth leads to power law growth where $L_c$ grows with a power law having an exponent somewhere between $1/5$ and $1/3$, depending on the relative sizes of $\Gamma_1$ and $\Gamma_2$.

\clearpage

\bibliography{Bibliography_Vira}

\begin{thebibliography}{98}%
\makeatletter
\providecommand \@ifxundefined [1]{%
 \@ifx{#1\undefined}
}%
\providecommand \@ifnum [1]{%
 \ifnum #1\expandafter \@firstoftwo
 \else \expandafter \@secondoftwo
 \fi
}%
\providecommand \@ifx [1]{%
 \ifx #1\expandafter \@firstoftwo
 \else \expandafter \@secondoftwo
 \fi
}%
\providecommand \natexlab [1]{#1}%
\providecommand \enquote  [1]{``#1''}%
\providecommand \bibnamefont  [1]{#1}%
\providecommand \bibfnamefont [1]{#1}%
\providecommand \citenamefont [1]{#1}%
\providecommand \href@noop [0]{\@secondoftwo}%
\providecommand \href [0]{\begingroup \@sanitize@url \@href}%
\providecommand \@href[1]{\@@startlink{#1}\@@href}%
\providecommand \@@href[1]{\endgroup#1\@@endlink}%
\providecommand \@sanitize@url [0]{\catcode `\\12\catcode `\$12\catcode `\&12\catcode `\#12\catcode `\^12\catcode `\_12\catcode `\%12\relax}%
\providecommand \@@startlink[1]{}%
\providecommand \@@endlink[0]{}%
\providecommand \url  [0]{\begingroup\@sanitize@url \@url }%
\providecommand \@url [1]{\endgroup\@href {#1}{\urlprefix }}%
\providecommand \urlprefix  [0]{URL }%
\providecommand \Eprint [0]{\href }%
\providecommand \doibase [0]{https://doi.org/}%
\providecommand \selectlanguage [0]{\@gobble}%
\providecommand \bibinfo  [0]{\@secondoftwo}%
\providecommand \bibfield  [0]{\@secondoftwo}%
\providecommand \translation [1]{[#1]}%
\providecommand \BibitemOpen [0]{}%
\providecommand \bibitemStop [0]{}%
\providecommand \bibitemNoStop [0]{.\EOS\space}%
\providecommand \EOS [0]{\spacefactor3000\relax}%
\providecommand \BibitemShut  [1]{\csname bibitem#1\endcsname}%
\let\auto@bib@innerbib\@empty
\bibitem [{\citenamefont {van Roy}\ and\ \citenamefont {Berx}(2008)}]{RoyBerx_cadherin_2008}%
  \BibitemOpen
  \bibfield  {author} {\bibinfo {author} {\bibfnamefont {F.}~\bibnamefont {van Roy}}\ and\ \bibinfo {author} {\bibfnamefont {G.}~\bibnamefont {Berx}},\ }\bibfield  {title} {\bibinfo {title} {The cell-cell adhesion molecule e-cadherin},\ }\href@noop {} {\bibfield  {journal} {\bibinfo  {journal} {Cellular and Molecular Life Sciences}\ }\textbf {\bibinfo {volume} {65}},\ \bibinfo {pages} {3756} (\bibinfo {year} {2008})}\BibitemShut {NoStop}%
\bibitem [{\citenamefont {Dumortier}\ \emph {et~al.}(2019)\citenamefont {Dumortier}, \citenamefont {Verge-Serandour}, \citenamefont {Tortorelli}, \citenamefont {Mielke}, \citenamefont {de~Plater}, \citenamefont {Turlier},\ and\ \citenamefont {Maître}}]{DumortierLVSTurlierMaitre2019}%
  \BibitemOpen
  \bibfield  {author} {\bibinfo {author} {\bibfnamefont {J.~G.}\ \bibnamefont {Dumortier}}, \bibinfo {author} {\bibfnamefont {M.~L.}\ \bibnamefont {Verge-Serandour}}, \bibinfo {author} {\bibfnamefont {A.~F.}\ \bibnamefont {Tortorelli}}, \bibinfo {author} {\bibfnamefont {A.}~\bibnamefont {Mielke}}, \bibinfo {author} {\bibfnamefont {L.}~\bibnamefont {de~Plater}}, \bibinfo {author} {\bibfnamefont {H.}~\bibnamefont {Turlier}},\ and\ \bibinfo {author} {\bibfnamefont {J.-L.}\ \bibnamefont {Maître}},\ }\bibfield  {title} {\bibinfo {title} {Hydraulic fracturing and active coarsening position the lumen of the mouse blastocyst},\ }\href@noop {} {\bibfield  {journal} {\bibinfo  {journal} {Science}\ }\textbf {\bibinfo {volume} {365}},\ \bibinfo {pages} {465} (\bibinfo {year} {2019})}\BibitemShut {NoStop}%
\bibitem [{\citenamefont {Grakoui}\ \emph {et~al.}(1999)\citenamefont {Grakoui}, \citenamefont {Bromley}, \citenamefont {Sumen}, \citenamefont {Davis}, \citenamefont {Shaw}, \citenamefont {Allen},\ and\ \citenamefont {Dustin}}]{grakoui1999}%
  \BibitemOpen
  \bibfield  {author} {\bibinfo {author} {\bibfnamefont {A.}~\bibnamefont {Grakoui}}, \bibinfo {author} {\bibfnamefont {S.~K.}\ \bibnamefont {Bromley}}, \bibinfo {author} {\bibfnamefont {C.}~\bibnamefont {Sumen}}, \bibinfo {author} {\bibfnamefont {M.~M.}\ \bibnamefont {Davis}}, \bibinfo {author} {\bibfnamefont {A.~S.}\ \bibnamefont {Shaw}}, \bibinfo {author} {\bibfnamefont {P.~M.}\ \bibnamefont {Allen}},\ and\ \bibinfo {author} {\bibfnamefont {M.~L.}\ \bibnamefont {Dustin}},\ }\bibfield  {title} {\bibinfo {title} {The immunological synapse: A molecular machine controlling t cell activation},\ }\href@noop {} {\bibfield  {journal} {\bibinfo  {journal} {Science}\ }\textbf {\bibinfo {volume} {285}},\ \bibinfo {pages} {221} (\bibinfo {year} {1999})}\BibitemShut {NoStop}%
\bibitem [{\citenamefont {Dustin}\ and\ \citenamefont {Cooper}(2000)}]{dustin_cooper_2000}%
  \BibitemOpen
  \bibfield  {author} {\bibinfo {author} {\bibfnamefont {M.~L.}\ \bibnamefont {Dustin}}\ and\ \bibinfo {author} {\bibfnamefont {J.~A.}\ \bibnamefont {Cooper}},\ }\bibfield  {title} {\bibinfo {title} {The immunological synapse and the actin cytoskeleton: molecular hardware for t cell signaling},\ }\href@noop {} {\bibfield  {journal} {\bibinfo  {journal} {Nature Immunology}\ }\textbf {\bibinfo {volume} {1}},\ \bibinfo {pages} {23–29} (\bibinfo {year} {2000})}\BibitemShut {NoStop}%
\bibitem [{\citenamefont {Qi}\ \emph {et~al.}(2001)\citenamefont {Qi}, \citenamefont {Groves},\ and\ \citenamefont {Chakraborty}}]{qi2001}%
  \BibitemOpen
  \bibfield  {author} {\bibinfo {author} {\bibfnamefont {S.~Y.}\ \bibnamefont {Qi}}, \bibinfo {author} {\bibfnamefont {J.~T.}\ \bibnamefont {Groves}},\ and\ \bibinfo {author} {\bibfnamefont {A.~K.}\ \bibnamefont {Chakraborty}},\ }\bibfield  {title} {\bibinfo {title} {Synaptic pattern formation during cellular recognition},\ }\href@noop {} {\bibfield  {journal} {\bibinfo  {journal} {Proceedings of the National Academy of Sciences}\ }\textbf {\bibinfo {volume} {98}},\ \bibinfo {pages} {6548} (\bibinfo {year} {2001})}\BibitemShut {NoStop}%
\bibitem [{\citenamefont {Dustin}(2010)}]{Dustin2010}%
  \BibitemOpen
  \bibfield  {author} {\bibinfo {author} {\bibfnamefont {M.~L.}\ \bibnamefont {Dustin}},\ }\bibinfo {title} {Insights into function of the immunological synapse from studies with supported planar bilayers}\ (\bibinfo  {publisher} {Springer Berlin Heidelberg},\ \bibinfo {address} {Berlin, Heidelberg},\ \bibinfo {year} {2010})\ pp.\ \bibinfo {pages} {1--24}\BibitemShut {NoStop}%
\bibitem [{\citenamefont {Le~Verge-Serandour}\ and\ \citenamefont {Turlier}(2021)}]{LeVerge-SerandourTurlier2021}%
  \BibitemOpen
  \bibfield  {author} {\bibinfo {author} {\bibfnamefont {M.}~\bibnamefont {Le~Verge-Serandour}}\ and\ \bibinfo {author} {\bibfnamefont {H.}~\bibnamefont {Turlier}},\ }\bibfield  {title} {\bibinfo {title} {A hydro-osmotic coarsening theory of biological cavity formation},\ }\href@noop {} {\bibfield  {journal} {\bibinfo  {journal} {PLOS Computational Biology}\ }\textbf {\bibinfo {volume} {17}},\ \bibinfo {pages} {1} (\bibinfo {year} {2021})}\BibitemShut {NoStop}%
\bibitem [{\citenamefont {Seifert}\ and\ \citenamefont {Lipowsky}(1990)}]{seifertAdhesionVesicles1990}%
  \BibitemOpen
  \bibfield  {author} {\bibinfo {author} {\bibfnamefont {U.}~\bibnamefont {Seifert}}\ and\ \bibinfo {author} {\bibfnamefont {R.}~\bibnamefont {Lipowsky}},\ }\bibfield  {title} {\bibinfo {title} {Adhesion of vesicles},\ }\href {https://doi.org/10.1103/PhysRevA.42.4768} {\bibfield  {journal} {\bibinfo  {journal} {Physical Review A}\ }\textbf {\bibinfo {volume} {42}},\ \bibinfo {pages} {4768} (\bibinfo {year} {1990})}\BibitemShut {NoStop}%
\bibitem [{\citenamefont {Smith}\ \emph {et~al.}(2003)\citenamefont {Smith}, \citenamefont {Sackmann},\ and\ \citenamefont {Seifert}}]{smithEffectsPullingForce2003}%
  \BibitemOpen
  \bibfield  {author} {\bibinfo {author} {\bibfnamefont {A.-S.}\ \bibnamefont {Smith}}, \bibinfo {author} {\bibfnamefont {E.}~\bibnamefont {Sackmann}},\ and\ \bibinfo {author} {\bibfnamefont {U.}~\bibnamefont {Seifert}},\ }\bibfield  {title} {\bibinfo {title} {Effects of a pulling force on the shape of a bound vesicle},\ }\href {https://doi.org/10.1209/epl/i2003-00499-9} {\bibfield  {journal} {\bibinfo  {journal} {Europhysics Letters}\ }\textbf {\bibinfo {volume} {64}},\ \bibinfo {pages} {281} (\bibinfo {year} {2003})}\BibitemShut {NoStop}%
\bibitem [{\citenamefont {Smith}\ \emph {et~al.}(2008)\citenamefont {Smith}, \citenamefont {Sengupta}, \citenamefont {Goennenwein}, \citenamefont {Seifert},\ and\ \citenamefont {Sackmann}}]{smithForceinducedGrowthAdhesion2008}%
  \BibitemOpen
  \bibfield  {author} {\bibinfo {author} {\bibfnamefont {A.-S.}\ \bibnamefont {Smith}}, \bibinfo {author} {\bibfnamefont {K.}~\bibnamefont {Sengupta}}, \bibinfo {author} {\bibfnamefont {S.}~\bibnamefont {Goennenwein}}, \bibinfo {author} {\bibfnamefont {U.}~\bibnamefont {Seifert}},\ and\ \bibinfo {author} {\bibfnamefont {E.}~\bibnamefont {Sackmann}},\ }\bibfield  {title} {\bibinfo {title} {Force-induced growth of adhesion domains is controlled by receptor mobility},\ }\href {https://doi.org/10.1073/pnas.0801706105} {\bibfield  {journal} {\bibinfo  {journal} {Proceedings of the National Academy of Sciences}\ }\textbf {\bibinfo {volume} {105}},\ \bibinfo {pages} {6906} (\bibinfo {year} {2008})}\BibitemShut {NoStop}%
\bibitem [{\citenamefont {Agrawal}(2011)}]{agrawalMechanicsMembraneMembrane2011}%
  \BibitemOpen
  \bibfield  {author} {\bibinfo {author} {\bibfnamefont {A.}~\bibnamefont {Agrawal}},\ }\bibfield  {title} {\bibinfo {title} {Mechanics of membrane--membrane adhesion},\ }\href {https://doi.org/10.1177/1081286511401364} {\bibfield  {journal} {\bibinfo  {journal} {Mathematics and Mechanics of Solids}\ }\textbf {\bibinfo {volume} {16}},\ \bibinfo {pages} {872} (\bibinfo {year} {2011})}\BibitemShut {NoStop}%
\bibitem [{\citenamefont {Hill}\ and\ \citenamefont {{Al-Amodi}}(2024)}]{hillContactlineBendingEnergy2024}%
  \BibitemOpen
  \bibfield  {author} {\bibinfo {author} {\bibfnamefont {R.~J.}\ \bibnamefont {Hill}}\ and\ \bibinfo {author} {\bibfnamefont {A.}~\bibnamefont {{Al-Amodi}}},\ }\bibfield  {title} {\bibinfo {title} {Contact-line bending energy controls phospholipid vesicle adhesion},\ }\href {https://doi.org/10.1098/rspa.2023.0545} {\bibfield  {journal} {\bibinfo  {journal} {Proceedings of the Royal Society A: Mathematical, Physical and Engineering Sciences}\ }\textbf {\bibinfo {volume} {480}},\ \bibinfo {pages} {20230545} (\bibinfo {year} {2024})}\BibitemShut {NoStop}%
\bibitem [{\citenamefont {Leong}\ and\ \citenamefont {Chiam}(2010)}]{leongAdhesiveDynamicsLubricated2010}%
  \BibitemOpen
  \bibfield  {author} {\bibinfo {author} {\bibfnamefont {F.~Y.}\ \bibnamefont {Leong}}\ and\ \bibinfo {author} {\bibfnamefont {K.-H.}\ \bibnamefont {Chiam}},\ }\bibfield  {title} {\bibinfo {title} {Adhesive dynamics of lubricated films},\ }\href {https://doi.org/10.1103/PhysRevE.81.041923} {\bibfield  {journal} {\bibinfo  {journal} {Physical Review E}\ }\textbf {\bibinfo {volume} {81}},\ \bibinfo {pages} {041923} (\bibinfo {year} {2010})}\BibitemShut {NoStop}%
\bibitem [{\citenamefont {Carlson}\ and\ \citenamefont {Mahadevan}(2015{\natexlab{a}})}]{carlson_mahadevan_synapse_2015}%
  \BibitemOpen
  \bibfield  {author} {\bibinfo {author} {\bibfnamefont {A.}~\bibnamefont {Carlson}}\ and\ \bibinfo {author} {\bibfnamefont {L.}~\bibnamefont {Mahadevan}},\ }\bibfield  {title} {\bibinfo {title} {Elastohydrodynamics and kinetics of protein patterning in the immunological synapse},\ }\bibfield  {journal} {\bibinfo  {journal} {PLOS Computational Biology}\ }\textbf {\bibinfo {volume} {11}},\ \href {https://doi.org/10.1371/journal.pcbi.1004481} {10.1371/journal.pcbi.1004481} (\bibinfo {year} {2015}{\natexlab{a}})\BibitemShut {NoStop}%
\bibitem [{\citenamefont {Carlson}\ and\ \citenamefont {Mahadevan}(2015{\natexlab{b}})}]{carlson2015_protadhesion_physfluids}%
  \BibitemOpen
  \bibfield  {author} {\bibinfo {author} {\bibfnamefont {A.}~\bibnamefont {Carlson}}\ and\ \bibinfo {author} {\bibfnamefont {L.}~\bibnamefont {Mahadevan}},\ }\bibfield  {title} {\bibinfo {title} {Protein mediated membrane adhesion},\ }\href@noop {} {\bibfield  {journal} {\bibinfo  {journal} {Physics of Fluids}\ }\textbf {\bibinfo {volume} {27}},\ \bibinfo {pages} {051901} (\bibinfo {year} {2015}{\natexlab{b}})}\BibitemShut {NoStop}%
\bibitem [{\citenamefont {Evans}\ \emph {et~al.}(1976)\citenamefont {Evans}, \citenamefont {Waugh},\ and\ \citenamefont {Melnik}}]{evansElasticAreaCompressibility1976}%
  \BibitemOpen
  \bibfield  {author} {\bibinfo {author} {\bibfnamefont {E.~A.}\ \bibnamefont {Evans}}, \bibinfo {author} {\bibfnamefont {R.}~\bibnamefont {Waugh}},\ and\ \bibinfo {author} {\bibfnamefont {L.}~\bibnamefont {Melnik}},\ }\bibfield  {title} {\bibinfo {title} {Elastic area compressibility modulus of red cell membrane},\ }\href {https://doi.org/10.1016/S0006-3495(76)85713-X} {\bibfield  {journal} {\bibinfo  {journal} {Biophysical Journal}\ }\textbf {\bibinfo {volume} {16}},\ \bibinfo {pages} {585} (\bibinfo {year} {1976})}\BibitemShut {NoStop}%
\bibitem [{\citenamefont {Oron}\ \emph {et~al.}(1997)\citenamefont {Oron}, \citenamefont {Davis},\ and\ \citenamefont {Bankoff}}]{OronBankoff1997}%
  \BibitemOpen
  \bibfield  {author} {\bibinfo {author} {\bibfnamefont {A.}~\bibnamefont {Oron}}, \bibinfo {author} {\bibfnamefont {S.}~\bibnamefont {Davis}},\ and\ \bibinfo {author} {\bibfnamefont {S.}~\bibnamefont {Bankoff}},\ }\bibfield  {title} {\bibinfo {title} {Long-scale evolution of thin liquid films},\ }\href@noop {} {\bibfield  {journal} {\bibinfo  {journal} {Reviews of Modern Physics}\ }\textbf {\bibinfo {volume} {69}},\ \bibinfo {pages} {931} (\bibinfo {year} {1997})}\BibitemShut {NoStop}%
\bibitem [{\citenamefont {Craster}\ and\ \citenamefont {Matar}(2009)}]{craster2009}%
  \BibitemOpen
  \bibfield  {author} {\bibinfo {author} {\bibfnamefont {R.}~\bibnamefont {Craster}}\ and\ \bibinfo {author} {\bibfnamefont {O.}~\bibnamefont {Matar}},\ }\bibfield  {title} {\bibinfo {title} {Dynamics and stability of thin liquid films},\ }\href@noop {} {\bibfield  {journal} {\bibinfo  {journal} {Reviews of Modern Physics}\ }\textbf {\bibinfo {volume} {81}} (\bibinfo {year} {2009})}\BibitemShut {NoStop}%
\bibitem [{\citenamefont {Zhang}\ and\ \citenamefont {Lister}(1999)}]{ZhangLister1999}%
  \BibitemOpen
  \bibfield  {author} {\bibinfo {author} {\bibfnamefont {W.~W.}\ \bibnamefont {Zhang}}\ and\ \bibinfo {author} {\bibfnamefont {J.~R.}\ \bibnamefont {Lister}},\ }\bibfield  {title} {\bibinfo {title} {Similarity solutions for van der waals rupture of a thin film on a solid substrate},\ }\href@noop {} {\bibfield  {journal} {\bibinfo  {journal} {Physics of Fluids}\ }\textbf {\bibinfo {volume} {11}},\ \bibinfo {pages} {2454} (\bibinfo {year} {1999})}\BibitemShut {NoStop}%
\bibitem [{\citenamefont {Mecke}\ and\ \citenamefont {Rauscher}(2005)}]{MeckeRauscher_2005}%
  \BibitemOpen
  \bibfield  {author} {\bibinfo {author} {\bibfnamefont {K.}~\bibnamefont {Mecke}}\ and\ \bibinfo {author} {\bibfnamefont {M.}~\bibnamefont {Rauscher}},\ }\bibfield  {title} {\bibinfo {title} {On thermal fluctuations in thin film flow},\ }\href@noop {} {\bibfield  {journal} {\bibinfo  {journal} {Journal of Physics: Condensed Matter}\ }\textbf {\bibinfo {volume} {17}},\ \bibinfo {pages} {S3515} (\bibinfo {year} {2005})}\BibitemShut {NoStop}%
\bibitem [{\citenamefont {Grün}\ \emph {et~al.}(2006)\citenamefont {Grün}, \citenamefont {Mecke},\ and\ \citenamefont {Rauscher}}]{grun2006}%
  \BibitemOpen
  \bibfield  {author} {\bibinfo {author} {\bibfnamefont {G.}~\bibnamefont {Grün}}, \bibinfo {author} {\bibfnamefont {K.}~\bibnamefont {Mecke}},\ and\ \bibinfo {author} {\bibfnamefont {M.}~\bibnamefont {Rauscher}},\ }\bibfield  {title} {\bibinfo {title} {Thin-film flow influenced by thermal noise},\ }\href@noop {} {\bibfield  {journal} {\bibinfo  {journal} {Journal of Statistical Physics}\ }\textbf {\bibinfo {volume} {122}},\ \bibinfo {pages} {1261} (\bibinfo {year} {2006})}\BibitemShut {NoStop}%
\bibitem [{\citenamefont {Nguyen}\ \emph {et~al.}(2014)\citenamefont {Nguyen}, \citenamefont {{Fuentes-Cabrera}}, \citenamefont {Fowlkes},\ and\ \citenamefont {Rack}}]{nguyenCoexistenceSpinodalInstability2014}%
  \BibitemOpen
  \bibfield  {author} {\bibinfo {author} {\bibfnamefont {T.~D.}\ \bibnamefont {Nguyen}}, \bibinfo {author} {\bibfnamefont {M.}~\bibnamefont {{Fuentes-Cabrera}}}, \bibinfo {author} {\bibfnamefont {J.~D.}\ \bibnamefont {Fowlkes}},\ and\ \bibinfo {author} {\bibfnamefont {P.~D.}\ \bibnamefont {Rack}},\ }\bibfield  {title} {\bibinfo {title} {Coexistence of spinodal instability and thermal nucleation in thin-film rupture: {{Insights}} from molecular levels},\ }\href {https://doi.org/10.1103/PhysRevE.89.032403} {\bibfield  {journal} {\bibinfo  {journal} {Physical Review E}\ }\textbf {\bibinfo {volume} {89}},\ \bibinfo {pages} {032403} (\bibinfo {year} {2014})}\BibitemShut {NoStop}%
\bibitem [{\citenamefont {Zhang}\ \emph {et~al.}(2019)\citenamefont {Zhang}, \citenamefont {Sprittles},\ and\ \citenamefont {Lockerby}}]{zhangMolecularSimulationThin2019}%
  \BibitemOpen
  \bibfield  {author} {\bibinfo {author} {\bibfnamefont {Y.}~\bibnamefont {Zhang}}, \bibinfo {author} {\bibfnamefont {J.~E.}\ \bibnamefont {Sprittles}},\ and\ \bibinfo {author} {\bibfnamefont {D.~A.}\ \bibnamefont {Lockerby}},\ }\bibfield  {title} {\bibinfo {title} {Molecular simulation of thin liquid films: {{Thermal}} fluctuations and instability},\ }\href {https://doi.org/10.1103/PhysRevE.100.023108} {\bibfield  {journal} {\bibinfo  {journal} {Physical Review E}\ }\textbf {\bibinfo {volume} {100}},\ \bibinfo {pages} {023108} (\bibinfo {year} {2019})}\BibitemShut {NoStop}%
\bibitem [{\citenamefont {Zhao}\ \emph {et~al.}(2022)\citenamefont {Zhao}, \citenamefont {Liu}, \citenamefont {Lockerby},\ and\ \citenamefont {Sprittles}}]{zhaoFluctuationdrivenDynamicsNanoscale2022}%
  \BibitemOpen
  \bibfield  {author} {\bibinfo {author} {\bibfnamefont {C.}~\bibnamefont {Zhao}}, \bibinfo {author} {\bibfnamefont {J.}~\bibnamefont {Liu}}, \bibinfo {author} {\bibfnamefont {D.~A.}\ \bibnamefont {Lockerby}},\ and\ \bibinfo {author} {\bibfnamefont {J.~E.}\ \bibnamefont {Sprittles}},\ }\bibfield  {title} {\bibinfo {title} {Fluctuation-driven dynamics in nanoscale thin-film flows: {{Physical}} insights from numerical investigations},\ }\href {https://doi.org/10.1103/PhysRevFluids.7.024203} {\bibfield  {journal} {\bibinfo  {journal} {Physical Review Fluids}\ }\textbf {\bibinfo {volume} {7}},\ \bibinfo {pages} {024203} (\bibinfo {year} {2022})}\BibitemShut {NoStop}%
\bibitem [{\citenamefont {Evans}(1974)}]{evansBendingResistanceChemically1974}%
  \BibitemOpen
  \bibfield  {author} {\bibinfo {author} {\bibfnamefont {E.~A.}\ \bibnamefont {Evans}},\ }\bibfield  {title} {\bibinfo {title} {Bending {{Resistance}} and {{Chemically Induced Moments}} in {{Membrane Bilayers}}},\ }\href {https://doi.org/10.1016/S0006-3495(74)85959-X} {\bibfield  {journal} {\bibinfo  {journal} {Biophysical Journal}\ }\textbf {\bibinfo {volume} {14}},\ \bibinfo {pages} {923} (\bibinfo {year} {1974})}\BibitemShut {NoStop}%
\bibitem [{\citenamefont {Deserno}(2015)}]{DesernoDiffCurv}%
  \BibitemOpen
  \bibfield  {author} {\bibinfo {author} {\bibfnamefont {M.}~\bibnamefont {Deserno}},\ }\bibfield  {title} {\bibinfo {title} {Fluid lipid membranes: From differential geometry to curvature stresses},\ }\href@noop {} {\bibfield  {journal} {\bibinfo  {journal} {Chemistry and Physics of Lipids}\ }\textbf {\bibinfo {volume} {185}},\ \bibinfo {pages} {11} (\bibinfo {year} {2015})},\ \bibinfo {note} {membrane mechanochemistry: From the molecular to the cellular scale}\BibitemShut {NoStop}%
\bibitem [{\citenamefont {Roman}\ and\ \citenamefont {Bico}(2010)}]{RomanBico2010}%
  \BibitemOpen
  \bibfield  {author} {\bibinfo {author} {\bibfnamefont {B.}~\bibnamefont {Roman}}\ and\ \bibinfo {author} {\bibfnamefont {J.}~\bibnamefont {Bico}},\ }\bibfield  {title} {\bibinfo {title} {Elasto-capillarity: Deforming an elastic structure with a liquid droplet},\ }\href@noop {} {\bibfield  {journal} {\bibinfo  {journal} {Journal of physics. Condensed matter : an Institute of Physics journal}\ }\textbf {\bibinfo {volume} {22}},\ \bibinfo {pages} {493101} (\bibinfo {year} {2010})}\BibitemShut {NoStop}%
\bibitem [{\citenamefont {Yuan}\ \emph {et~al.}(2021)\citenamefont {Yuan}, \citenamefont {Alimohamadi}, \citenamefont {Bakka}, \citenamefont {Trementozzi}, \citenamefont {Day}, \citenamefont {Fawzi}, \citenamefont {Rangamani},\ and\ \citenamefont {Stachowiak}}]{yuanMembraneBendingProtein2021}%
  \BibitemOpen
  \bibfield  {author} {\bibinfo {author} {\bibfnamefont {F.}~\bibnamefont {Yuan}}, \bibinfo {author} {\bibfnamefont {H.}~\bibnamefont {Alimohamadi}}, \bibinfo {author} {\bibfnamefont {B.}~\bibnamefont {Bakka}}, \bibinfo {author} {\bibfnamefont {A.~N.}\ \bibnamefont {Trementozzi}}, \bibinfo {author} {\bibfnamefont {K.~J.}\ \bibnamefont {Day}}, \bibinfo {author} {\bibfnamefont {N.~L.}\ \bibnamefont {Fawzi}}, \bibinfo {author} {\bibfnamefont {P.}~\bibnamefont {Rangamani}},\ and\ \bibinfo {author} {\bibfnamefont {J.~C.}\ \bibnamefont {Stachowiak}},\ }\bibfield  {title} {\bibinfo {title} {Membrane bending by protein phase separation},\ }\href {https://doi.org/10.1073/pnas.2017435118} {\bibfield  {journal} {\bibinfo  {journal} {Proceedings of the National Academy of Sciences}\ }\textbf {\bibinfo {volume} {118}},\ \bibinfo {pages} {e2017435118} (\bibinfo {year} {2021})}\BibitemShut {NoStop}%
\bibitem [{\citenamefont {Liese}\ and\ \citenamefont {Carlson}(2021)}]{lieseMembraneShapeRemodeling2021}%
  \BibitemOpen
  \bibfield  {author} {\bibinfo {author} {\bibfnamefont {S.}~\bibnamefont {Liese}}\ and\ \bibinfo {author} {\bibfnamefont {A.}~\bibnamefont {Carlson}},\ }\bibfield  {title} {\bibinfo {title} {Membrane shape remodeling by protein crowding},\ }\href {https://doi.org/10.1016/j.bpj.2021.04.029} {\bibfield  {journal} {\bibinfo  {journal} {Biophysical Journal}\ }\textbf {\bibinfo {volume} {120}},\ \bibinfo {pages} {2482} (\bibinfo {year} {2021})}\BibitemShut {NoStop}%
\bibitem [{\citenamefont {Aarts}(2004)}]{aartsDirectVisualObservation2004}%
  \BibitemOpen
  \bibfield  {author} {\bibinfo {author} {\bibfnamefont {D.~G. A.~L.}\ \bibnamefont {Aarts}},\ }\bibfield  {title} {\bibinfo {title} {Direct {{Visual Observation}} of {{Thermal Capillary Waves}}},\ }\href {https://doi.org/10.1126/science.1097116} {\bibfield  {journal} {\bibinfo  {journal} {Science}\ }\textbf {\bibinfo {volume} {304}},\ \bibinfo {pages} {847} (\bibinfo {year} {2004})}\BibitemShut {NoStop}%
\bibitem [{\citenamefont {{Delgado-Buscalioni}}\ \emph {et~al.}(2008)\citenamefont {{Delgado-Buscalioni}}, \citenamefont {Chacon},\ and\ \citenamefont {Tarazona}}]{delgado-buscalioniHydrodynamicsNanoscopicCapillary2008}%
  \BibitemOpen
  \bibfield  {author} {\bibinfo {author} {\bibfnamefont {R.}~\bibnamefont {{Delgado-Buscalioni}}}, \bibinfo {author} {\bibfnamefont {E.}~\bibnamefont {Chacon}},\ and\ \bibinfo {author} {\bibfnamefont {P.}~\bibnamefont {Tarazona}},\ }\bibfield  {title} {\bibinfo {title} {Hydrodynamics of {{Nanoscopic Capillary Waves}}},\ }\href {https://doi.org/10.1103/PhysRevLett.101.106102} {\bibfield  {journal} {\bibinfo  {journal} {Physical Review Letters}\ }\textbf {\bibinfo {volume} {101}},\ \bibinfo {pages} {106102} (\bibinfo {year} {2008})}\BibitemShut {NoStop}%
\bibitem [{\citenamefont {Guo}\ \emph {et~al.}(2014)\citenamefont {Guo}, \citenamefont {Ehrlicher}, \citenamefont {Jensen}, \citenamefont {Renz}, \citenamefont {Moore}, \citenamefont {Goldman}, \citenamefont {Lippincott-Schwartz}, \citenamefont {Mackintosh},\ and\ \citenamefont {Weitz}}]{GuoWeitz2014}%
  \BibitemOpen
  \bibfield  {author} {\bibinfo {author} {\bibfnamefont {M.}~\bibnamefont {Guo}}, \bibinfo {author} {\bibfnamefont {A.}~\bibnamefont {Ehrlicher}}, \bibinfo {author} {\bibfnamefont {M.}~\bibnamefont {Jensen}}, \bibinfo {author} {\bibfnamefont {M.}~\bibnamefont {Renz}}, \bibinfo {author} {\bibfnamefont {J.}~\bibnamefont {Moore}}, \bibinfo {author} {\bibfnamefont {R.}~\bibnamefont {Goldman}}, \bibinfo {author} {\bibfnamefont {J.}~\bibnamefont {Lippincott-Schwartz}}, \bibinfo {author} {\bibfnamefont {F.}~\bibnamefont {Mackintosh}},\ and\ \bibinfo {author} {\bibfnamefont {D.}~\bibnamefont {Weitz}},\ }\bibfield  {title} {\bibinfo {title} {Probing the stochastic, motor-driven properties of the cytoplasm using force spectrum microscopy},\ }\href@noop {} {\bibfield  {journal} {\bibinfo  {journal} {Cell}\ }\textbf {\bibinfo {volume} {158}},\ \bibinfo {pages} {822} (\bibinfo {year} {2014})}\BibitemShut {NoStop}%
\bibitem [{\citenamefont {Gupta}\ and\ \citenamefont {Guo}(2017)}]{GuptaGuo2017}%
  \BibitemOpen
  \bibfield  {author} {\bibinfo {author} {\bibfnamefont {S.}~\bibnamefont {Gupta}}\ and\ \bibinfo {author} {\bibfnamefont {M.}~\bibnamefont {Guo}},\ }\bibfield  {title} {\bibinfo {title} {Equilibrium and out-of-equilibrium mechanics of living mammalian cytoplasm},\ }\href@noop {} {\bibfield  {journal} {\bibinfo  {journal} {Journal of the Mechanics and Physics of Solids}\ }\textbf {\bibinfo {volume} {107}} (\bibinfo {year} {2017})}\BibitemShut {NoStop}%
\bibitem [{\citenamefont {Sprittles}\ \emph {et~al.}(2023)\citenamefont {Sprittles}, \citenamefont {Liu}, \citenamefont {Lockerby},\ and\ \citenamefont {Grafke}}]{SprittlesJBLGrafke2023}%
  \BibitemOpen
  \bibfield  {author} {\bibinfo {author} {\bibfnamefont {J.}~\bibnamefont {Sprittles}}, \bibinfo {author} {\bibfnamefont {J.}~\bibnamefont {Liu}}, \bibinfo {author} {\bibfnamefont {D.}~\bibnamefont {Lockerby}},\ and\ \bibinfo {author} {\bibfnamefont {T.}~\bibnamefont {Grafke}},\ }\bibfield  {title} {\bibinfo {title} {Rogue nanowaves: A route to film rupture},\ }\href@noop {} {\bibfield  {journal} {\bibinfo  {journal} {Physical Review Fluids}\ }\textbf {\bibinfo {volume} {8}} (\bibinfo {year} {2023})}\BibitemShut {NoStop}%
\bibitem [{\citenamefont {Liu}\ \emph {et~al.}(2024)\citenamefont {Liu}, \citenamefont {Sprittles},\ and\ \citenamefont {Grafke}}]{liuMeanFirstPassage2024}%
  \BibitemOpen
  \bibfield  {author} {\bibinfo {author} {\bibfnamefont {J.}~\bibnamefont {Liu}}, \bibinfo {author} {\bibfnamefont {J.~E.}\ \bibnamefont {Sprittles}},\ and\ \bibinfo {author} {\bibfnamefont {T.}~\bibnamefont {Grafke}},\ }\bibfield  {title} {\bibinfo {title} {Mean first passage times and {{Eyring}}--{{Kramers}} formula for fluctuating hydrodynamics},\ }\href {https://doi.org/10.1088/1742-5468/ad8075} {\bibfield  {journal} {\bibinfo  {journal} {Journal of Statistical Mechanics: Theory and Experiment}\ }\textbf {\bibinfo {volume} {2024}},\ \bibinfo {pages} {103206} (\bibinfo {year} {2024})}\BibitemShut {NoStop}%
\bibitem [{\citenamefont {Janes}\ \emph {et~al.}(2012)\citenamefont {Janes}, \citenamefont {Nievergall},\ and\ \citenamefont {Lackmann}}]{janesConceptsConsequencesEph2012}%
  \BibitemOpen
  \bibfield  {author} {\bibinfo {author} {\bibfnamefont {P.~W.}\ \bibnamefont {Janes}}, \bibinfo {author} {\bibfnamefont {E.}~\bibnamefont {Nievergall}},\ and\ \bibinfo {author} {\bibfnamefont {M.}~\bibnamefont {Lackmann}},\ }\bibfield  {title} {\bibinfo {title} {Concepts and consequences of {{Eph}} receptor clustering},\ }\href {https://doi.org/10.1016/j.semcdb.2012.01.001} {\bibfield  {journal} {\bibinfo  {journal} {Seminars in Cell \& Developmental Biology}\ }\bibinfo {series} {Signalling via {{Eph Receptors}} and {{Ephrins}}},\ \textbf {\bibinfo {volume} {23}},\ \bibinfo {pages} {43} (\bibinfo {year} {2012})}\BibitemShut {NoStop}%
\bibitem [{\citenamefont {Case}\ \emph {et~al.}(2019)\citenamefont {Case}, \citenamefont {Ditlev},\ and\ \citenamefont {Rosen}}]{caseRegulationTransmembraneSignaling2019}%
  \BibitemOpen
  \bibfield  {author} {\bibinfo {author} {\bibfnamefont {L.~B.}\ \bibnamefont {Case}}, \bibinfo {author} {\bibfnamefont {J.~A.}\ \bibnamefont {Ditlev}},\ and\ \bibinfo {author} {\bibfnamefont {M.~K.}\ \bibnamefont {Rosen}},\ }\bibfield  {title} {\bibinfo {title} {Regulation of {{Transmembrane Signaling}} by {{Phase Separation}}},\ }\href {https://doi.org/10.1146/annurev-biophys-052118-115534} {\bibfield  {journal} {\bibinfo  {journal} {Annual Review of Biophysics}\ }\textbf {\bibinfo {volume} {48}},\ \bibinfo {pages} {465} (\bibinfo {year} {2019})}\BibitemShut {NoStop}%
\bibitem [{\citenamefont {Shelby}\ \emph {et~al.}(2023)\citenamefont {Shelby}, \citenamefont {{Castello-Serrano}}, \citenamefont {Wisser}, \citenamefont {Levental},\ and\ \citenamefont {Veatch}}]{shelbyMembranePhaseSeparation2023}%
  \BibitemOpen
  \bibfield  {author} {\bibinfo {author} {\bibfnamefont {S.~A.}\ \bibnamefont {Shelby}}, \bibinfo {author} {\bibfnamefont {I.}~\bibnamefont {{Castello-Serrano}}}, \bibinfo {author} {\bibfnamefont {K.~C.}\ \bibnamefont {Wisser}}, \bibinfo {author} {\bibfnamefont {I.}~\bibnamefont {Levental}},\ and\ \bibinfo {author} {\bibfnamefont {S.~L.}\ \bibnamefont {Veatch}},\ }\bibfield  {title} {\bibinfo {title} {Membrane phase separation drives responsive assembly of receptor signaling domains},\ }\href {https://doi.org/10.1038/s41589-023-01268-8} {\bibfield  {journal} {\bibinfo  {journal} {Nature Chemical Biology}\ }\textbf {\bibinfo {volume} {19}},\ \bibinfo {pages} {750} (\bibinfo {year} {2023})}\BibitemShut {NoStop}%
\bibitem [{\citenamefont {Dinet}\ \emph {et~al.}(2023)\citenamefont {Dinet}, \citenamefont {Torres-Sánchez}, \citenamefont {Lanfranco}, \citenamefont {Michele}, \citenamefont {Arroyo},\ and\ \citenamefont {Staykova}}]{DinetArroyoStaykova2023}%
  \BibitemOpen
  \bibfield  {author} {\bibinfo {author} {\bibfnamefont {C.}~\bibnamefont {Dinet}}, \bibinfo {author} {\bibfnamefont {A.}~\bibnamefont {Torres-Sánchez}}, \bibinfo {author} {\bibfnamefont {R.}~\bibnamefont {Lanfranco}}, \bibinfo {author} {\bibfnamefont {L.}~\bibnamefont {Michele}}, \bibinfo {author} {\bibfnamefont {M.}~\bibnamefont {Arroyo}},\ and\ \bibinfo {author} {\bibfnamefont {M.}~\bibnamefont {Staykova}},\ }\bibfield  {title} {\bibinfo {title} {Patterning and dynamics of membrane adhesion under hydraulic stress},\ }\href@noop {} {\bibfield  {journal} {\bibinfo  {journal} {Nature Communications}\ }\textbf {\bibinfo {volume} {14}} (\bibinfo {year} {2023})}\BibitemShut {NoStop}%
\bibitem [{\citenamefont {Allen}\ and\ \citenamefont {Cahn}(1979)}]{AllenCahn1979}%
  \BibitemOpen
  \bibfield  {author} {\bibinfo {author} {\bibfnamefont {S.~M.}\ \bibnamefont {Allen}}\ and\ \bibinfo {author} {\bibfnamefont {J.~W.}\ \bibnamefont {Cahn}},\ }\bibfield  {title} {\bibinfo {title} {A microscopic theory for antiphase boundary motion and its application to antiphase domain coarsening},\ }\href@noop {} {\bibfield  {journal} {\bibinfo  {journal} {Acta Metallurgica}\ }\textbf {\bibinfo {volume} {27}},\ \bibinfo {pages} {1085} (\bibinfo {year} {1979})}\BibitemShut {NoStop}%
\bibitem [{\citenamefont {Komura}\ \emph {et~al.}(1985)\citenamefont {Komura}, \citenamefont {Osamura}, \citenamefont {Fujii},\ and\ \citenamefont {Takeda}}]{KomuraTakeda1985}%
  \BibitemOpen
  \bibfield  {author} {\bibinfo {author} {\bibfnamefont {S.}~\bibnamefont {Komura}}, \bibinfo {author} {\bibfnamefont {K.}~\bibnamefont {Osamura}}, \bibinfo {author} {\bibfnamefont {H.}~\bibnamefont {Fujii}},\ and\ \bibinfo {author} {\bibfnamefont {T.}~\bibnamefont {Takeda}},\ }\bibfield  {title} {\bibinfo {title} {Time evolution of the structure function of quenched al-zn and al-zn-mg alloys},\ }\href@noop {} {\bibfield  {journal} {\bibinfo  {journal} {Phys. Rev. B}\ }\textbf {\bibinfo {volume} {31}},\ \bibinfo {pages} {1278} (\bibinfo {year} {1985})}\BibitemShut {NoStop}%
\bibitem [{\citenamefont {Bray}(1994)}]{Bray1994}%
  \BibitemOpen
  \bibfield  {author} {\bibinfo {author} {\bibfnamefont {A.}~\bibnamefont {Bray}},\ }\bibfield  {title} {\bibinfo {title} {Theory of phase-ordering kinetics},\ }\href@noop {} {\bibfield  {journal} {\bibinfo  {journal} {Advances in Physics}\ }\textbf {\bibinfo {volume} {43}},\ \bibinfo {pages} {357} (\bibinfo {year} {1994})}\BibitemShut {NoStop}%
\bibitem [{\citenamefont {Livet}\ \emph {et~al.}(2001)\citenamefont {Livet}, \citenamefont {Bley}, \citenamefont {Caudron}, \citenamefont {Geissler}, \citenamefont {Abernathy}, \citenamefont {Detlefs}, \citenamefont {Grübel},\ and\ \citenamefont {Sutton}}]{LivetSutton2001}%
  \BibitemOpen
  \bibfield  {author} {\bibinfo {author} {\bibfnamefont {F.}~\bibnamefont {Livet}}, \bibinfo {author} {\bibfnamefont {F.}~\bibnamefont {Bley}}, \bibinfo {author} {\bibfnamefont {R.}~\bibnamefont {Caudron}}, \bibinfo {author} {\bibfnamefont {E.}~\bibnamefont {Geissler}}, \bibinfo {author} {\bibfnamefont {D.}~\bibnamefont {Abernathy}}, \bibinfo {author} {\bibfnamefont {C.}~\bibnamefont {Detlefs}}, \bibinfo {author} {\bibfnamefont {G.}~\bibnamefont {Grübel}},\ and\ \bibinfo {author} {\bibfnamefont {M.}~\bibnamefont {Sutton}},\ }\bibfield  {title} {\bibinfo {title} {Kinetic evolution of unmixing in an alli alloy using x-ray intensity fluctuation spectroscopy},\ }\href@noop {} {\bibfield  {journal} {\bibinfo  {journal} {Physical review. E, Statistical, nonlinear, and soft matter physics}\ }\textbf {\bibinfo {volume} {63}},\ \bibinfo {pages} {036108} (\bibinfo {year} {2001})}\BibitemShut {NoStop}%
\bibitem [{\citenamefont {Su}\ \emph {et~al.}(2024)\citenamefont {Su}, \citenamefont {Ho}, \citenamefont {Gettel}, \citenamefont {Rowland}, \citenamefont {Keating},\ and\ \citenamefont {Parikh}}]{SuParikh2024}%
  \BibitemOpen
  \bibfield  {author} {\bibinfo {author} {\bibfnamefont {W.-C.}\ \bibnamefont {Su}}, \bibinfo {author} {\bibfnamefont {J.~C.~S.}\ \bibnamefont {Ho}}, \bibinfo {author} {\bibfnamefont {D.~L.}\ \bibnamefont {Gettel}}, \bibinfo {author} {\bibfnamefont {A.~T.}\ \bibnamefont {Rowland}}, \bibinfo {author} {\bibfnamefont {C.~D.}\ \bibnamefont {Keating}},\ and\ \bibinfo {author} {\bibfnamefont {A.~N.}\ \bibnamefont {Parikh}},\ }\bibfield  {title} {\bibinfo {title} {Kinetic control of shape deformations and membrane phase separation inside giant vesicles},\ }\href@noop {} {\bibfield  {journal} {\bibinfo  {journal} {Nature Chemistry}\ }\textbf {\bibinfo {volume} {16}},\ \bibinfo {pages} {54} (\bibinfo {year} {2024})}\BibitemShut {NoStop}%
\bibitem [{\citenamefont {Derrida}\ \emph {et~al.}(1991)\citenamefont {Derrida}, \citenamefont {Godr\`eche},\ and\ \citenamefont {Yekutieli}}]{DerridaYekutelia1991}%
  \BibitemOpen
  \bibfield  {author} {\bibinfo {author} {\bibfnamefont {B.}~\bibnamefont {Derrida}}, \bibinfo {author} {\bibfnamefont {C.}~\bibnamefont {Godr\`eche}},\ and\ \bibinfo {author} {\bibfnamefont {I.}~\bibnamefont {Yekutieli}},\ }\bibfield  {title} {\bibinfo {title} {Scale-invariant regimes in one-dimensional models of growing and coalescing droplets},\ }\href@noop {} {\bibfield  {journal} {\bibinfo  {journal} {Phys. Rev. A}\ }\textbf {\bibinfo {volume} {44}},\ \bibinfo {pages} {6241} (\bibinfo {year} {1991})}\BibitemShut {NoStop}%
\bibitem [{\citenamefont {Otto}\ \emph {et~al.}(2006)\citenamefont {Otto}, \citenamefont {Rump},\ and\ \citenamefont {Slepcev}}]{OttoRumpSlepcev2006}%
  \BibitemOpen
  \bibfield  {author} {\bibinfo {author} {\bibfnamefont {F.}~\bibnamefont {Otto}}, \bibinfo {author} {\bibfnamefont {T.}~\bibnamefont {Rump}},\ and\ \bibinfo {author} {\bibfnamefont {D.}~\bibnamefont {Slepcev}},\ }\bibfield  {title} {\bibinfo {title} {Coarsening rates for a droplet model: Rigorous upper bounds},\ }\href@noop {} {\bibfield  {journal} {\bibinfo  {journal} {SIAM Journal on Mathematical Analysis}\ }\textbf {\bibinfo {volume} {38}},\ \bibinfo {pages} {503} (\bibinfo {year} {2006})}\BibitemShut {NoStop}%
\bibitem [{\citenamefont {Gratton}\ and\ \citenamefont {Witelski}(2008)}]{GrattonWitelski2008}%
  \BibitemOpen
  \bibfield  {author} {\bibinfo {author} {\bibfnamefont {M.~B.}\ \bibnamefont {Gratton}}\ and\ \bibinfo {author} {\bibfnamefont {T.~P.}\ \bibnamefont {Witelski}},\ }\bibfield  {title} {\bibinfo {title} {Coarsening of unstable thin films subject to gravity},\ }\href@noop {} {\bibfield  {journal} {\bibinfo  {journal} {Phys. Rev. E}\ }\textbf {\bibinfo {volume} {77}},\ \bibinfo {pages} {016301} (\bibinfo {year} {2008})}\BibitemShut {NoStop}%
\bibitem [{\citenamefont {Lal}\ \emph {et~al.}(2020)\citenamefont {Lal}, \citenamefont {Lurio}, \citenamefont {Liang}, \citenamefont {Narayanan}, \citenamefont {Darling},\ and\ \citenamefont {Sutton}}]{LalSutton2020}%
  \BibitemOpen
  \bibfield  {author} {\bibinfo {author} {\bibfnamefont {J.}~\bibnamefont {Lal}}, \bibinfo {author} {\bibfnamefont {L.}~\bibnamefont {Lurio}}, \bibinfo {author} {\bibfnamefont {D.}~\bibnamefont {Liang}}, \bibinfo {author} {\bibfnamefont {S.}~\bibnamefont {Narayanan}}, \bibinfo {author} {\bibfnamefont {S.}~\bibnamefont {Darling}},\ and\ \bibinfo {author} {\bibfnamefont {M.}~\bibnamefont {Sutton}},\ }\bibfield  {title} {\bibinfo {title} {Universal dynamics of coarsening during polymer-polymer thin-film spinodal dewetting kinetics},\ }\href@noop {} {\bibfield  {journal} {\bibinfo  {journal} {Physical Review E}\ }\textbf {\bibinfo {volume} {102}} (\bibinfo {year} {2020})}\BibitemShut {NoStop}%
\bibitem [{\citenamefont {Williams}\ and\ \citenamefont {Davis}(1982)}]{WilliamsDavis1982}%
  \BibitemOpen
  \bibfield  {author} {\bibinfo {author} {\bibfnamefont {M.~B.}\ \bibnamefont {Williams}}\ and\ \bibinfo {author} {\bibfnamefont {S.~H.}\ \bibnamefont {Davis}},\ }\bibfield  {title} {\bibinfo {title} {Nonlinear theory of film rupture},\ }\href@noop {} {\bibfield  {journal} {\bibinfo  {journal} {Journal of Colloid and Interface Science}\ }\textbf {\bibinfo {volume} {90}},\ \bibinfo {pages} {220} (\bibinfo {year} {1982})}\BibitemShut {NoStop}%
\bibitem [{\citenamefont {Carlson}\ and\ \citenamefont {Mahadevan}(2016)}]{carlson2016adhesiontouchdown}%
  \BibitemOpen
  \bibfield  {author} {\bibinfo {author} {\bibfnamefont {A.}~\bibnamefont {Carlson}}\ and\ \bibinfo {author} {\bibfnamefont {L.}~\bibnamefont {Mahadevan}},\ }\bibfield  {title} {\bibinfo {title} {Similarity and singularity in adhesive elastohydrodynamic touchdown},\ }\href@noop {} {\bibfield  {journal} {\bibinfo  {journal} {Physics of Fluids}\ }\textbf {\bibinfo {volume} {28}},\ \bibinfo {pages} {011702} (\bibinfo {year} {2016})}\BibitemShut {NoStop}%
\bibitem [{\citenamefont {Fenz}\ \emph {et~al.}(2017)\citenamefont {Fenz}, \citenamefont {Bihr}, \citenamefont {Schmidt}, \citenamefont {Merkel}, \citenamefont {Seifert}, \citenamefont {Sengupta},\ and\ \citenamefont {Smith}}]{FenzAna-Suncana2017}%
  \BibitemOpen
  \bibfield  {author} {\bibinfo {author} {\bibfnamefont {S.}~\bibnamefont {Fenz}}, \bibinfo {author} {\bibfnamefont {T.}~\bibnamefont {Bihr}}, \bibinfo {author} {\bibfnamefont {D.}~\bibnamefont {Schmidt}}, \bibinfo {author} {\bibfnamefont {R.}~\bibnamefont {Merkel}}, \bibinfo {author} {\bibfnamefont {U.}~\bibnamefont {Seifert}}, \bibinfo {author} {\bibfnamefont {K.}~\bibnamefont {Sengupta}},\ and\ \bibinfo {author} {\bibfnamefont {A.-S.}\ \bibnamefont {Smith}},\ }\bibfield  {title} {\bibinfo {title} {Membrane fluctuations mediate lateral interaction between cadherin bonds},\ }\href@noop {} {\bibfield  {journal} {\bibinfo  {journal} {Nature Physics}\ }\textbf {\bibinfo {volume} {13}} (\bibinfo {year} {2017})}\BibitemShut {NoStop}%
\bibitem [{\citenamefont {Batchelor}(2000)}]{batchelor_2000}%
  \BibitemOpen
  \bibfield  {author} {\bibinfo {author} {\bibfnamefont {G.~K.}\ \bibnamefont {Batchelor}},\ }\href@noop {} {\emph {\bibinfo {title} {An Introduction to Fluid Dynamics}}},\ Cambridge Mathematical Library\ (\bibinfo  {publisher} {Cambridge University Press},\ \bibinfo {year} {2000})\BibitemShut {NoStop}%
\bibitem [{\citenamefont {Landau}\ and\ \citenamefont {Lifshitz}(1987)}]{landau_lifshitz_fluids}%
  \BibitemOpen
  \bibfield  {author} {\bibinfo {author} {\bibfnamefont {L.~D.}\ \bibnamefont {Landau}}\ and\ \bibinfo {author} {\bibfnamefont {E.~M.}\ \bibnamefont {Lifshitz}},\ }\href@noop {} {\emph {\bibinfo {title} {Course of theoretical physics: Fluid mechanics}}},\ \bibinfo {edition} {2nd}\ ed.,\ Vol.~\bibinfo {volume} {6}\ (\bibinfo  {publisher} {Elsevier Science \& Technology},\ \bibinfo {year} {1987})\BibitemShut {NoStop}%
\bibitem [{\citenamefont {Durán-Olivencia}\ \emph {et~al.}(2019)\citenamefont {Durán-Olivencia}, \citenamefont {Gvalani}, \citenamefont {Kalliadasis},\ and\ \citenamefont {Pavliotis}}]{duran2019}%
  \BibitemOpen
  \bibfield  {author} {\bibinfo {author} {\bibfnamefont {M.~A.}\ \bibnamefont {Durán-Olivencia}}, \bibinfo {author} {\bibfnamefont {R.~S.}\ \bibnamefont {Gvalani}}, \bibinfo {author} {\bibfnamefont {S.}~\bibnamefont {Kalliadasis}},\ and\ \bibinfo {author} {\bibfnamefont {G.~A.}\ \bibnamefont {Pavliotis}},\ }\bibfield  {title} {\bibinfo {title} {Instability, rupture and fluctuations in thin liquid films: Theory and computations},\ }\href@noop {} {\bibfield  {journal} {\bibinfo  {journal} {Journal of Statistical Physics}\ }\textbf {\bibinfo {volume} {174}},\ \bibinfo {pages} {579–604} (\bibinfo {year} {2019})}\BibitemShut {NoStop}%
\bibitem [{\citenamefont {Davidovitch}\ \emph {et~al.}(2005)\citenamefont {Davidovitch}, \citenamefont {Moro},\ and\ \citenamefont {Stone}}]{davidovitch2005}%
  \BibitemOpen
  \bibfield  {author} {\bibinfo {author} {\bibfnamefont {B.}~\bibnamefont {Davidovitch}}, \bibinfo {author} {\bibfnamefont {E.}~\bibnamefont {Moro}},\ and\ \bibinfo {author} {\bibfnamefont {H.~A.}\ \bibnamefont {Stone}},\ }\bibfield  {title} {\bibinfo {title} {Spreading of viscous fluid drops on a solid substrate assisted by thermal fluctuations},\ }\href@noop {} {\bibfield  {journal} {\bibinfo  {journal} {Phys. Rev. Lett.}\ }\textbf {\bibinfo {volume} {95}},\ \bibinfo {pages} {244505} (\bibinfo {year} {2005})}\BibitemShut {NoStop}%
\bibitem [{\citenamefont {Cates}(2022)}]{CatesLesHouches}%
  \BibitemOpen
  \bibfield  {author} {\bibinfo {author} {\bibfnamefont {M.~E.}\ \bibnamefont {Cates}},\ }\bibfield  {title} {\bibinfo {title} {Active field theories},\ }in\ \href@noop {} {\emph {\bibinfo {booktitle} {Active Matter and Nonequilibrium Statistical Physics: Lecture Notes of the Les Houches Summer School: Volume 112, September 2018}}}\ (\bibinfo  {publisher} {Oxford University Press},\ \bibinfo {year} {2022})\BibitemShut {NoStop}%
\bibitem [{\citenamefont {Glasner}(2008)}]{Glasner2008}%
  \BibitemOpen
  \bibfield  {author} {\bibinfo {author} {\bibfnamefont {K.~B.}\ \bibnamefont {Glasner}},\ }\bibfield  {title} {\bibinfo {title} {Ostwald ripening in thin film equations},\ }\href {https://doi.org/10.1137/080713732} {\bibfield  {journal} {\bibinfo  {journal} {SIAM Journal on Applied Mathematics}\ }\textbf {\bibinfo {volume} {69}},\ \bibinfo {pages} {473} (\bibinfo {year} {2008})}\BibitemShut {NoStop}%
\bibitem [{\citenamefont {Gnann}\ and\ \citenamefont {Petrache}(2018)}]{GnannPetrache2018}%
  \BibitemOpen
  \bibfield  {author} {\bibinfo {author} {\bibfnamefont {M.}~\bibnamefont {Gnann}}\ and\ \bibinfo {author} {\bibfnamefont {M.}~\bibnamefont {Petrache}},\ }\bibfield  {title} {\bibinfo {title} {The navier-slip thin-film equation for 3d fluid films: Existence and uniqueness},\ }\href@noop {} {\bibfield  {journal} {\bibinfo  {journal} {Journal of Differential Equations}\ }\textbf {\bibinfo {volume} {265}} (\bibinfo {year} {2018})}\BibitemShut {NoStop}%
\bibitem [{\citenamefont {Zhang}\ \emph {et~al.}(2020)\citenamefont {Zhang}, \citenamefont {Sprittles},\ and\ \citenamefont {Lockerby}}]{ZhangSprittlesLockerby2020}%
  \BibitemOpen
  \bibfield  {author} {\bibinfo {author} {\bibfnamefont {Y.}~\bibnamefont {Zhang}}, \bibinfo {author} {\bibfnamefont {J.~E.}\ \bibnamefont {Sprittles}},\ and\ \bibinfo {author} {\bibfnamefont {D.~A.}\ \bibnamefont {Lockerby}},\ }\bibfield  {title} {\bibinfo {title} {Nanoscale thin-film flows with thermal fluctuations and slip},\ }\href@noop {} {\bibfield  {journal} {\bibinfo  {journal} {Phys. Rev. E}\ }\textbf {\bibinfo {volume} {102}},\ \bibinfo {pages} {053105} (\bibinfo {year} {2020})}\BibitemShut {NoStop}%
\bibitem [{\citenamefont {Landau}\ and\ \citenamefont {Lifshitz}(1986)}]{landau_lifshitz_elasticity}%
  \BibitemOpen
  \bibfield  {author} {\bibinfo {author} {\bibfnamefont {L.~D.}\ \bibnamefont {Landau}}\ and\ \bibinfo {author} {\bibfnamefont {E.~M.}\ \bibnamefont {Lifshitz}},\ }\href@noop {} {\emph {\bibinfo {title} {Course of theoretical physics: Theory of elasticity}}},\ \bibinfo {edition} {3rd}\ ed.,\ Vol.~\bibinfo {volume} {7}\ (\bibinfo  {publisher} {Butterworth-Heinemann},\ \bibinfo {year} {1986})\BibitemShut {NoStop}%
\bibitem [{\citenamefont {Howell}\ \emph {et~al.}(2008)\citenamefont {Howell}, \citenamefont {Kozyreff},\ and\ \citenamefont {Ockendon}}]{Howell_Kozyreff_Ockendon_2008}%
  \BibitemOpen
  \bibfield  {author} {\bibinfo {author} {\bibfnamefont {P.}~\bibnamefont {Howell}}, \bibinfo {author} {\bibfnamefont {G.}~\bibnamefont {Kozyreff}},\ and\ \bibinfo {author} {\bibfnamefont {J.}~\bibnamefont {Ockendon}},\ }\href@noop {} {\emph {\bibinfo {title} {Applied Solid Mechanics}}},\ Cambridge Texts in Applied Mathematics\ (\bibinfo  {publisher} {Cambridge University Press},\ \bibinfo {year} {2008})\BibitemShut {NoStop}%
\bibitem [{\citenamefont {Evans}(1985)}]{Evans1985}%
  \BibitemOpen
  \bibfield  {author} {\bibinfo {author} {\bibfnamefont {E.}~\bibnamefont {Evans}},\ }\bibfield  {title} {\bibinfo {title} {Detailed mechanics of membrane-membrane adhesion and separation. i. continuum of molecular cross-bridges},\ }\href@noop {} {\bibfield  {journal} {\bibinfo  {journal} {Biophysical journal}\ }\textbf {\bibinfo {volume} {48}},\ \bibinfo {pages} {175} (\bibinfo {year} {1985})}\BibitemShut {NoStop}%
\bibitem [{\citenamefont {Kodio}\ \emph {et~al.}(2017)\citenamefont {Kodio}, \citenamefont {Griffiths},\ and\ \citenamefont {Vella}}]{KodioVella2017}%
  \BibitemOpen
  \bibfield  {author} {\bibinfo {author} {\bibfnamefont {O.}~\bibnamefont {Kodio}}, \bibinfo {author} {\bibfnamefont {I.~M.}\ \bibnamefont {Griffiths}},\ and\ \bibinfo {author} {\bibfnamefont {D.}~\bibnamefont {Vella}},\ }\bibfield  {title} {\bibinfo {title} {Lubricated wrinkles: Imposed constraints affect the dynamics of wrinkle coarsening},\ }\href@noop {} {\bibfield  {journal} {\bibinfo  {journal} {Phys. Rev. Fluids}\ }\textbf {\bibinfo {volume} {2}},\ \bibinfo {pages} {014202} (\bibinfo {year} {2017})}\BibitemShut {NoStop}%
\bibitem [{\citenamefont {Audoly}\ and\ \citenamefont {Pomeau}(2008)}]{audolyetPomeau}%
  \BibitemOpen
  \bibfield  {author} {\bibinfo {author} {\bibfnamefont {B.}~\bibnamefont {Audoly}}\ and\ \bibinfo {author} {\bibfnamefont {Y.}~\bibnamefont {Pomeau}},\ }\bibinfo {title} {Elasticty and geometry: from hair curls to the nonlinear response of shells}\ (\bibinfo {year} {2008})\ p.\ \bibinfo {pages} {Oxford University Press}\BibitemShut {NoStop}%
\bibitem [{\citenamefont {Bell}\ \emph {et~al.}(1984)\citenamefont {Bell}, \citenamefont {Dembo},\ and\ \citenamefont {Bongrand}}]{BellDemboBongrand}%
  \BibitemOpen
  \bibfield  {author} {\bibinfo {author} {\bibfnamefont {G.}~\bibnamefont {Bell}}, \bibinfo {author} {\bibfnamefont {M.}~\bibnamefont {Dembo}},\ and\ \bibinfo {author} {\bibfnamefont {P.}~\bibnamefont {Bongrand}},\ }\bibfield  {title} {\bibinfo {title} {Cell adhesion. competition between nonspecific repulsion and specific bonding},\ }\href@noop {} {\bibfield  {journal} {\bibinfo  {journal} {Biophysical Journal}\ }\textbf {\bibinfo {volume} {45}},\ \bibinfo {pages} {1051} (\bibinfo {year} {1984})}\BibitemShut {NoStop}%
\bibitem [{\citenamefont {Dhaliwal}\ \emph {et~al.}(2024)\citenamefont {Dhaliwal}, \citenamefont {Pedersen}, \citenamefont {Kadri}, \citenamefont {Miquelard-Garnier}, \citenamefont {Sollogoub}, \citenamefont {Peixinho}, \citenamefont {Salez},\ and\ \citenamefont {Carlson}}]{dhaliwal2024instability}%
  \BibitemOpen
  \bibfield  {author} {\bibinfo {author} {\bibfnamefont {V.}~\bibnamefont {Dhaliwal}}, \bibinfo {author} {\bibfnamefont {C.}~\bibnamefont {Pedersen}}, \bibinfo {author} {\bibfnamefont {K.}~\bibnamefont {Kadri}}, \bibinfo {author} {\bibfnamefont {G.}~\bibnamefont {Miquelard-Garnier}}, \bibinfo {author} {\bibfnamefont {C.}~\bibnamefont {Sollogoub}}, \bibinfo {author} {\bibfnamefont {J.}~\bibnamefont {Peixinho}}, \bibinfo {author} {\bibfnamefont {T.}~\bibnamefont {Salez}},\ and\ \bibinfo {author} {\bibfnamefont {A.}~\bibnamefont {Carlson}},\ }\bibfield  {title} {\bibinfo {title} {Instability and rupture of sheared viscous liquid nanofilms},\ }\href@noop {} {\bibfield  {journal} {\bibinfo  {journal} {Physical Review Fluids}\ }\textbf {\bibinfo {volume} {9}},\ \bibinfo {pages} {024201} (\bibinfo {year} {2024})}\BibitemShut {NoStop}%
\bibitem [{\citenamefont {Carlson}(2018)}]{carlson_2018}%
  \BibitemOpen
  \bibfield  {author} {\bibinfo {author} {\bibfnamefont {A.}~\bibnamefont {Carlson}},\ }\bibfield  {title} {\bibinfo {title} {Fluctuation assisted spreading of a fluid filled elastic blister},\ }\href {https://doi.org/10.1017/jfm.2018.288} {\bibfield  {journal} {\bibinfo  {journal} {Journal of Fluid Mechanics}\ }\textbf {\bibinfo {volume} {846}},\ \bibinfo {pages} {1076–1087} (\bibinfo {year} {2018})}\BibitemShut {NoStop}%
\bibitem [{\citenamefont {Dhaliwal}(2025)}]{GithubVira}%
  \BibitemOpen
  \bibfield  {author} {\bibinfo {author} {\bibfnamefont {V.}~\bibnamefont {Dhaliwal}},\ }\href {https://github.com/vir-a/ViscMembProt} {\bibinfo {title} {Viscmembprot --- {GitHub}}} (\bibinfo {year} {2025}),\ \bibinfo {note} {accessed: 2025-05-06}\BibitemShut {NoStop}%
\bibitem [{\citenamefont {Logg}\ \emph {et~al.}(2012)\citenamefont {Logg}, \citenamefont {Mardal}, \citenamefont {Wells} \emph {et~al.}}]{LoggMardalEtAl2012}%
  \BibitemOpen
  \bibfield  {author} {\bibinfo {author} {\bibfnamefont {A.}~\bibnamefont {Logg}}, \bibinfo {author} {\bibfnamefont {K.-A.}\ \bibnamefont {Mardal}}, \bibinfo {author} {\bibfnamefont {G.~N.}\ \bibnamefont {Wells}}, \emph {et~al.},\ }\href@noop {} {\emph {\bibinfo {title} {Automated Solution of Differential Equations by the Finite Element Method}}},\ edited by\ \bibinfo {editor} {\bibfnamefont {A.}~\bibnamefont {Logg}}, \bibinfo {editor} {\bibfnamefont {K.-A.}\ \bibnamefont {Mardal}},\ and\ \bibinfo {editor} {\bibfnamefont {G.~N.}\ \bibnamefont {Wells}}\ (\bibinfo  {publisher} {Springer},\ \bibinfo {year} {2012})\BibitemShut {NoStop}%
\bibitem [{\citenamefont {Oliphant}(2006)}]{OliphantNumpy2006}%
  \BibitemOpen
  \bibfield  {author} {\bibinfo {author} {\bibfnamefont {T.}~\bibnamefont {Oliphant}},\ }\href@noop {} {\emph {\bibinfo {title} {Guide to NumPy}}}\ (\bibinfo {year} {2006})\BibitemShut {NoStop}%
\bibitem [{\citenamefont {Helfrich}\ and\ \citenamefont {Servuss}(1984)}]{helfrichUndulationsStericInteraction1984}%
  \BibitemOpen
  \bibfield  {author} {\bibinfo {author} {\bibfnamefont {W.}~\bibnamefont {Helfrich}}\ and\ \bibinfo {author} {\bibfnamefont {R.~M.}\ \bibnamefont {Servuss}},\ }\bibfield  {title} {\bibinfo {title} {Undulations, steric interaction and cohesion of fluid membranes},\ }\href {https://doi.org/10.1007/BF02452208} {\bibfield  {journal} {\bibinfo  {journal} {Il Nuovo Cimento D}\ }\textbf {\bibinfo {volume} {3}},\ \bibinfo {pages} {137} (\bibinfo {year} {1984})}\BibitemShut {NoStop}%
\bibitem [{\citenamefont {Rautu}\ \emph {et~al.}(2017)\citenamefont {Rautu}, \citenamefont {Orsi}, \citenamefont {Michele}, \citenamefont {Rowlands}, \citenamefont {Cicuta},\ and\ \citenamefont {Turner}}]{rautuRoleOpticalProjection2017}%
  \BibitemOpen
  \bibfield  {author} {\bibinfo {author} {\bibfnamefont {S.~A.}\ \bibnamefont {Rautu}}, \bibinfo {author} {\bibfnamefont {D.}~\bibnamefont {Orsi}}, \bibinfo {author} {\bibfnamefont {L.~D.}\ \bibnamefont {Michele}}, \bibinfo {author} {\bibfnamefont {G.}~\bibnamefont {Rowlands}}, \bibinfo {author} {\bibfnamefont {P.}~\bibnamefont {Cicuta}},\ and\ \bibinfo {author} {\bibfnamefont {M.~S.}\ \bibnamefont {Turner}},\ }\bibfield  {title} {\bibinfo {title} {The role of optical projection in the analysis of membrane fluctuations},\ }\href {https://doi.org/10.1039/C7SM00108H} {\bibfield  {journal} {\bibinfo  {journal} {Soft Matter}\ }\textbf {\bibinfo {volume} {13}},\ \bibinfo {pages} {3480} (\bibinfo {year} {2017})}\BibitemShut {NoStop}%
\bibitem [{\citenamefont {Eyring}(1935)}]{Eyring1935}%
  \BibitemOpen
  \bibfield  {author} {\bibinfo {author} {\bibfnamefont {H.}~\bibnamefont {Eyring}},\ }\bibfield  {title} {\bibinfo {title} {The activated complex in chemical reactions},\ }\href@noop {} {\bibfield  {journal} {\bibinfo  {journal} {The Journal of Chemical Physics}\ }\textbf {\bibinfo {volume} {3}},\ \bibinfo {pages} {107} (\bibinfo {year} {1935})}\BibitemShut {NoStop}%
\bibitem [{\citenamefont {Kramers}(1940)}]{KRAMERS1940}%
  \BibitemOpen
  \bibfield  {author} {\bibinfo {author} {\bibfnamefont {H.}~\bibnamefont {Kramers}},\ }\bibfield  {title} {\bibinfo {title} {Brownian motion in a field of force and the diffusion model of chemical reactions},\ }\href@noop {} {\bibfield  {journal} {\bibinfo  {journal} {Physica}\ }\textbf {\bibinfo {volume} {7}},\ \bibinfo {pages} {284} (\bibinfo {year} {1940})}\BibitemShut {NoStop}%
\bibitem [{\citenamefont {Hohenberg}\ and\ \citenamefont {Halperin}(1977)}]{HohenbergHalperin1977}%
  \BibitemOpen
  \bibfield  {author} {\bibinfo {author} {\bibfnamefont {P.~C.}\ \bibnamefont {Hohenberg}}\ and\ \bibinfo {author} {\bibfnamefont {B.~I.}\ \bibnamefont {Halperin}},\ }\bibfield  {title} {\bibinfo {title} {Theory of dynamic critical phenomena},\ }\href@noop {} {\bibfield  {journal} {\bibinfo  {journal} {Rev. Mod. Phys.}\ }\textbf {\bibinfo {volume} {49}},\ \bibinfo {pages} {435} (\bibinfo {year} {1977})}\BibitemShut {NoStop}%
\bibitem [{\citenamefont {Furukawa}(1985)}]{Furukawa1985}%
  \BibitemOpen
  \bibfield  {author} {\bibinfo {author} {\bibfnamefont {H.}~\bibnamefont {Furukawa}},\ }\bibfield  {title} {\bibinfo {title} {A dynamic scaling assumption for phase separation},\ }\href@noop {} {\bibfield  {journal} {\bibinfo  {journal} {Advances in Physics}\ }\textbf {\bibinfo {volume} {34}},\ \bibinfo {pages} {703} (\bibinfo {year} {1985})}\BibitemShut {NoStop}%
\bibitem [{\citenamefont {Camley}\ and\ \citenamefont {Brown}(2011)}]{CamleyBrown2011}%
  \BibitemOpen
  \bibfield  {author} {\bibinfo {author} {\bibfnamefont {B.~A.}\ \bibnamefont {Camley}}\ and\ \bibinfo {author} {\bibfnamefont {F.~L.~H.}\ \bibnamefont {Brown}},\ }\bibfield  {title} {\bibinfo {title} {Dynamic scaling in phase separation kinetics for quasi-two-dimensional membranes},\ }\href@noop {} {\bibfield  {journal} {\bibinfo  {journal} {The Journal of Chemical Physics}\ }\textbf {\bibinfo {volume} {135}},\ \bibinfo {pages} {225106} (\bibinfo {year} {2011})}\BibitemShut {NoStop}%
\bibitem [{\citenamefont {McLeish}\ \emph {et~al.}(2003)\citenamefont {McLeish}, \citenamefont {Cates}, \citenamefont {Higgins}, \citenamefont {Olmsted},\ and\ \citenamefont {Bray}}]{Bray2003}%
  \BibitemOpen
  \bibfield  {author} {\bibinfo {author} {\bibfnamefont {T.~C.~B.}\ \bibnamefont {McLeish}}, \bibinfo {author} {\bibfnamefont {M.~E.}\ \bibnamefont {Cates}}, \bibinfo {author} {\bibfnamefont {J.~S.}\ \bibnamefont {Higgins}}, \bibinfo {author} {\bibfnamefont {P.~D.}\ \bibnamefont {Olmsted}},\ and\ \bibinfo {author} {\bibfnamefont {A.~J.}\ \bibnamefont {Bray}},\ }\bibfield  {title} {\bibinfo {title} {Coarsening dynamics of phase-separating systems},\ }\href {https://royalsocietypublishing.org/doi/abs/10.1098/rsta.2002.1164} {\bibfield  {journal} {\bibinfo  {journal} {Philosophical Transactions of the Royal Society of London. Series A: Mathematical, Physical and Engineering Sciences}\ }\textbf {\bibinfo {volume} {361}},\ \bibinfo {pages} {781} (\bibinfo {year} {2003})}\BibitemShut {NoStop}%
\bibitem [{\citenamefont {Shinozaki}\ and\ \citenamefont {Oono}(1993)}]{ShinozakiOono1993}%
  \BibitemOpen
  \bibfield  {author} {\bibinfo {author} {\bibfnamefont {A.}~\bibnamefont {Shinozaki}}\ and\ \bibinfo {author} {\bibfnamefont {Y.}~\bibnamefont {Oono}},\ }\bibfield  {title} {\bibinfo {title} {Spinodal decomposition in 3-space},\ }\href@noop {} {\bibfield  {journal} {\bibinfo  {journal} {Phys. Rev. E}\ }\textbf {\bibinfo {volume} {48}},\ \bibinfo {pages} {2622} (\bibinfo {year} {1993})}\BibitemShut {NoStop}%
\bibitem [{\citenamefont {Sung}\ \emph {et~al.}(1996)\citenamefont {Sung}, \citenamefont {Karim}, \citenamefont {Douglas},\ and\ \citenamefont {Han}}]{SungHan1995}%
  \BibitemOpen
  \bibfield  {author} {\bibinfo {author} {\bibfnamefont {L.}~\bibnamefont {Sung}}, \bibinfo {author} {\bibfnamefont {A.}~\bibnamefont {Karim}}, \bibinfo {author} {\bibfnamefont {J.~F.}\ \bibnamefont {Douglas}},\ and\ \bibinfo {author} {\bibfnamefont {C.~C.}\ \bibnamefont {Han}},\ }\bibfield  {title} {\bibinfo {title} {Dimensional crossover in the phase separation kinetics of thin polymer blend films},\ }\href {https://doi.org/10.1103/PhysRevLett.76.4368} {\bibfield  {journal} {\bibinfo  {journal} {Phys. Rev. Lett.}\ }\textbf {\bibinfo {volume} {76}},\ \bibinfo {pages} {4368} (\bibinfo {year} {1996})}\BibitemShut {NoStop}%
\bibitem [{\citenamefont {Tateno}\ and\ \citenamefont {Tanaka}(2021)}]{TatenoTanaka2021}%
  \BibitemOpen
  \bibfield  {author} {\bibinfo {author} {\bibfnamefont {M.}~\bibnamefont {Tateno}}\ and\ \bibinfo {author} {\bibfnamefont {H.}~\bibnamefont {Tanaka}},\ }\bibfield  {title} {\bibinfo {title} {Power-law coarsening in network-forming phase separation governed by mechanical relaxation},\ }\href@noop {} {\bibfield  {journal} {\bibinfo  {journal} {Nature Communications}\ }\textbf {\bibinfo {volume} {12}},\ \bibinfo {pages} {912} (\bibinfo {year} {2021})}\BibitemShut {NoStop}%
\bibitem [{\citenamefont {Saiseau}\ \emph {et~al.}(2024)\citenamefont {Saiseau}, \citenamefont {Truong}, \citenamefont {Guérin}, \citenamefont {Delabre},\ and\ \citenamefont {Delville}}]{SaiseauDelville2024}%
  \BibitemOpen
  \bibfield  {author} {\bibinfo {author} {\bibfnamefont {R.}~\bibnamefont {Saiseau}}, \bibinfo {author} {\bibfnamefont {H.}~\bibnamefont {Truong}}, \bibinfo {author} {\bibfnamefont {T.}~\bibnamefont {Guérin}}, \bibinfo {author} {\bibfnamefont {U.}~\bibnamefont {Delabre}},\ and\ \bibinfo {author} {\bibfnamefont {J.-P.}\ \bibnamefont {Delville}},\ }\bibfield  {title} {\bibinfo {title} {Decay dynamics of a single spherical domain in near-critical phase-separated conditions},\ }\href@noop {} {\bibfield  {journal} {\bibinfo  {journal} {Physical Review Letters}\ }\textbf {\bibinfo {volume} {133}} (\bibinfo {year} {2024})}\BibitemShut {NoStop}%
\bibitem [{\citenamefont {Vladimirova}\ \emph {et~al.}(1998)\citenamefont {Vladimirova}, \citenamefont {Malagoli},\ and\ \citenamefont {Mauri}}]{VladimirovaMauri1998}%
  \BibitemOpen
  \bibfield  {author} {\bibinfo {author} {\bibfnamefont {N.}~\bibnamefont {Vladimirova}}, \bibinfo {author} {\bibfnamefont {A.}~\bibnamefont {Malagoli}},\ and\ \bibinfo {author} {\bibfnamefont {R.}~\bibnamefont {Mauri}},\ }\bibfield  {title} {\bibinfo {title} {Diffusion-driven phase separation of deeply quenched mixtures},\ }\href@noop {} {\bibfield  {journal} {\bibinfo  {journal} {Phys. Rev. E}\ }\textbf {\bibinfo {volume} {58}},\ \bibinfo {pages} {7691} (\bibinfo {year} {1998})}\BibitemShut {NoStop}%
\bibitem [{\citenamefont {Tiribocchi}\ \emph {et~al.}(2015)\citenamefont {Tiribocchi}, \citenamefont {Wittkowski}, \citenamefont {Marenduzzo},\ and\ \citenamefont {Cates}}]{TiribocchiCates2015}%
  \BibitemOpen
  \bibfield  {author} {\bibinfo {author} {\bibfnamefont {A.}~\bibnamefont {Tiribocchi}}, \bibinfo {author} {\bibfnamefont {R.}~\bibnamefont {Wittkowski}}, \bibinfo {author} {\bibfnamefont {D.}~\bibnamefont {Marenduzzo}},\ and\ \bibinfo {author} {\bibfnamefont {M.~E.}\ \bibnamefont {Cates}},\ }\bibfield  {title} {\bibinfo {title} {Active model h: Scalar active matter in a momentum-conserving fluid},\ }\href@noop {} {\bibfield  {journal} {\bibinfo  {journal} {Phys. Rev. Lett.}\ }\textbf {\bibinfo {volume} {115}},\ \bibinfo {pages} {188302} (\bibinfo {year} {2015})}\BibitemShut {NoStop}%
\bibitem [{\citenamefont {Wagner}(1961)}]{Wagner1961}%
  \BibitemOpen
  \bibfield  {author} {\bibinfo {author} {\bibfnamefont {C.}~\bibnamefont {Wagner}},\ }\bibfield  {title} {\bibinfo {title} {Theorie der alterung von niederschl{\"a}gen durch uml{\"o}sen (ostwald-reifung)},\ }\href@noop {} {\bibfield  {journal} {\bibinfo  {journal} {Zeitschrift f{\"u}r Elektrochemie, Berichte der Bunsengesellschaft f{\"u}r physikalische Chemie}\ }\textbf {\bibinfo {volume} {65}},\ \bibinfo {pages} {581} (\bibinfo {year} {1961})}\BibitemShut {NoStop}%
\bibitem [{\citenamefont {Lifshitz}\ and\ \citenamefont {Slyozov}(1961)}]{LIFSHITZSlyozov1961}%
  \BibitemOpen
  \bibfield  {author} {\bibinfo {author} {\bibfnamefont {I.}~\bibnamefont {Lifshitz}}\ and\ \bibinfo {author} {\bibfnamefont {V.}~\bibnamefont {Slyozov}},\ }\bibfield  {title} {\bibinfo {title} {The kinetics of precipitation from supersaturated solid solutions},\ }\href@noop {} {\bibfield  {journal} {\bibinfo  {journal} {Journal of Physics and Chemistry of Solids}\ }\textbf {\bibinfo {volume} {19}},\ \bibinfo {pages} {35} (\bibinfo {year} {1961})}\BibitemShut {NoStop}%
\bibitem [{\citenamefont {Chou}\ and\ \citenamefont {D'Orsogna}(2014)}]{ChouDorsogna2014}%
  \BibitemOpen
  \bibfield  {author} {\bibinfo {author} {\bibfnamefont {T.}~\bibnamefont {Chou}}\ and\ \bibinfo {author} {\bibfnamefont {M.~R.}\ \bibnamefont {D'Orsogna}},\ }\bibinfo {title} {First passage problems in biology},\ in\ \href@noop {} {\emph {\bibinfo {booktitle} {First-Passage Phenomena and Their Applications}}}\ (\bibinfo  {publisher} {World Scientific},\ \bibinfo {year} {2014})\ Chap.\ \bibinfo {chapter} {Chapter 1}, pp.\ \bibinfo {pages} {306--345}\BibitemShut {NoStop}%
\bibitem [{\citenamefont {{L H Tanner}}(1979)}]{lhtannerSpreadingSiliconeOil1979}%
  \BibitemOpen
  \bibfield  {author} {\bibinfo {author} {\bibnamefont {{L H Tanner}}},\ }\bibfield  {title} {\bibinfo {title} {The spreading of silicone oil drops on horizontal surfaces},\ }\href {https://doi.org/10.1088/0022-3727/12/9/009} {\bibfield  {journal} {\bibinfo  {journal} {Journal of Physics D: Applied Physics}\ }\textbf {\bibinfo {volume} {12}},\ \bibinfo {pages} {1473} (\bibinfo {year} {1979})}\BibitemShut {NoStop}%
\bibitem [{\citenamefont {Lister}\ \emph {et~al.}(2013)\citenamefont {Lister}, \citenamefont {Peng},\ and\ \citenamefont {Neufeld}}]{listerViscousControlPeeling2013}%
  \BibitemOpen
  \bibfield  {author} {\bibinfo {author} {\bibfnamefont {J.~R.}\ \bibnamefont {Lister}}, \bibinfo {author} {\bibfnamefont {G.~G.}\ \bibnamefont {Peng}},\ and\ \bibinfo {author} {\bibfnamefont {J.~A.}\ \bibnamefont {Neufeld}},\ }\bibfield  {title} {\bibinfo {title} {Viscous {{Control}} of {{Peeling}} an {{Elastic Sheet}} by {{Bending}} and {{Pulling}}},\ }\href {https://doi.org/10.1103/PhysRevLett.111.154501} {\bibfield  {journal} {\bibinfo  {journal} {Physical Review Letters}\ }\textbf {\bibinfo {volume} {111}},\ \bibinfo {pages} {154501} (\bibinfo {year} {2013})}\BibitemShut {NoStop}%
\bibitem [{\citenamefont {Pedersen}\ \emph {et~al.}(2019)\citenamefont {Pedersen}, \citenamefont {Niven}, \citenamefont {Salez}, \citenamefont {Dalnoki-Veress},\ and\ \citenamefont {Carlson}}]{pedersen_niven_salez_dalnoki-veress_carlson_2019}%
  \BibitemOpen
  \bibfield  {author} {\bibinfo {author} {\bibfnamefont {C.}~\bibnamefont {Pedersen}}, \bibinfo {author} {\bibfnamefont {J.~F.}\ \bibnamefont {Niven}}, \bibinfo {author} {\bibfnamefont {T.}~\bibnamefont {Salez}}, \bibinfo {author} {\bibfnamefont {K.}~\bibnamefont {Dalnoki-Veress}},\ and\ \bibinfo {author} {\bibfnamefont {A.}~\bibnamefont {Carlson}},\ }\bibfield  {title} {\bibinfo {title} {Asymptotic regimes in elastohydrodynamic and stochastic leveling on a viscous film},\ }\href@noop {} {\bibfield  {journal} {\bibinfo  {journal} {Physical Review Fluids}\ }\textbf {\bibinfo {volume} {4}} (\bibinfo {year} {2019})}\BibitemShut {NoStop}%
\bibitem [{\citenamefont {Limary}\ and\ \citenamefont {Green}(2003)}]{LimaryGreen2003}%
  \BibitemOpen
  \bibfield  {author} {\bibinfo {author} {\bibfnamefont {R.}~\bibnamefont {Limary}}\ and\ \bibinfo {author} {\bibfnamefont {P.~F.}\ \bibnamefont {Green}},\ }\bibfield  {title} {\bibinfo {title} {Dynamics of droplets on the surface of a structured fluid film: Late-stage coarsening},\ }\href@noop {} {\bibfield  {journal} {\bibinfo  {journal} {Langmuir}\ }\textbf {\bibinfo {volume} {19}},\ \bibinfo {pages} {2419} (\bibinfo {year} {2003})}\BibitemShut {NoStop}%
\bibitem [{\citenamefont {Mani}\ \emph {et~al.}(2012)\citenamefont {Mani}, \citenamefont {Gopinath},\ and\ \citenamefont {Mahadevan}}]{ManiMaha2012}%
  \BibitemOpen
  \bibfield  {author} {\bibinfo {author} {\bibfnamefont {M.}~\bibnamefont {Mani}}, \bibinfo {author} {\bibfnamefont {A.}~\bibnamefont {Gopinath}},\ and\ \bibinfo {author} {\bibfnamefont {L.}~\bibnamefont {Mahadevan}},\ }\bibfield  {title} {\bibinfo {title} {How things get stuck: Kinetics, elastohydrodynamics, and soft adhesion},\ }\href@noop {} {\bibfield  {journal} {\bibinfo  {journal} {Phys. Rev. Lett.}\ }\textbf {\bibinfo {volume} {108}},\ \bibinfo {pages} {226104} (\bibinfo {year} {2012})}\BibitemShut {NoStop}%
\bibitem [{\citenamefont {Grafke}\ \emph {et~al.}(2024)\citenamefont {Grafke}, \citenamefont {Sch{\"a}fer},\ and\ \citenamefont {{Vanden-Eijnden}}}]{grafkeSharpAsymptoticEstimates2024}%
  \BibitemOpen
  \bibfield  {author} {\bibinfo {author} {\bibfnamefont {T.}~\bibnamefont {Grafke}}, \bibinfo {author} {\bibfnamefont {T.}~\bibnamefont {Sch{\"a}fer}},\ and\ \bibinfo {author} {\bibfnamefont {E.}~\bibnamefont {{Vanden-Eijnden}}},\ }\bibfield  {title} {\bibinfo {title} {Sharp asymptotic estimates for expectations, probabilities, and mean first passage times in stochastic systems with small noise},\ }\href {https://doi.org/10.1002/cpa.22177} {\bibfield  {journal} {\bibinfo  {journal} {Communications on Pure and Applied Mathematics}\ }\textbf {\bibinfo {volume} {77}},\ \bibinfo {pages} {2268} (\bibinfo {year} {2024})}\BibitemShut {NoStop}%
\bibitem [{\citenamefont {Gardiner}(2009)}]{gardiner2009stochastic}%
  \BibitemOpen
  \bibfield  {author} {\bibinfo {author} {\bibfnamefont {C.}~\bibnamefont {Gardiner}},\ }\href@noop {} {\emph {\bibinfo {title} {Stochastic Methods: A Handbook for the Natural and Social Sciences}}},\ Springer Series in Synergetics\ (\bibinfo  {publisher} {Springer Berlin Heidelberg},\ \bibinfo {year} {2009})\BibitemShut {NoStop}%
\bibitem [{\citenamefont {Freidlin}\ \emph {et~al.}(2012)\citenamefont {Freidlin}, \citenamefont {Szucs},\ and\ \citenamefont {Wentzell}}]{freidlinRandomPerturbationsDynamical2012}%
  \BibitemOpen
  \bibfield  {author} {\bibinfo {author} {\bibfnamefont {M.}~\bibnamefont {Freidlin}}, \bibinfo {author} {\bibfnamefont {J.}~\bibnamefont {Szucs}},\ and\ \bibinfo {author} {\bibfnamefont {A.}~\bibnamefont {Wentzell}},\ }\href@noop {} {\emph {\bibinfo {title} {Random Perturbations of Dynamical Systems}}},\ Grundlehren Der Mathematischen Wissenschaften\ (\bibinfo  {publisher} {Springer New York},\ \bibinfo {year} {2012})\BibitemShut {NoStop}%
\bibitem [{Note1()}]{Note1}%
  \BibitemOpen
  \bibinfo {note} {For calculation of the functional derivative, see~\protect \citep {SprittlesJBLGrafke2023,liuMeanFirstPassage2024}.}\BibitemShut {Stop}%
\bibitem [{\citenamefont {Boas}(2006)}]{Boas2006}%
  \BibitemOpen
  \bibfield  {author} {\bibinfo {author} {\bibfnamefont {M.~L.}\ \bibnamefont {Boas}},\ }\href {https://cds.cern.ch/record/913305} {\emph {\bibinfo {title} {Mathematical methods in the physical sciences; 3rd ed.}}}\ (\bibinfo  {publisher} {Wiley},\ \bibinfo {address} {Hoboken, NJ},\ \bibinfo {year} {2006})\BibitemShut {NoStop}%
\bibitem [{\citenamefont {Furukawa}(2000)}]{Furukawa1999}%
  \BibitemOpen
  \bibfield  {author} {\bibinfo {author} {\bibfnamefont {H.}~\bibnamefont {Furukawa}},\ }\bibfield  {title} {\bibinfo {title} {Spinodal decomposition of two-dimensional fluid mixtures: A spectral analysis of droplet growth},\ }\href@noop {} {\bibfield  {journal} {\bibinfo  {journal} {Phys. Rev. E}\ }\textbf {\bibinfo {volume} {61}},\ \bibinfo {pages} {1423} (\bibinfo {year} {2000})}\BibitemShut {NoStop}%
\end{thebibliography}%

\end{document}